\documentclass[lineno]{jfm}

\usepackage{graphicx}
\usepackage{newtxtext}
\usepackage{newtxmath}
\usepackage{natbib}
\usepackage{hyperref}
\hypersetup{
    colorlinks = true,
    urlcolor   = blue,
    citecolor  = black,
}

\newcommand{\RomanNumeralCaps}[1]
\linenumbers

\usepackage{siunitx}
\usepackage{bm}
\usepackage{xcolor}
\usepackage{booktabs}
\usepackage{caption}
\usepackage{subcaption}
\usepackage{changepage}
\usepackage{multirow}
\usepackage{rotating}
\usepackage{adjustbox}
\usepackage{comment}
\usepackage{float}
\usepackage{tabularray}
\usepackage{todonotes}

\usepackage{tikz}
\usepackage{svg}
\usepackage{pgfplots}
\graphicspath{{figures/}}
\usetikzlibrary{decorations.markings}
\newlength\fwidth
\usetikzlibrary{backgrounds}
\usepgfplotslibrary{polar}
\usetikzlibrary{shapes.geometric}
\usetikzlibrary[pgfplots.groupplots]

\def\pgftransform@angle{0}

\definecolor{darkturquoise0191191}{RGB}{0,191,191}
\definecolor{gray}{RGB}{128,128,128}
\definecolor{green01270}{RGB}{0,127,0}
\definecolor{orange}{HTML}{e79856}
\definecolor{burgundy}{HTML}{93003a}
\definecolor{G-1.4-30}{HTML}{d7191c}
\definecolor{G-0.6-30}{HTML}{8dd3c7}
\definecolor{G-1-30-rp6k}{HTML}{fdae61}
\definecolor{G-1-30}{HTML}{211a2b}
\definecolor{G-1-50}{HTML}{2c7bb6}
\definecolor{eg16}{RGB}{0,127,0}
\definecolor{eg6}{HTML}{FF00FF}
\definecolor{G-2.4-70}{HTML}{949f36}
\definecolor{SIMcol}{rgb}{0.00000,0.44700,0.74100}%
\definecolor{STcol}{rgb}{0.85000,0.32500,0.09800}%
\definecolor{Acol}{rgb}{0.49400,0.18400,0.55600}%
\definecolor{Zcol}{rgb}{0.30100,0.74500,0.93300}%
\definecolor{KKcol}{rgb}{0.46600,0.67400,0.18800}%
\definecolor{NMcol}{rgb}{0,0,0}%
\definecolor{Kcol}{rgb}{0.63500,0.07800,0.18400}%
\definecolor{Effcol}{rgb}{0.289,0.402,0.2539} 
\definecolor{Dcol}{RGB}{128,128,128}
\definecolor{Ngocol}{HTML}{211a2b}
\definecolor{pink}{HTML}{bc5090}
\definecolor{orange2}{HTML}{ffa600}
\definecolor{gray}{RGB}{128,128,128}
\definecolor{darkblue}{HTML}{003f5c}

\newcommand{\urms}{u_\mathrm{rms}}
\newcommand{\diff}{\mathrm{d}}
\newcommand{\wrms}{w_{r,\mathrm{rms}}}
\newcommand{\tke}{k}
\newcommand{\kolmlen}{\eta}
\newcommand{\tdr}{\varepsilon}
\newcommand{\kinvisc}{\nu_f}
\newcommand{\dynvisc}{\mu_f}

\title{An integrated multi-size collision model for flotation}

\author{Benedikt Tiedemann
  \corresp{\email{benedikt.tiedemann@tu-dresden.de}},
  Moritz Kreuseler
 \and Jochen Fröhlich}

\affiliation{Institute of Fluid Mechanics, TU Dresden, George-Bähr-Straße 3c, 01062 Dresden, Germany}

\begin{document}
\maketitle

\begin{abstract}
The accuracy obtained with CFD and process simulations of flotation critically depends on the quality and robustness of the underlying models for the non-resolved sub-processes. An important issue in flotation is the collision between particles and air bubbles. Many models have been developed, but their accuracy for applications in flotation is limited. In particular, the significant size difference between particles and bubbles and their intricate coupling to the turbulent flow field pose severe challenges. The present paper first reviews presently employed collision models, highlighting their advantages and disadvantages when applied to flotation. On this basis, the "Integrated Multi-Size Collision model" (IMSC) is proposed. After a detailed evaluation, it combines existing approaches from various sources and introduces new developments designed to address present shortcomings. The model is validated by own DNS data as well as data from the literature.
It is shown that, overall, the IMSC provides better predictions for the collision rate in typical flotation conditions than presently employed collision models and covers the entire parameter range of the flotation process very well. Using the available data, some of the underlying modelling assumptions are validated. Finally, a comprehensive overview of the model is provided for further use in Euler-Euler frameworks or process simulations.

\end{abstract}

\section{Introduction}

Froth flotation is a key process in the production of raw materials such as copper, gold, and rare earths needed for a vast range of technological products.
This physico-chemical process is presently almost the only one used to separate these valuable minerals from other gangue materials \citep{Yarar2000, Nguyen2004, Fuerstenau2007}.
Small and finely ground ore particles are fed into large flotation cells that are filled with water.
In the case of mechanical flotation cells, air bubbles are injected, and the slurry is agitated by a mechanical rotor.
The separation is achieved by the difference in hydrophobicity between the valuable minerals and the gangue material, with hydrophobic particles attaching to air bubbles and being transported to a surface froth, where they are recovered.
Unwanted, hydrophilic material does not attach and settles to the bottom of the flotation cell.
Chemical reagents serve to enhance the disparity in hydrophobicity among the particles, and to adjust the properties of the froth thereby optimising the operational conditions of the process \citep{Nguyen2004}.

The industrial application of the flotation process presents significant challenges.
For example, the ore grades encountered in mining today range from single-digit percentages to even lower values \citep{Crowson2012, Calvo2016}.
Furthermore, the demand for raw materials is increasing due to the shift towards greener and more environmentally friendly technologies \citep{Vidal2013}.
In particular, the availability of 17 minerals, including copper, silver, and titanium, has been identified as critical to the success of this transition \citep{WBG2017, Hund2020, Tabelin2021}.
Considering the transition to a more environmentally friendly process, the impact of the flotation process on the environment itself constitutes a significant issue due to its energy consumption \citep{Lelinski2011, Tabosa2016} and waste handling \citep{Phiri2021, Grieco2021} with their respective environmental impact.
It is therefore of great importance to increase the efficiency and the sustainability of the flotation process.

To achieve this goal, improved flotation equipment and processes are required.
A valuable instrument in the conceptualisation of flotation equipment and the examination of hydrodynamic process conditions are Euler-Euler simulation frameworks.
In the past, these frameworks have been extensively employed in numerous flotation studies \citep{Koh2003, Koh2006, Koh2007, Fayed2015, Wang2018, Shi2022, Zuerner2024, Draw2025}.
However, the accuracy of such simulations depends on the accuracy of the simulation models employed.

One of the major sub-processes in flotation is the collision process, especially between particles and bubbles \citep{Dai2000, Nguyen2004}, as it is directly related to the flotation performance.
Before a particle can attach to a bubble, it must first collide with the bubble, so that the number of attached particles captured by the bubbles is significantly influenced by the number of particle-bubble collisions \citep{Duan2003, You2017}.
The collision frequency is a function of the local relative velocity of the collision partners \citep{Saffman1956}.
As these local relative velocities are not resolved in large-scale simulations, a submodel is required, which is hence critical for the accuracy of the entire simulation.

Modelling collisions in flotation poses particular challenges due to the specific parameter regime encountered.
A variety of models to predict the collision frequency exist, with comprehensive reviews compiled by \citet{Dai2000}, \citet{Hassanzadeh2018}, and \citet{Kostoglou2020a}.
However, as will be shown later, many of the existing collision models are not suitable for the application in typical conditions of flotation.
This is primarily due to their design for different situations, the exclusion of swarm effects, the exclusion of effects resulting from size differences between particles and bubbles, and the lack of coverage of size ranges present in flotation.
Furthermore, some models have mathematical inconsistencies.

In light of these observations, this contribution puts forward a new model, the "Integrated Multi-Size Collision model" (IMSC), that predicts the collision frequency with a particular emphasis on the flotation process.
To achieve a good representation of the collision rates, detailed data from Direct Numerical Simulations (DNS) of the collision process in flotation is used.
These data are employed to validate the basic modelling assumptions, to select appropriate sub-models for each sub-process involved, and to validate the entire collision model.

The paper is laid out as follows.
First, the collision process and some important previous models from the literature are discussed in section \ref{sec:currModels}.
Thereafter, the DNS data used is presented in section \ref{sec:numMethod}.
The IMSC is then derived in section \ref{sec:IMSC}.
Finally, the overall model is validated with own DNS data and data from the literature in sections \ref{sec:fines} and \ref{sec:further_valid}.
Appendix \ref{sec:summary} provides a concise overview of the model for the purpose of implementation.

\section{Current collision models}
\label{sec:currModels}

\subsection{Mechanisms creating relative motion}
\label{sec:MechRelVels}

The occurrence of collisions is defined as the moment at which the surfaces of the two collision partners come into sufficiently close contact \citep{Saffman1956, Sundaram1997, Nguyen2004}.
For the sake of brevity and generality, mineral particles and bubbles are collectively referred to as collision partners or dispersed elements.
Unlike some literature on multiphase flows, the term 'particle' is used here to describe mineral particles in flotation processes, excluding bubbles.
The index $p$ is used for particles, while the index $b$ is used for bubbles.
A collision partner from either group is denoted by the index $\alpha$.
Two unspecified collision partners from either class are distinguished by $i$ and $j$.
Complying with the flotation literature, large particles are addressed as \textit{coarse}, while small particles are termed \textit{fine}.

Collision can only occur if there is a relative velocity between the two collision partners.
Several mechanisms have been identified in the multiphase flow literature that cause relative motion between collision partners in turbulent flow \citep{Saffman1956, Abrahamson1975,Kostoglou2006, Meyer2011, Kostoglou2020a}.
Here, the nomenclature of \citet{Kostoglou2006} is employed.

\underline{Mechanism I (Shear)} describes the motion of small, inertialess dispersed elements perfectly following the fluid streamlines.
In this case, the relative velocity of the collision partners is caused by the fluid shear, i.e. the velocity gradient of the fluid.
However, as this mechanism predominantly concerns small dispersed elements and, since the fluid velocity between two points is decorrelated at large scales \citep{Sawford1991}, its efficacy is limited to the viscous subscale.

\underline{Mechanism II (Accelerative drift)} describes the drift of larger and heavier dispersed elements from the fluid streamlines due to their inertia.
The dispersed elements primarily interact with the large-scale fluid eddies, and their velocity is only partially correlated or fully uncorrelated with the fluid velocity.

The motion of dispersed elements can be classified into these mechanisms based on their Stokes number, which describes the ratio of the response time of a dispersed element, $\tau_\alpha$, to the characteristic scale of the fluid.
For a given dispersed element of phase $\alpha$, the Stokes number is given by
\begin{flalign}
	\label{eq:St_alpha}
	&&	St_\alpha=\frac{\tau_\alpha}{\tau_f}=\frac{d_\alpha^2(\rho_\alpha/\rho_f+c_{AM})}{18\nu\tau_\eta} && \alpha=i,j.
\end{flalign}
\noindent In this expression, the Kolmogorov time-scale $\tau_\eta=(\nu/\varepsilon)^{1/2}$ is used for the fluid response time, $\tau_f$, with the turbulent dissipation rate
\begin{equation}
	\varepsilon = 2\nu \langle \mathsfbi{S}\mathsfbi{S} \rangle,
\end{equation}
\noindent where $\mathsfbi{S}$ is the Reynolds stress tensor, and $\nu$ is the molecular kinematic viscosity.
The added mass is taken into account by the second term in the nominator of (\ref{eq:St_alpha}), with the default value of $c_{AM}=0.5$ for single spherical dispersed elements \citep{Meyer2011, Kostoglou2006}.
Mechanism I describes the limit of $St_\alpha=0$, while Mechanism II assumes very large but finite Stokes numbers \citep{Abrahamson1975, Meyer2011}.
Typical real-world Stokes numbers in flotation for particles are in the range of $St_p\approx0.1$ to $St_p\approx8$ \citep{Yuu1984, Chan2023}.
Consequently, the motion of these particles is determined by a combination of these mechanisms.

\underline{Gravity} is a third mechanism causing relative motion between collision partners, introducing a deterministic component to the relative velocity between collision partners of different masses in cases of different sizes or densities.

\subsection{Requirements imposed on collision models in flotation}
\label{sec:requirements}

The conditions of flotation are highly challenging for collision models.
To be considered suitable for application in the context of flotation, a collision model must meet the following requirements.

\begin{enumerate}
	\item The model should include all relevant mechanisms causing relative motion between bubbles and particles. As highlighted above, these are turbulent shear (Mechanism I), inertial drift of the collision partners (Mechanism II), and deterministic motion due to gravity.
	\item Especially in the case of particle-bubble collisions, there is a significant size difference between the collision partners. The larger of the two, usually the bubble, distorts the surrounding flow field. This, in turn, affects the motion of the smaller particles and alters the resulting collision frequency \citep{Kostoglou2020a}. The collision model should take these flow field distortions into account.
	\item A general collision model should be able to cover the broad range of particle and bubble diameters encountered in flotation, extending from the sub-micron scale of particles to the millimetre range of bubbles, as well as encompassing the variability in the diameter of particles, which can range from single-digit micrometers to several hundred micrometers \citep{Deglon2000, Ostadrahimi2020}.
	\item Flotation involves dense swarms of bubbles and particles. Therefore, the model should account for swarm effects and not just individual bubbles and particles.
	\item The model is required to be mathematically consistent. This entails, for instance, that the various components of the model should be formulated within a common frame of reference and the employed sub-models should be valid for the specified parameter range.
\end{enumerate}

Most of the available models in the literature, however, were developed for purposes other than flotation and do not meet all the requirements listed above.
Detailed reviews of existing collision models were compiled by \citet{Nguyen2016}, \citet{Hassanzadeh2018}, and \citet{Kostoglou2020a} for example.
Here, for convenience a brief overview of some important models relevant to the subsequent introduction of the IMSC is provided.

\newpage

\subsection{Models for limiting cases}

\subsubsection{Fine, inertialess elements in turbulent flow}

One of the earliest models for the case of small droplets with $St=0$ was devised by \citet{Saffman1956}.
This model represents a limiting case for the motion of dispersed elements caused solely by Mechanism I.
The collision frequency per unit volume, $Z_{ij}$, represents the number of collisions between collision partners of two classes $i$ and $j$ \citep{Saffman1956, Duan2003}.
As the collision frequency depends on the number of suspended elements of the classes $i$ and $j$ it is usually normalised by these numbers, defining the collision kernel $\Gamma_{ij}$ \citep{Saffman1956}, with
\begin{equation}
	\Gamma_{ij}= \frac{Z_{ij}}{N_iN_j},
\end{equation}
\noindent where the total number of dispersed elements of classes $i$ and $j$ present in the domain is denoted by  $N_i$ and $N_j$, respectively.
The collision kernel is equivalent to the flux of dispersed elements from class $j$ into a sphere of radius $r_c$ around the centre of a dispersed element from class $i$, which is termed collision radius. In mathematical terms, this reads \citep{Saffman1956}
\begin{equation}
	\label{eq:colkern_general}
	\Gamma_{ij} = \int_{\theta=0}^{2\pi} \int_{\phi = 0}^{\pi} r_c^2 \sin(\phi) \ w_{r}^{(+)} \mathrm{d} \phi \ \mathrm{d} \theta.
\end{equation}
\noindent The collision radius $r_c$ is defined as the distance between the centres of the collision partners at the time of collision.
In the case of spherical collision partners, it is the sum of the radii of the respective elements $r_c=r_i+r_j$, the distance at which the surfaces of the collision partners touch.
Furthermore, $w_{r}$ is the radial component of the relative velocity between the collision partners.
For the sake of convenience, positive values of $w_r$ are denoted by $w_r^{(+)}=H(w_r)w_r$, where $H$ is the Heaviside function.

Due to continuity, incompressibility, and under the assumption of homogeneous isotopic turbulence, \citet{Saffman1956} simplified (\ref{eq:colkern_general}) to
\begin{equation}
	\Gamma_{ij} = 2\pi r_c^2 w_{r,\mathrm{rms}}, \label{eq:colkern_short}
\end{equation}
\noindent where $w_{r,\mathrm{rms}}$ is the root mean square (rms) of the radial relative velocity between the collision partners.
In the literature, this approach is referred to  as the spherical collision kernel.
While (\ref{eq:colkern_short}) is an exact solution of (\ref{eq:colkern_general}) under the given conditions, modelling by Saffman and Turner was performed by setting $w_{r,\mathrm{rms}}$ equal to the rms-value of the fluid velocity in the Euler-Euler approach, rather than using microscopic values in the vicinity of the bubble.
Employing a Gaussian normal distribution of the velocities of the collision partners and an analytical relation for the fluid shear in a turbulent flow developed by \citet{Taylor1935}, Saffman and Turner arrived at their final model

\begin{equation}
	\Gamma_{ij}^{(ST)} = r_c^3\sqrt{\frac{8\pi\varepsilon}{15\nu}}. \label{eq:Saffman}
\end{equation}

Whilst the aforementioned model excludes the effects of gravity, Saffman and Turner also proposed a model incorporating this effect \citep{Saffman1956}.
However, it was demonstrated that the proposed formulation is erroneous \citep{Dodin2002, Kostoglou2020a, Chan2023}, which was then remediated by \citet{Dodin2002}.

\subsubsection{Coarse elements with high Stokes number}

In contrast to the previous model of \citet{Saffman1956}, \citet{Abrahamson1975} proposed a model for the case of very high but finite Stokes numbers, $St_\alpha\rightarrow\infty$.
Abrahamson assumed small but heavy collision partners, whose motion is completely determined by their inertia (Mechanism II), and which interact only with large-scale turbulence.

In contrast to the spherical formulation of the collision kernel used by \citet{Saffman1956}, Abrahamson employed a cylindrical formulation of the collision kernel.
This formulation, first introduced by \citet{Sutherland1948}, assumes that all dispersed elements approaching the bubble in a critical streamtube collide with the bubble.
This formulation, however, is only applicable under very limited conditions.
While correct in the case of the model of \citet{Abrahamson1975}, the cylindrical collision kernel has been shown to be inaccurate in more general cases than this \citep{Wang1998, Kostoglou2020a}.

Assuming a Gaussian distribution of the particle velocities $v_\alpha$, the collision kernel according to Abrahamson is
\begin{equation}
	\Gamma_{ij}^{(A)}= \sqrt{8\pi} \ r_c^2 \sqrt{\langle v_i^2 \rangle + \langle v_j^2 \rangle}.
\end{equation}
The velocities of the collision partners are obtained by combining the energy spectra for the particle equation of motion and the fluid velocity using
\begin{flalign}
	&& \langle v_\alpha^2 \rangle = \frac{a_\alpha T_L + b_\alpha^2}{a_\alpha T_L +1}, && \alpha=i,j
\end{flalign}
\noindent where $a_\alpha$ and $b_\alpha$ are the reciprocal particle relaxation time and the density coefficient, respectively, defined later in (\ref{eq:a}) and (\ref{eq:b}), while $T_L=0.466 \ k/\varepsilon$ is the integral fluid time scale, with $k$ the turbulent kinetic energy.
The influence of gravity can be incorporated into this model by shifting the mean of the Gaussian probability distribution for the vertical velocity component.
However, the formulation proposed by Abrahmason is flawed due to an incorrect integration \citep{Kostoglou2020a}.

\subsection{Models based on decomposition of the relative velocity}

The two models of \citet{Saffman1956} and \citet{Abrahamson1975} discussed above are only valid for the limiting cases of $St_\alpha\rightarrow 0$ and $St_\alpha\rightarrow\infty$, respectively.
Since real collision partners cover intermediate Stokes numbers, \citet{Yuu1984} introduced an approach to fill this gap.
The basic assumption is that the total relative velocity can be decomposed into a contribution from Mechanism I (shear) and a contribution from Mechanism II (inerta) acting independently of each other. The magnitude of the total relative velocity can then be modelled as
\begin{equation}
	w=\sqrt{\langle w_I^2 \rangle + \langle w_{II}^2 \rangle},
\end{equation}
\noindent where $w_I$ and $w_{II}$ are the relative velocities of the collision partners caused by Mechanism I and Mechanism II, respectively.

\citet{Yuu1984} linearised the velocity of the collision partners around their centres using a Taylor expansion as highlighted in figure \ref{fig:Linearization}.
For modelling Mechanism I, the gradient for the fluid velocity by \citet{Taylor1935} was employed, as Saffman and Turner did.
The contribution of Mechanism II was modelled analogously to \citet{Abrahamson1975}.
Due to the extent of the model, the full set of equations is not reproduced here.
This model constitutes a major step in the field of collision modelling.

Nevertheless, some important points about the model of \citet{Yuu1984} need to be reconsidered, as outlined in the literature \citep{NgoCong2018, Kostoglou2020a}.
First, the cylindrical collision kernel description was used.
This is not applicable to the case considered in this model where the background fluid turbulence is strong.
Second, in order to be able to combine the different effects that cause relative particle motion, it is important to ensure that reference frames are used consistently.
This is not met by the model of \citet{Yuu1984}, as it alternates between an Eulerian description for Mechanism II and a Lagrangian description for parts of Mechanism I.
\citet{NgoCong2018} introduced a model that closely follows the approach by Yuu with the appropriate adjustments.

Besides the model developed by \citet{NgoCong2018}, the decomposition approach by Yuu has been widely used in modelling.
A notable example was published by \citet{Kruis1997}.
These authors utilised an autocorrelation function of \citet{Williams1980} for the fluid, thereby extending the validity of their model beyond the viscous sub-scale.
Furthermore, their sub-model for Mechanism II of large collision partners employs a different approach, as proposed by \citet{Williams1983}, based on the difference between the fluid and the particle velocity.
However, the inconsistent use of reference frames and the cylindrical collision kernel were taken over from \citet{Yuu1984}.
In addition, the contribution accounting for Mechanism I is directly proportional to $r_\alpha$, resulting in a monotonous increase of the collision kernel with the radii of the collision partners.

\subsection{Further models}

In addition to the models referenced above, Zaichik and co-workers developed a somewhat different collision model published in several versions \citep{Zaichik2006, Zaichik2010}.
As in other models, they assumed a Gaussian probability distribution of the relative velocity of the collision partners.
The overall model was obtained by integrating the corresponding 1D probability density function representing the velocity of the dispersed elements.
A two-scale bi-exponential fluid autocorrelation function from \citet{Sawford1991} was used to correlate the particle velocity with the fluid velocity.
Furthermore, the longitudinal fluid structure function proposed by \citet{Borgas2004} was employed to achieve a correlation between the particle velocities.
In contrast to the Yuu family of models, \citet{Zaichik2010} did not distinguish between the contributions of Mechanisms I and II.
Instead, they adopted a holistic approach, utilising the relative velocity of the collision partners throughout.
A retrospective decomposition into the contributions for $w_I$ and $w_{II}$ was introduced, though, to facilitate comparisons.
It can be shown that the model proposed by Zaichik reduces to the ones of Saffman and Turner, and Abrahamson for the cases of $St\rightarrow 0$ and $St\rightarrow\infty$, respectively \citep{Zaichik2010}.

The collision models discussed so far assume collision partners of similar size and ignore the effects of the dispersed elements on the fluid.
As a remedy, \citet{Kostoglou2020b} developed an approach specifically for the case of fine particles in flotation.
It is motivated by the observation that in scenarios where the diameters of the collision partners differ significantly, the smaller partner is found to be profoundly affected by the flow field disturbances induced by the larger partner.
As with earlier models, the total relative velocity is decomposed into the influences caused by the individual mechanisms of relative particle motion.
\citet{Kostoglou2020b} assumed that the motion of the particles is governed exclusively by Mechanism I, and that of the bubbles is governed exclusively by Mechanism II.
The disturbance of the flow field around the bubbles is approximated in this model based on an analytical description by \citet{Nguyen1999}.
Furthermore, corrections were introduced to account for deviations from Stokes drag with non-vanishing Reynolds numbers.

Beyond the central approaches discussed here, a wide range of other models can be found in the literature, including those of \citet{Smoluchowski1917}, \citet{Bloom2002}, and \citet{Wang2005}.
They are not further discussed here, as their main concepts are already included in the models presented above or are not applicable to the conditions of flotation. 

To account for the flow distortion of bubbles significantly larger than the particles, in the flotation literature often a modelling approach using a so-called collision efficiency $E_c$ is employed describing the ratio of the actual collision rate, $Z_{pb}$, to a nominal collision rate $Z_{pb}^\prime$, resulting in $Z_{pb}=E_cZ_{pb}^\prime$.
An overview of models based on this concept can be found in \citet{Dai2000}.
However, as pointed out by \citet{Nguyen2016} and \citet{Kostoglou2020a}, in more complex and turbulent cases, no single definition of a nominal collision rate, $Z_{pb}^\prime$, exists.
Therefore, the present contribution directly targets the overall collision frequency $Z_{ij}$ and does not perform a split into a geometric collision rate and a collision efficiency.

\section{Integrated Multi-Size Collision model}
\label{sec:IMSC}

\subsection{Basic structure}

	\begin{figure}
		\centering
		\includegraphics[width=\textwidth]{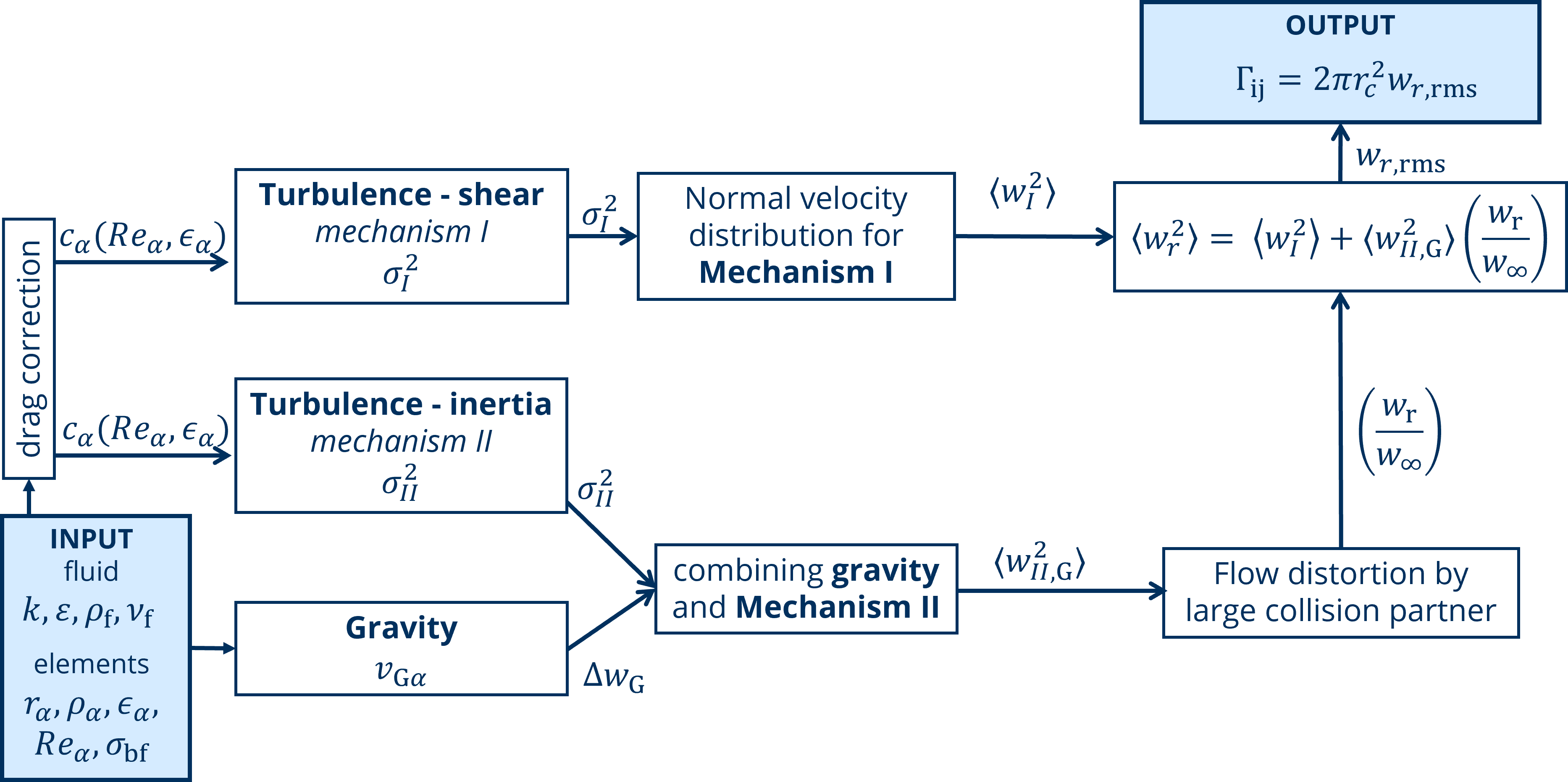}
		\caption{Schematic overview of the framework of the IMSC indicating input and output quantities, as well as the different components of the model together with the main data determined in intermediate steps. Nomenclature introduced in the text.}
		\label{fig:overviewIMSC}
	\end{figure}

From the discussion above, it is to be concluded that a significant proportion of the models previously discussed are unable to simultaneously satisfy all the listed requirements.
Moreover, numerous models exhibit substantial discrepancies between their results and those obtained from the DNS, as previously highlighted in the literature \citep{Nguyen2016, Kostoglou2020a, Chan2023, Tiedemann2024b}.
Consequently, a new model, designated as the "Integrated Multi-Size Collision model" (IMSC), is proposed here.
It integrates established concepts from the literature together with new components and is applicable to a wide range of parameters for collision partners of varying sizes.
The general approach is outlined in figure \ref{fig:overviewIMSC}.

The input parameters of the IMSC contain two groups.
Some immediately result from the physical system, such as the size and density of the dispersed elements, $r_\alpha$ and $\rho_\alpha$, as well as the fluid density and viscosity $\rho_f$ and $\nu_f$, and the surface tension $\sigma_{bf}$
The others are available in the context of an Euler-Euler framework, like the turbulent kinetic energy $k$ and the dissipation rate $\varepsilon$.
The volume fractions $\epsilon_\alpha$ are also computed with such an approach, together with the velocity of all components, resulting in the Reynolds numbers of the dispersed elements, $Re_\alpha$.
The desired output of the model is the collision kernel, which in the first place is used for particle-bubble collisions, but can also be evaluated for particle-particle, as well as for bubble-bubble collisions, if desired.
In a large Euler-Euler simulation, which does not resolve individual dispersed elements nor fine-scale fluid turbulence, the conditions vary over large scales in space and time.
The modelling discussed here is local and addresses local quantities.
The physical system considered can, hence, be interpreted as the one inside a single computational cell of an Euler-Euler simulation.

The IMSC is based on the spherical collision kernel given in (\ref{eq:colkern_short}).
The spherical collision kernel consists of two main components, the collision radius $r_c$, and the radial component of the relative velocity between the collision partners, $w_{r,\mathrm{rms}}$.
As the collision radius is directly specified by the given system, i.e. the radii of the collision partners, modelling of the collision kernel is effectively reduced to modelling the radial relative velocity between the collision partners, $w_{r,\mathrm{rms}}$.

Following the decomposition approach of \citet{Yuu1984}, the radial component of the relative velocity, $w_r$ is decomposed into the contributions from Mechanism I, Mechanism II, and gravity.
For both components, a stochastic approach based on a Gaussian distribution of the velocity of the collision partners is followed.
The modelling thus focuses on the description of the variance $\sigma^2$ of the velocity distribution \citep{Yuu1984, Kruis1997, NgoCong2018}.
It is well known that real-world bubbles and particles in turbulent flow do not fully obey a Gaussian velocity distribution \citep{Wang1996, Angriman2020}.
However, the literature suggests that a reasonable approximation for the first two moments of the velocity distribution of the dispersed elements is close to those of a Gaussian distribution \citep{Wang1996}.
The suitability of this approximation is confirmed in section \ref{sec:velo_distribution} below.
Gravity causes a deterministic velocity component that is independent of the fluid motion.
As Mechanism II describes the motion of the dispersed elements relative to the fluid, its velocity distribution is combined with that obtained by gravity.
In case of a substantial size difference between the collision partners, the disturbances of the fluid flow of the larger collision partner are taken into account using a correction factor, $w_r/w_\infty$, with $w_\infty$ describing the radial component of the relative velocity between the two collision partners at large distance.

In the following sections the IMSC is readily devised.
A summary of equations and implementation is given in section \ref{sec:summary_IMSC} and appendix \ref{sec:summary} below.

\subsection{Modelling assumptions}
\label{sec:assumptions}

Based on the requirements of the flotation process set out in section \ref{sec:requirements} and based on modelling assumptions made in the literature, the following assumptions are made to construct the IMSC.

\begin{enumerate}
	\item  Homogeneous and isotropic fluid turbulence is assumed. In section \ref{sec:isotropy} the applicability of this assumption is discussed.
	\item  The single-phase fluid is incompressible, i.e. the single-phase fluid velocity field is solenoidal.
	\item  The distribution of the dispersed elements is locally homogeneous, i.e. there is no preferential concentration. In the case that a deviation from this assumption exists in the system considered, the resulting collision kernel $\Gamma_{ij}$ can be multiplied by the locally applicable radial distribution function evaluated at the collision radius $g(r_c)$ \citep{Kostoglou2020a, Chan2023}, so that
	\begin{equation}
		\label{eq:correctionRDF}
		\Gamma_{ij} = \Gamma_{ij}(g=1) \; g(r_c).
	\end{equation}
	\item  The presence of the dispersed elements does not affect the fluid, unless otherwise stated.
	\item  All elements are assumed to be rigid, monodisperse spherical bodies within each designated class.
	\item  Contact and collision forces between collision partners are ignored.
\end{enumerate}

\subsection{Decomposition of turbulent motion}
\label{sec:decomposition}

\begin{figure}
	\centering
	\def\svgwidth{0.5\textwidth}
	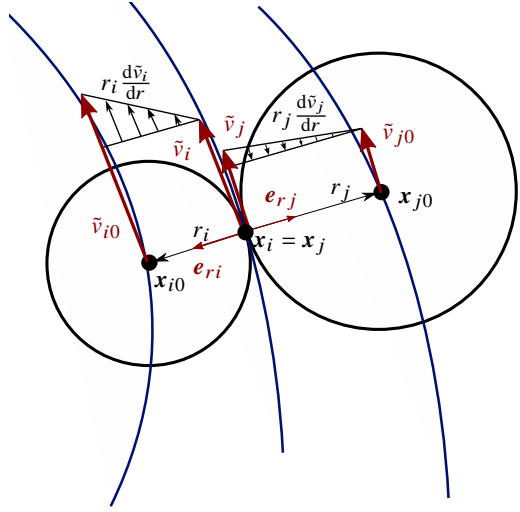
	\caption{Linearization of velocity of the collision partners for decomposition of turbulence-induced motion after \citet{Yuu1984} and \citet{NgoCong2018}. Nomenclature introduced in the text.}
	\label{fig:Linearization}
\end{figure}

In turbulent flow, the motion of dispersed elements is governed by Mechanism I (shear) and Mechanism II (inertia-induced drift).
Both mechanisms are assumed to act independently and uncorrelated \citep{Yuu1984, Kostoglou2020b, Kostoglou2020a} so that they can be modelled independently.
A decomposition of the overall radial relative velocity $w_{r,\mathrm{rms}}$ is possible, therefore.
This decomposition of the relative velocity into its components resulting from Mechanisms I and II is performed according to \citet{Yuu1984} and \citet{NgoCong2018}.
The principal idea is highlighted in figure \ref{fig:Linearization} and laid out in the following.
Note that the IMSC is a stochastic description of the collision kernel.
Hence, all quantities discussed below are generally of a stochastic nature.
In individual cases, instantaneous quantities are used to illustrate the underlying concepts.

The collision of two representative elements of classes $i$ and $j$ is considered with their centres located at $\bm{x}_{i0}$ and $\bm{x}_{j0}$, respectively.
The contact point of their surfaces is located at $\bm{x}=\bm{x}_i=\bm{x}_j$, with
\begin{flalign}
	&& \bm{x}_\alpha=\bm{x}_{\alpha0}-r_\alpha\bm{e}_{r\alpha}, && \alpha=i,j
\end{flalign}
\noindent where $\bm{e}_{r\alpha}$ is the unit normal vector connecting the centre of the dispersed element and the contact point.

In a spatially fixed frame of reference, each dispersed element has a velocity of $\bm{\Tilde{v}}_{\alpha 0}$ at its centre, with the tilde indicating that a fixed frame of reference is used.
Variables without a tilde refer to a frame of reference moving with the mean fluid velocity taken over the entire control volume considered.
The relative velocity between the centres of the collision partners $i$ and $j$ is
\begin{equation}
	\label{eq:rel_vel}
	\bm{w}_{ij}= \bm{\Tilde{w}}_{ij}=\bm{\Tilde{v}}_{j0}-\bm{\Tilde{v}}_{i0}.
\end{equation}
\noindent with the components $\bm{w}_{ij}=(w_x, w_y, w_z)^\mathrm{T}$.
As the collision partners are assumed to be rigid bodies, the velocity magnitudes\footnote{Vectors are denoted in bold face, the same quantity in light face describes the magnitude of this vector.} at their centres can be obtained as a function of their velocities at the point of contact \citep{Yuu1984, NgoCong2018}
\begin{flalign}
	\label{eq:linerization}
	&& \Tilde{v}_{i0} = \Tilde{v}_{i} \pm r_i \left. \frac{\diff \Tilde{v}_i}{\diff r} \right\rvert_{\bm{x}_i} && 
	\Tilde{v}_{j0} = \Tilde{v}_{j} \mp r_j  \left. \frac{\diff \Tilde{v}_j}{\diff r} \right\rvert_{\bm{x}_j} && i,j=p,b.
\end{flalign}
\noindent Here, and in the subsequent steps, the velocity of the collision partners is described in a moving frame of reference.

To obtain a statistical description of the individual velocities of the collision partners, it is assumed that their velocity and their relative velocity follow a Gaussian probability distribution.
This is done in accordance with the literature and for ease of modelling  \citep{Saffman1956, Yuu1984, Wang1998}.
Due to the assumed isotropy of the fluid motion, the probability distributions of the individual components of the relative velocity vector $\bm{w}_{ij}$ are equal.
Hence, a single velocity distribution for the relative velocity can be defined, $P(w)=P(w_x)=P(w_y)=P(w_z)$ given as
\begin{equation}
	\label{eq:PDF}
	P(w) = \frac{1}{\sqrt{2\pi \sigma^2}} \exp\left(-\frac{w^2}{2\sigma^2} \right),
\end{equation}
\noindent where $\sigma^2$ is the variance of the distribution.
Hence, the resulting one-dimensional mean square relative velocity is given by $\langle w^2 \rangle=\sigma_{ij}^2$.
Since both Mechanisms I and II act independently, the total variance $\sigma^2$ is a superposition of the variances originating from each mechanism.
Averaging in time and space together with (\ref{eq:rel_vel}) and (\ref{eq:linerization}) results in \citep{Yuu1984, NgoCong2018}
\begin{equation}
	\label{eq:DecompositionRadRelVel}
	\sigma_{ij}^2=\langle w^2 \rangle=\Tilde{\sigma_{I}}^2+\Tilde{\sigma_{II}}^2,
\end{equation}
\noindent with
\begin{flalign}
	\label{eq:MechI_origin}
	&& \Tilde{\sigma_{I}}^2=\biggl\langle r_i^2 \left( \left. \frac{\diff \Tilde{v}_i}{\diff r} \right\rvert_{\bm{x}_i} \right)^2 \biggr\rangle +\biggl\langle r_j^2 \left( \left. \frac{\diff \Tilde{v}_j}{\diff r} \right\rvert_{\bm{x}_j} \right)^2 \biggr\rangle + 2 \biggl\langle r_ir_j \left. \frac{\diff \Tilde{v}_i}{\diff r} \right\rvert_{\bm{x}_i} \left. \frac{\diff \Tilde{v}_j}{\diff r} \right\rvert_{\bm{x}_j} \biggr\rangle && i,j=b,p
\end{flalign}
\noindent and
\begin{equation}
	\label{eq:MechII_origin}
	\Tilde{\sigma}_{II}=\langle\Tilde{v}_{i}^2 \rangle + 		\langle\Tilde{v}_{j}^2 \rangle  -2\langle\Tilde{v}_{i}\Tilde{v}_{j} \rangle_{II}.
\end{equation}
Altogether, the radial component of the relative velocity in the collision kernel in (\ref{eq:colkern_short}) can be expressed as \citep{Wang1998}
\begin{equation}
	\label{eq:wrms_raw}
	w_{r,\mathrm{rms}} = \sqrt{\frac{2}{\pi}} \sigma=\sqrt{\frac{2}{\pi}\Tilde{\sigma}_I^2 +  \frac{2}{\pi}\Tilde{\sigma}_{II}^2}= \sqrt{\langle w_I^2 \rangle  +\langle w_{II}^2 \rangle}.
\end{equation}
Further details on the decomposition of the relative velocity and the associated linearisation of the velocity of the dispersed elements can be found in \citet{Yuu1984}, \citet{Kruis1997}, and \citet{NgoCong2018}.
It should be emphasised that the relative velocity is the same in a stationary reference frame as in a frame moving with the fluid (\ref{eq:rel_vel}).
However, it is important that a consistent reference frame is used for the subsequent derivations of the contributions from Mechanisms I and II \citep{Kostoglou2020b, Kostoglou2020a}.

\subsection{Influence of Mechanism I}

Mechanism I describes the relative velocity of two collision partners due to fluid shear.
In isotropic turbulence, the mean fluid velocity is $\langle u \rangle=0$.
Therefore, the fluctuations are $u^\prime=u$.
The fluctuations can statistically be described by $u_\mathrm{rms}=\sqrt{2k/3}$, where $k$ is the turbulent kinetic energy.
Expanding the right-hand side of (\ref{eq:MechI_origin}) with $\langle u^2\rangle$ and considering homogeneous and isotropic turbulence, $\sigma_{I}$ can be rearranged to \citep{Yuu1984, NgoCong2018}
\begin{flalign}
	\label{eq:MechI}
	&& \Tilde{\sigma}_I^2 = \left( r_i^2  \frac{\langle \Tilde{v}^2_i\rangle}{\urms^2} + r_j^2  \frac{\langle \Tilde{v}^2_j\rangle}{\urms^2 } + r_ir_j  \frac{\langle \Tilde{v}_i\Tilde{v}_j\rangle}{\urms^2 } \right) \biggl\langle \left( \frac{\diff u}{\diff r} \right)^2  \biggr\rangle && i,j=b,p.
\end{flalign}
There are three principal contributions to the model in this equation.
First, the two terms containing the variance of the velocity of the collision partners relative to the fluctuations of the fluid velocity, $\langle \Tilde{v}_\alpha^2 \rangle / \urms^2$, which describes the degree of coupling between the motion of the respective collision partner and the motion of the fluid.
Second, the correlation of the velocities of the collision partners $\langle \Tilde{v}_i\Tilde{v}_j \rangle / \urms^2$.
This is a measure of the degree to which the motion of these two is coupled via the fluid.
Third, the mean squared turbulent fluid shear gradient $\langle (\diff u / \diff r )^2 \rangle$.

The individual sub-models for these three contributions are presented in the following sections.
It should be noted that the variances of the motion of the collision partners in a fixed and in a relative frame of reference are equal, for example $\sigma_{I}=\Tilde{\sigma}_{I}$.

\subsubsection{Coupling of collision partner and fluid velocity}
\label{sec:particle_velocity}

Based on the Basset-Boussinesq-Oseen equation with extensions by \citet{Tchen1947} and \citet{Maxey1983}, the motion of one collision partner can be described by
\begin{equation}
	\label{eq:eom_basics}
	m_\alpha\frac{\diff \bm{v}_\alpha}{\diff t}=\bm{F}_D+\bm{F}_P+\bm{F}_{AM}+\bm{F}_B+\bm{F}_V,
\end{equation}
\noindent where the right-hand side assembles forces due to drag, pressure, added mass, Basset history term and volume forces.
The deterministic velocity contribution due to gravity will be considered later, so the volume force $\bm{F}_V$ is removed here.
Under the assumption of spherical collision partners and not considering the Basset history term, equation (\ref{eq:eom_basics}) can be rearranged to \citep{Abrahamson1975}
\begin{equation}
	\label{eq:eom}
	\bm{\dot{v} }_\alpha + a_\alpha \bm{v}_\alpha = b_\alpha \dot{\bm{u}} + a_\alpha \bm{u},
\end{equation}
\noindent with the reciprocal element relaxation time
\begin{equation}
	a_\alpha=\frac{1}{\tau_\alpha}=\frac{9\dynvisc c_\alpha c_\epsilon^{-2}}{r_\alpha^2(2\rho_\alpha + \rho_f)} \label{eq:a}
\end{equation}
\noindent and the density coefficient
\begin{equation}
	\label{eq:b}
	b_\alpha = \frac{3\rho_f}{2\rho_\alpha + \rho_f},
\end{equation}
\noindent where $c_\alpha$ and $c_\epsilon$ are correction factors for deviations from Stokes drag and for the presence of a swarm of dispersed elements, respectively.

Particles and bubbles move with Reynolds numbers generally larger than unity, so that the drag correction by \citet{Schiller1935} is used.
This formulation is applicable until $Re_\alpha\approx130$ \citep{Clift1978}.
While particles generally remain within this limit, bubbles reach larger Reynolds numbers due to their larger size and higher density difference, so that a further extension is required.
\citet{Karamanev1992} demonstrated that beyond $Re_b\approx 130$ the drag coefficient of a bubble is constant with $C_D \approx0.95$. 
Based on these considerations, $c_\alpha$ in (\ref{eq:a}) is set here to
\begin{equation}
	\label{eq:drag_correction}
	c_{\alpha} = \begin{cases}
		1+0.15 Re_\alpha^{0.687} & \hspace{0.5cm} Re_\alpha<136 \\
		0.95 \frac{Re_\alpha}{24} & \hspace{0.5cm} Re_\alpha\ge 136,
	\end{cases}
\end{equation}
\noindent with
\begin{flalign}
    && Re_\alpha = \frac{2r_\alpha (\Tilde{v}_\alpha- u_\mathrm{rms}) }{\kinvisc} && \alpha=b,p
\end{flalign}
\noindent and the threshold value $Re_\alpha=136$ to warrant a continuous function $c_a(Re_\alpha)$.
The ratio of the drag experienced by an individual dispersed element in a swarm compared to the drag experienced by a single dispersed element without the swarm is expressed as $c_\epsilon^{-2}$ in (\ref{eq:a}).
For bubbles, the correction of \citet{Garnier2002} is utilised here.
For particles, the relation of \citet{Richardson1954} is employed. This results in
\begin{equation}
	\label{eq:swarm correction}
	c_{\epsilon,\alpha} = \begin{cases}
		1-\epsilon_\alpha^{1/3} & \hspace{0.5cm} \alpha=b\\
		(1-\epsilon_\alpha)^{n} & \hspace{0.5cm} \alpha=p,\\
	\end{cases}
\end{equation}
\noindent with the volume fraction $\epsilon_\alpha$ of the respective phase and the exponent $n\in\mathbb{R}$.
The latter is typically fitted to experimental data and depends on the liquid, the particles used, their surface properties, and the Reynolds number.
Typical values range from $n=2$ to $n=5$ \citep{Richardson1954}.
The values for spherical particles in water employed here are given in table \ref{tab:model} of appendix \ref{sec:summary}.

It is important to note that the correlations (\ref{eq:swarm correction}) were designed for two-phase flows with either bubbles or particles present.
However, in flotation, dense three-phase flows containing particles and bubbles are present.
In these flows, the presence of the other dispersed phase hinders each dispersed phase.
Nonetheless, the literature on swarm corrections for such three-phase systems is scarce.

Especially for particles and bubbles of similar size, the authors assume this effect to be relevant. 
For the case of fine particles, their simulation data show that the effect of the particles present on the bubble velocity and the subsequent collision kernel is small.
All cases presented in \citet{Tiedemann2024a, Tiedemann2024b} were also simulated without particles and only bubbles present.
The influence of the particles on the bubble Reynolds number was found to be less than $\qty{1}{\percent}$ in all cases.
Furthermore, cases G-1-30-ep7.5 and G-1-30-ep5.0 with lower particle volume fractions confirm these findings, while also showing a marginal impact of the particle volume fraction on the particle-bubble collision kernel.
As highlighted later in section \ref{sec:grav}, the correlation (\ref{eq:swarm correction}) in combination with an existing model by \citet{Rodrigue2001} for the rise velocity of bubbles provides a very good match of the simulation data.

Nonetheless, even in the case of fine particles, the significantly larger bubbles take up some space from the overall volume, hence increasing the effective concentration of the particles in the fluid domain.
This increase in effective concentration could be considered by, for example, defining a corrected volume fraction of phase $i$ as
\begin{flalign}
    && \epsilon_{i,corr} = \frac{\sum V_i}{V_\Omega-\sum V_j}, && i,j=b,p
\end{flalign}
where $V_\Omega$ is the overall control volume and $\sum V_i$ and $\sum V_j$ are the total volume occupied by phase $i$ and $j$, respectively.
Employing this correction for a particle volume fraction of $\epsilon_p=\qty{10}{\percent}$ and a bubble volume fraction of $\epsilon_p=\qty{8.8}{\percent}$, as used in the simulations, results in $\epsilon_{p,corr}=\qty{10.9}{\percent}$, instead of $\epsilon_{p,corr}=\qty{10}{\percent}$ which is less than $1/10$ of the value.
The influence on (\ref{eq:swarm correction}) is, therefore, limited, and the ultimate effect on the overall collision kernel is small.
In light of the other approximations made in the IMSC, this additional complexity is, therefore, not retained here, but could readily be employed if desired.

For further use, (\ref{eq:eom}) is transformed to Fourier space in time
\begin{equation}
	E_\alpha(\omega) = \frac{a_\alpha^2+b_\alpha^2\omega^2}{a_\alpha^2+\omega^2}E_\mathrm{f}(\omega), \label{eq:Ealpha}
\end{equation}
\noindent with the frequency $\omega=\kappa u_\mathrm{rms}$, the wavenumber $\kappa$, and the fluid energy spectrum $E_f(\omega)$ \citep{Yuu1984, Hinze1975, Kruis1997, NgoCong2018}.
Integrating the energy spectrum from $\omega=0$ to $\omega\rightarrow\infty$ provides the mean squared velocity of the dispersed element $\langle \Tilde{v}_\alpha^2 \rangle$ and the correlation of the velocities of the collision partners $\langle \Tilde{v}_i\Tilde{v}_j \rangle$.
Detailed formulations of the energy spectra can be found in \citet{Yuu1984} and \citet{Hinze1975}.

\subsubsection{Description of the fluid energy spectrum}

In (\ref{eq:Ealpha}) a fluid energy spectrum with respect to time is required. 
Here, the modelled fluid energy spectrum $E_\mathrm{f}(\omega)$ is based on a single-phase parabolic exponential autocorrelation function $f(r)$, given by \citet{Williams1980} and used by \citet{Kruis1997}
\begin{equation}
	\label{eq:correlation_function}
	f^\mathrm{(PE)}(r) = \frac{\gamma}{\gamma -1 } \left( \exp \left( -\frac{r}{L}\right) - \frac{1}{\gamma}\exp \left( -\gamma\frac{r}{L}\right) \right),
\end{equation}
\noindent leading to
\begin{equation}
	\label{eq:tkespectrumPE}
	E_{\mathrm{f}}^{\mathrm{(PE)}}(\omega) = \urms^2 \frac{2}{\pi} \frac{\gamma}{\gamma -1 } \left(\frac{T_{\mathrm{L}}}{1+T_{\mathrm{L}}^2\omega^2} -\frac{T_{\mathrm{L}}}{\gamma^2+T_{\mathrm{L}}^2\omega^2}  \right),
\end{equation}
\noindent where $\gamma$ is the ratio of the integral length scale $L$ to the Taylor scale $\lambda$,
\begin{equation}
	\gamma = \frac{L}{\lambda},
\end{equation}
\noindent and related to turbulence quantities by
\begin{align}
	\lambda &= \sqrt{\frac{10\kinvisc\tke}{\tdr}} \\					 
	L &= T_L \urms.
\end{align}
The integral time scale $T_L$ is approximated by
\begin{equation}
	\label{eq: tL}
	T_L =  \frac{2(Re_\lambda+32)}{7\sqrt{15} }\sqrt{\frac{\kinvisc}{\tdr}},
\end{equation}
\noindent as proposed by \citet{Sawford1991}, used by \citet{Zaichik2010}, and validated by \citet{Yeung1989}, with the Taylor Reynolds number $Re_\lambda$
\begin{equation}
	\label{eq: Re_taylor}
	Re_\lambda = \sqrt{\frac{2\tke}{3}}\frac{\lambda}{\kinvisc}.
\end{equation}
Integrating (\ref{eq:Ealpha}) using this spectrum provides the Lagrangian velocity variance of the dispersed elements and their velocity correlation, which can be inserted into (\ref{eq:MechI})
\begin{flalign}
	\frac{\langle \Tilde{v}_\alpha^2 \rangle }{ \urms^2 } &= \frac{\gamma}{\gamma-1} \left( \frac{T_{\mathrm{L}} a_\alpha + b_\alpha^2}{T_{\mathrm{L}} a_\alpha +1} - \frac{T_{\mathrm{L}} a_\alpha + \gamma b_\alpha^2}{\gamma(T_{\mathrm{L}} a_\alpha +\gamma)} \right) && \alpha=b,p \label{eq:particle_velocity} \\
	\frac{\langle \Tilde{v}_i\Tilde{v}_j \rangle }{  \urms^2 } &= \frac{\gamma}{\gamma -1} (I_1 - I_2)  &&  i,j=b,p. \label{eq:particle_velocity_correlation}
\end{flalign}
\noindent The definition of the constants $I_1$ and $I_2$ is given in table \ref{tab:model}.
Other descriptions of the fluid energy spectrum, such as those based on a two-scale bi-exponential autocorrelation function of \citet{Sawford1991}, which was used by \citet{Zaichik2010}, were found to give less accurate results of the overall collision kernel compared to the simulation data presented in sections \ref{sec:fines} and \ref{sec:further_valid} below.

\subsubsection{Spatial structure of the fluid velocity}
\label{sec:structure_func}

\begin{figure}
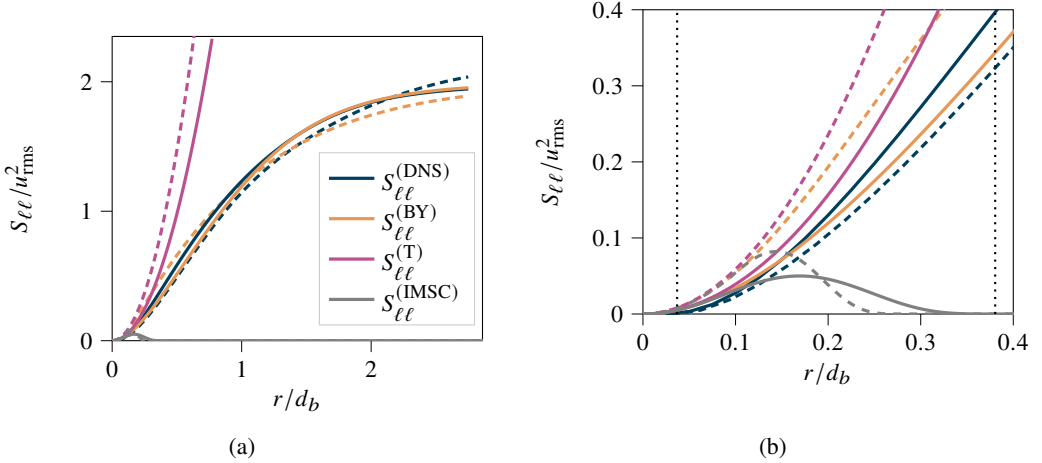

	
	\centering
	\begin{subfigure}{0.48\textwidth}
    	\setlength\fwidth{0.3\textwidth}
    	\input{figures/structure_function_comparison.tex}	
		\caption{}
		\label{fig:StructureFunctionMod}
	\end{subfigure}
	\hfill
	\begin{subfigure}{0.48\textwidth}
        \setlength\fwidth{0.3\textwidth}
		\input{figures/structure_function_comparison_snapshot.tex}
		\caption{}
		\label{fig:StructureFunctionModSnapshot}
	\end{subfigure}

	\caption{Comparison of models for the longitudinal fluid structure function $S_\mathrm{\ell\ell}^\mathrm{(IMSC)}$ (\ref{eq:blend_sin}) against DNS data for the present three-phase flow. Solid lines relate to case R53-1-30, dashed lines to the gravity-driven case G-1-30 (table \ref{tab:sim_cases}). a) Data for $r/d_b$ up to $3$, b) zoom on small radii. The vertical dotted lines represent the Kolmogorov length scale $\eta$ and the cut-off length scale for $S_\mathrm{\ell\ell}^\mathrm{(IMSC)}$, $r_\lambda$, respectively, evaluated for R53-1-30.}
	\label{fig:StructureFunctionModOverall}
	
\end{figure}

The remaining component for the description of Mechanism I in (\ref{eq:MechI}) is the fluid shear gradient.
For two points separated by a small distance $r$, the shear gradient can be linearised.
This linearisation can be approximated by the longitudinal fluid structure function \citep{Pope2000}	
\begin{equation}
    \label{eq:structurefunctionraw}
    r^2 \biggl\langle \left(\frac{\diff u}{\diff r}\right)^2 \biggr\rangle \approx S_{\mathrm{\ell\ell}}(r)=\langle (\bm{u}(\bm{x})-\bm{u}(\bm{x}+\bm{r}))^2\rangle.
\end{equation}
Evaluating (\ref{eq:MechI}) with the fluid structure function leads to
\begin{flalign}
	\label{eq:MechI_with_Sll}
	&& \Tilde{\sigma}_I^2 = \frac{\langle \Tilde{v}^2_i\rangle}{\urms^2}S_\mathrm{\ell\ell}(r_i) + \frac{\langle \Tilde{v}^2_j\rangle}{\urms^2 }S_\mathrm{\ell\ell}(r_j) +  \frac{\langle \Tilde{v}_i\Tilde{v}_j\rangle}{\urms^2 }S_\mathrm{\ell\ell}(r_c) && i,j=b,p,
\end{flalign}
where the approximation of the structure function is evaluated at the radii of the collision partners, $r_i$ and $r_j$, respectively.
Different models for the longitudinal fluid structure function exist in the literature.
Figure \ref{fig:StructureFunctionMod} provides an exemplary comparison between the models that will be introduced in the following, evaluated for the representative DNS data, cases R53-1-30 and G-1-30 introduced in section \ref{sec:numMethod} below, and the respective longitudinal structure function evaluated from the DNS, $S_{\mathrm{\ell\ell}}^\mathrm{(DNS)}$.
The choice of a suitable submodel is exemplarily highlighted based on this comparison.
An exact analytical solution of the longitudinal structure function for the dissipative subrange, i.e. $r<\eta$, commonly used in the literature, is the one given by \citet{Taylor1935} 
\begin{equation}
    \label{eq:Sll_Taylor}
    S_{\ell\ell}^{(T)} = r^2 \frac{\varepsilon}{15\nu_f}.
\end{equation}
For distances outside the dissipative subrange, this approximation is invalid since it increases monotonically with $r$, as can be seen in figure \ref{fig:StructureFunctionMod}.
This was remedied by \citet{Borgas2004}, who porposed a model for the longitudinal structure function applicable for all $r$, reading
\begin{equation}
    \label{eqn:BorgasSll}
    S^{\mathrm{(BY)}}_{\mathrm{\ell\ell}}(r) = 2Re_\lambda \sqrt{\frac{\varepsilon \kinvisc}{15}} \left[1- \exp\left(-\frac{r}          {30^{3/4}\kolmlen}\right)\right]^{4/3} \left[\frac{15^3 r^4}{15^3 r^4 + \kolmlen^4 Re_\lambda^6}\right]^{1/6} \hspace{0.5cm}.
\end{equation}
Equation (\ref{eqn:BorgasSll}) reproduces the Taylor gradient for $r<\eta$ and provides a good fit to the DNS data.
For case R53-1-30 under background turbulence, the model by Borgas and Yeung provides a good approximation of the longitudinal fluid structure function.
Only minor deviations for $0.2d_b<r<1.2d_b$ exist.
These are most likely caused by the presence of bubbles in the three-phase DNS or particle swarm effects, causing a higher fluid velocity gradient on the length scales of the dispersed elements.
For smaller and larger distances, there is almost no deviation between the DNS data and $S^{\mathrm{(BY)}}_{\mathrm{\ell\ell}}$.
The gravity-driven case G-1-30 shows larger differences between the longitudinal structure function obtained from the DNS results and the model by Borgas and Yeung.
In the case of fine particles, i.e. for small $r$, the DNS results are lower than the modelling prediction.
For intermediate distances, the differences decrease while increasing again for $r>1.5d_b$.
The model was designed for single-phase, homogeneous, isotropic turbulence, but the investigated case is gravity-driven.
Hence, the largest flow structures are of the size of the bubble diameter and do not obey homogeneous and isotropic turbulence.
Overall, this is not too relevant for the IMSC, as Mechanism I of bubbles for gravity-driven cases is small, and the bubble velocity is dominated by Mechanism II and gravity for this case.

It is important to note that, in general, for the IMSC, it is assumed that the dispersed phases do not alter the fluid flow field, i.e. their influence is negligible.
However, the presence of larger bubbles or coarse particles locally destroys the small-scale fluid velocity structures, because for $r>>\eta$ they are far beyond the viscous subrange and more firmly in the inertial subrange \citep{Nguyen2004, Kostoglou2020b}.
It also means that the motion of large, dispersed elements with diameters well outside the viscous subrange is not affected by the small-scale fluid motion.
Thus, the motion of the large dispersed elements and the fluid motion become uncorrelated in turbulent flow, resulting in no contribution from Mechanism I to the relative motion.

A simple way to account for this effect is to assume zero fluid velocity at the bubble surface, hence $S_\mathrm{\ell\ell}(r>>\eta)=0$ as proposed by \citet{Kostoglou2020b}.
However, as the IMSC is designed to handle a wide range of particle diameters and turbulence parameters, this strict treatment may not be applicable in all cases.
To address this, a blending function between the two limiting cases $r<<\eta$ and $r>>\eta$ is proposed here, defining the longitudinal fluid structure function as
\begin{equation}
    \label{eq:blend_sin}
    S_{\mathrm{\ell\ell}}^{\mathrm{(IMSC)}}(r)  = 
    \begin{cases}
        S_{\mathrm{\ell\ell}}^{\mathrm{(BY)}} & \hspace{0.5cm},r\leq r_\eta\\
        S_{\mathrm{\ell\ell}}^{\mathrm{(BY)}}\left(r\cos^2\left( \frac{\pi}{2}\frac{r-r_\eta}{r_\lambda-r_\eta} \right) \right)	 & \hspace{0.5cm}, r_\eta<r\leq r_\lambda\\
        0 & \hspace{0.5cm},r>r_\lambda\\
    \end{cases}
\end{equation}
\noindent with $r_\eta =1.5\eta$ and the cut-off length $r_\lambda=0.6\lambda+0.1L$.
The prefactors were chosen to best match the collision kernels in the available DNS data presented in Sections \ref{sec:fines} and \ref{sec:further_valid} below.
The overall behaviour of the modified longitudinal structure function for the two cases R53-1-30 and G-1-30 using fine particles is depicted in figure \ref{fig:StructureFunctionModSnapshot} in comparison with the other longitudinal structure functions listed above focusing on radii in the vicinity of the respective particle diameters.
It was tried to introduce an additional factor in (\ref{eqn:BorgasSll}) to obtain a model better matching the longitudinal fluid structure function derived from the simulations.
All simulation cases listed in table \ref{tab:sim_cases} were considered in this attempt, but it turned out that the results were not general enough to be used in the final model.

The fluid velocity at two widely separated points becomes uncorrelated.
Following an approach by \citet{Zaichik2010}, the IMSC accounts for this by multiplying the correlation term with the fluid correlation function $f(r_c)$ as set out in (\ref{eq:correlation_function}).	
Inserting all the components in (\ref{eq:MechI_with_Sll}) ultimately results in
\begin{equation}
    \label{eq:MechI_complete}
    \begin{split}
        \Tilde{\sigma}_I^2 = \sigma_I^2 = & S_{\mathrm{\ell\ell}}^{\mathrm{(IMSC)}}(r_i) \frac{\langle \Tilde{v}_i^2\rangle}{\urms^2} + S_{\mathrm{\ell\ell}}^{\mathrm{(IMSC)}}(r_j) \frac{\langle \Tilde{v}_j^2\rangle}{\urms^2}  \\
        &+ S_{\mathrm{\ell\ell}}^{\mathrm{(IMSC)}}(r_c)f(r_c) \frac{\langle \Tilde{v}_i\Tilde{v}_j\rangle}{\urms^2} \hspace{3cm}i,j=b,p.
    \end{split}
\end{equation}
If $St=0$ for both collision partners, the contribution of Mechanism I reduces to the formulation of \citet{Saffman1956}.
If $St=0$, the reciprocal relaxation time $a_\alpha \rightarrow \infty$. Hence, both constants $I_1$ and $I_2$ equal zero.

\subsection{Modelling of Mechanism II}

This section is concerned with modelling $\Tilde{\sigma}_{II}$ via the decomposition in (\ref{eq:MechII_origin}).
Given the assumptions listed in section \ref{sec:assumptions}, it follows that the correlation term vanishes, i.e. $\langle\Tilde{v}_{i}\Tilde{v}_{j} \rangle_{II}=0$.
Since Mechanism II pertains to the inertia-induced drift of the dispersed elements and assumes no feedback to the fluid, the movements of the two collision partners due to Mechanism II are uncorrelated.

While Mechanism I results from the motion of the collision partners with the fluid, Mechanism II involves their motion relative to the fluid.
Consequently, to ensure a consistent reference frame with the description of Mechanism I, the original equation of Mechanism II (\ref{eq:MechII_origin}), with the third term being dropped there, is transferred to a reference frame moving with the fluid velocity $\bm{u}$, reading
\begin{equation}
	\label{eq:MechII}
	\begin{split}
		\sigma_{II}^2 &= \biggl\langle(\Tilde{v}_i-u)^2\biggr\rangle +  \biggl\langle(\Tilde{v}_j-u)^2\biggr\rangle \\
		&= \langle\Tilde{v}_{i}^2 \rangle + \langle\Tilde{v}_{j}^2 \rangle - 2 \langle \Tilde{v}_i u \rangle- 2 \langle \Tilde{v}_j u \rangle + 2 \langle u^2 \rangle.
	\end{split}
\end{equation}
To obtain the correlation of the velocity of the dispersed elements with the fluid velocity, $\langle \Tilde{v}_\alpha u \rangle$, the relation in (\ref{eq:particle_velocity_correlation}) is modified.
The relation in (\ref{eq:particle_velocity_correlation}) describes the correlation of the velocity of two collision partners from classes $i$ and $j$, i.e. $\langle \Tilde{v}_i \Tilde{v}_j \rangle$.
It may be assumed that one of the collision partners is a tracer without inertia perfectly following the fluid streamlines \citep{Kostoglou2020b}.
This tracer represents the fluid motion, hence, its velocity magnitude is $u$.
For such a tracer, the relaxation time (\ref{eq:a}) and the density coefficient (\ref{eq:b}) are $\tau_j=0$, $a_j\rightarrow\infty$, and $b_j=1$, respectively.
Inserting into (\ref{eq:particle_velocity_correlation}) set to
\begin{flalign}
	\label{eq:correlation_particle_tracer}
	&& \frac{ \langle \Tilde{v}_\alpha u \rangle }{\urms^2} 
	=   \frac{\gamma}{\gamma-1} \biggl( \frac{T_\mathrm{L}a_\alpha+b_\alpha}{T_\mathrm{L}a_\alpha+1} - 
	\frac{T_\mathrm{L}a_\alpha+b_\alpha\gamma}{\gamma(T_\mathrm{L}a_\alpha+\gamma)} \biggr) && \alpha=i,j.
\end{flalign}
\noindent Lastly, the variance of the fluid velocity is given by
\begin{equation}
	\langle u^2 \rangle = \urms^2 = \frac{2}{3}k.
\end{equation}
For the limiting case of $St=0$, the dispersed elements perfectly follow the fluid streamlines, hence no deviation from them exists.
This is captured by the IMSC, as in this case $\langle\Tilde{v}^2_{i} \rangle = \langle\Tilde{v}^2_{j} \rangle = \langle u^2 \rangle$.
Hence, from (\ref{eq:MechII}) it follows that $\sigma_{II}^2=0$.
Hence, not only the contribution of Mechanism I, but the entire IMSC reduces to the formulation of \citet{Saffman1956}, (\ref{eq:Saffman}), in the limit of $St=0$.

\subsection{Modelling of gravity}
\label{sec:grav}

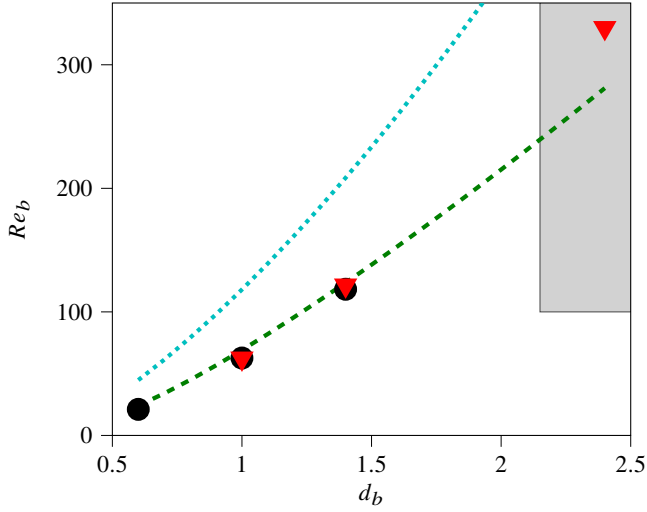
\begin{figure}
	\centering
	\setlength\fwidth{0.5\textwidth}
\begin{tikzpicture}

\begin{axis}[
tick align=outside,
tick pos=left,
x grid style={gray},
xlabel={\(\displaystyle d_b\)},
xmin=0.50, xmax=2.5,
xtick style={color=black},
y grid style={gray},
ylabel={\(\displaystyle Re_b\)},
ymin=0, ymax=350,
ytick style={color=black}
]
\draw[fill=gray!50,opacity=.7] (axis cs:2.15,100) -- (axis cs:2.15,350) -- (axis cs:2.5,350) -- (axis cs:2.5,100) -- cycle ;

\addplot [ultra thick, black, mark=*, mark size=3.5, mark options={solid}, only marks]
table {%
0.6 21.1038098262396
1 62.6501521156542
1.4 118.312933997209
};\label{plt:Redp30}
\addplot [ultra thick, red, dash pattern=on 5pt off 5pt, mark=triangle*, mark size=3.5, mark options={solid,rotate=180}, only marks]
table {%
1 62.3777957137061
1.4 121.498259815879
2.4 329.893294125848
};\label{plt:Redp50}
\addplot [ultra thick, green01270, dashed]
table {%
0.6 24.1048590848956
0.7 33.9859522531542
0.8 44.9234968248363
0.9 56.6921545318024
1 69.1275884560705
1.1 82.11238757846
1.2 95.5628955978405
1.3 109.419016713078
1.4 123.636921122112
1.5 138.184007843244
1.6 153.035477333098
1.7 168.172007591495
1.8 183.578173736544
1.9 199.241365750324
2 215.151040430655
2.1 231.29819870735
2.2 247.675016074949
2.3 264.274577991227
2.4 281.0906879431
};\label{plt:Rodrigue}

\addplot [ultra thick, darkturquoise0191191, dotted]
table {%
0.6 44.7727458354633
0.7 61.2851886674445
0.8 79.0590601981731
0.9 98.0147546899907
1 118.086299545644
1.1 139.21784510237
1.2 161.361304951206
1.3 184.47470252521
1.4 208.520973107989
1.5 233.467071489062
1.6 259.283291668572
1.7 285.94273785332
1.8 313.420906012591
1.9 341.695347921881
2 370.745397877193
2.1 400.551947791903
2.2 425.848357798223
2.3 443.597473571319
2.4 461.255731604257
};\label{plt:Clift}


\end{axis}

\end{tikzpicture}	
	\caption{Bubble Reynolds number as a function of bubble diameter $d_b$ for particle diameters of $d_p=\SI{30}{\mu m}$ (\ref{plt:Redp30}) and $d_p=\SI{50}{\mu m}$ (\ref{plt:Redp50}), both with $\epsilon_b=\qty{8.8}{\percent}$, as obtained in the DNS. The rise velocity according to the models by \citet{Rodrigue2001} (\ref{plt:Rodrigue}) and \citet{Clift1978} (\ref{plt:Clift}), both corrected for the presence of a bubble swarm according to \citet{Garnier2002}, are shown for reference. The shaded area approximately marks the regime of non-spherical bubbles [from \citet{Tiedemann2024b}].}
	\label{fig:Re_b}
\end{figure}

So far, the IMSC takes into account the stochastic motion of the collision partners due to the influence of turbulence.
However, the presence of gravity introduces a deterministic component to the relative velocity, caused by the difference in the rate of rise or sedimentation of different types of dispersed elements in a polydisperse suspension.
\begin{equation}
    \label{eq:delta_w_G}
    \Delta \Tilde{w}_{\mathrm{G}} = \Tilde{v}_{\mathrm{G}i} -  \Tilde{v}_{\mathrm{G}j},
\end{equation}
\noindent where $\Tilde{v}_{\mathrm{G}\alpha}$ is the gravity-induced velocity of a class $\alpha$ of dispersed elements. 
Several models in the literature describe the rise and sedimentation velocity of single particles and bubbles.
A model for the rise velocity of a single bubble in quiescent flow is given by \citet{Rodrigue2001} reading
\begin{equation}
    \label{eqn:Rodrigue_rise_velo}
    v_{\mathrm{G}b}^{\mathrm{(R)}} = Ve \left( \frac{d_b^2 \rho_f}{\sigma_{bf} \nu_f} \right)^{-1/3},
\end{equation}
\noindent where $Ve$ is the velocity number defined as
\begin{equation}
    Ve = \frac{Fl/12}{1+0.049Fl^{3/4}},
\end{equation}
with the flow number
\begin{equation}
    Fl = g\left( \frac{d_b^8\rho_f}{\sigma_{bf}\nu_f^4} \right)^{1/3}, 
\end{equation}
where $d_b=2r_b$ is the bubble diameter and $\sigma_{bf}$ is the surface tension of the bubble.
For a single particle in quiescent flow, the model of \citet{Nguyen2004} is used here, reading
\begin{equation}
    \label{eq:Nguyen_settle_velo}
    v_{\mathrm{G}p}^{\mathrm{(NS)}} = \frac{2 r_p^2 (\rho_p-\rho_f) g}{9 \kinvisc \rho_f}
    \left(
    1 + \frac{Ar_p}{96}(1+0.079Ar_p^{0.749})^{-0.755}
    \right)^{-1},
\end{equation}
\noindent with the Archimedes number
\begin{equation}
    Ar_\alpha=\frac{8r_\alpha^3 g (\rho_\alpha-\rho_f)}{\rho_f \nu^2}.
\end{equation}
To correct for the influence of swarm effects, the same models by \citet{Garnier2002} and \citet{Richardson1954}, (\ref{eq:swarm correction}), as used in section \ref{sec:particle_velocity}, are employed here to determine the correction factor $c_\epsilon$, giving
\begin{equation}
    v_{\mathrm{G}\alpha} = 
    \begin{cases}
        v_{\mathrm{G}b}^{\mathrm{(R)}}c_{\epsilon b} \hspace{0.7cm} \alpha=b\\
        v_{\mathrm{G}p}^{\mathrm{(NS)}}c_{\epsilon p} \hspace{0.5cm} \alpha=p,
    \end{cases}
\end{equation}
\noindent for bubbles and particles, respectively.
The remarks in section \ref{sec:particle_velocity} above concerning a suitable choice of the swarm corrections apply here as well. 
	
In addition to the validation of these models in the literature \citep{Rodrigue2001, Nguyen2004}, a brief comparison with the simulation data is made here.
Only the simulations in which gravity is the primary driver of particle and bubble motion are utilised, as in these cases the velocity component due to gravity can be clearly distinguished.
Figure \ref{fig:Re_b} shows the bubble Reynolds number as a function of $d_b$ for different $d_p$.
Obviously, changes in $d_p$ have little influence on the movement of the bubbles for the fine particles considered here.
The data are compared to the model by \citet{Rodrigue2001} supplied with the swarm correction of \citet{Garnier2002}.
The agreement is very good up to a bubble diameter of $d_b=\SI{1.4}{\milli\meter}$.
For $d_b = \SI{2.4}{mm}$, a discrepancy of $\SI{15}{\percent}$ can be observed between the modelled rise velocity and the simulation result.
This particular case, characterised by a bubble Reynolds number $Re=330$, is situated at the threshold of the assumption of spherical bubbles according to the regime map of \citet{Clift1978}.
While in the simulations rigid bubbles with $Eo=0$ are assumed, in reality $Eo\approx0.77$ for this case.
Overall, this causes a difference from the simulation results and the model by \citet{Rodrigue2001}.
This data point was included in the Figure for further reference.
Finally, a comparison is also made with the model of \citet{Clift1978} for the bubble rise velocity using the same swarm correction of \citet{Garnier2002}. 
As it provides worse results, it is not used in the IMSC.

Using the decomposition of the relative velocity introduced earlier, $w$ is split into a contribution from the motion of the collision partners with the fluid (Mechanism I) and a contribution from the motion of the collision partners relative to the fluid (Mechanism II).
As gravity causes a motion relative to the fluid, a combined effect of Mechanism II and gravity, $w_{II,\mathrm{G}}$, is used following a method by \citet{Dodin2002}.
Thus, (\ref{eq:wrms_raw}) changes to
\begin{equation}
    \label{eq:wrms_gravity}
    w_{r,\mathrm{rms}} =\sqrt{\frac{2}{\pi}\sigma_I^2 +  \langle w_{II,\mathrm{G}}^2 \rangle}.
\end{equation}
Similar to the definition of the spherical collision kernel in (\ref{eq:colkern_general}), the flux over the collision sphere of radius $r_c$ due to $w_{II,\mathrm{G}}$ is
\begin{equation}
    F_G = \sqrt{\langle w_{II,\mathrm{G}}^2 \rangle} 4\pi r_c^2 
    =  \int_{0}^{2\pi} \int_{0}^{\pi} r_c^2 \sin(\phi) \cos(\theta) \sqrt{w_{II,\mathrm{G},r}^2(\phi,\theta)}\, \diff \phi \diff \theta,
\end{equation}
\noindent where $w_{II,\mathrm{G},r}$ is the radial component of $w_{II,\mathrm{G}}$.
Assuming a Gaussian probability distribution for the relative velocity yields
\begin{equation}
    \label{eq: MechIIGrav}
    \begin{split}
        \sqrt{\langle w_{II,\mathrm{G}}^2 \rangle} = \int_{\phi = 0}^{\pi/2} \int_{w=-\infty}^{\infty} \sin(\phi) 
        &(w-\cos (\phi) \Delta w_{\mathrm{G}} )\\
        & \frac{1}{\sqrt{2\pi\sigma^2_{II}}} \exp \left( - \frac{w^2}{2\sigma^2_{II}} \right) \diff w \diff \phi,     
    \end{split}
\end{equation}
\noindent with $\Delta w_G=\Delta \Tilde{w}_{\mathrm{G}}$, defined in (\ref{eq:delta_w_G}), since the relative velocity is the same in a fixed and a moving frame of reference.
This integral is inaccessible for analytical treatment \citep{Abrahamson1975}, so that numerical integration must be performed.
Suitable limits of integration for $w$ are $[-3\sigma_{II}, 3\sigma_{II}]$, instead of $[-\infty, \infty]$.

\subsection{Fluid disturbance by large bubbles}

So far, only collisions between similarly sized collision partners have been considered, such as those involving pairs of bubbles or pairs of particles.
It has been assumed that the presence of the collision partners and their interaction with the fluid do not significantly alter the relative velocity between them.
However, this assumption is not valid when the collision partners exhibit a substantial disparity in size, as is often the case in flotation.
Bubbles significantly alter the fluid field around them and, hence, the radial relative velocity of small particles approaching them.
The method discussed in the following can also be employed for pairs of bubbles, pairs of particles  or pairs of particles and bubbles of vastly different sizes.
The index $j$ refers to the larger collision partner, and $i$ to the smaller partner.

\begin{figure}
	
	\centering
		\def\svgwidth{0.4\textwidth}
		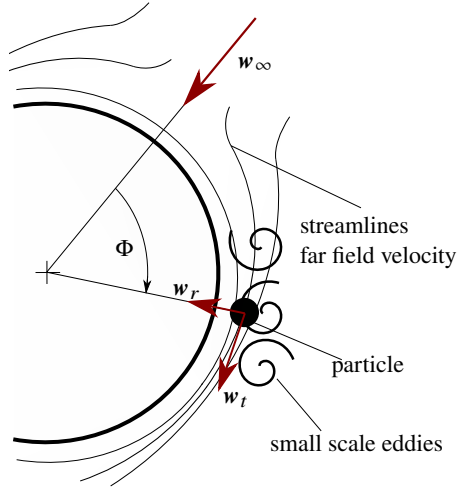
		\caption{Schematic of the deflection of a small particle by the flow field modulations around a large bubble [after \citet{Nguyen1999}]. }
		\label{fig:disturbance_large_sketch}
\end{figure}

\begin{figure}
    \centering
	\begin{subfigure}{0.5\textwidth}
\begin{tikzpicture}

\begin{axis}[
log basis y={10},
tick align=outside,
tick pos=left,
width=0.951\textwidth,
height=0.8510\textwidth,
x grid style={gray},
xmin=-0.029, xmax=1.049,
xtick style={color=black},
y grid style={gray},
ymin=0.0200887975557657, ymax=4.06980200445468,
xmin=0.0, xmax=0.6,
ymode=log,
ytick style={color=black},
xlabel={$r_p/r_b$},
ylabel={$\Gamma_{pb} \ \tau_\kolmlen/r_c^3$},
]
\begin{scope}[on background layer]
    \fill[green,opacity=0.2] ({rel axis cs:0,0}) rectangle ({rel axis cs:0.17,1});
    \fill[blue,opacity=0.2] ({rel axis cs:0.5,0}) rectangle ({rel axis cs:1,1});
    \fill[red,opacity=0.2] ({rel axis cs:0.17,0}) rectangle ({rel axis cs:0.5,1});
\end{scope}
\node[draw           ] at (0.54\textwidth,-0.3\textwidth) {\small no modulation};
\node[draw           ] at (0.24\textwidth,-0.3\textwidth) {\small transition};
\node[draw, rotate=90] at (0.05\textwidth,-0.05\textwidth) {\small modulation $w_r/w_\infty$};
\addplot [color=NMcol, line width=1.0pt]
  table[row sep=crcr]{%
0.02 0.0255740950988856\\
0.04 0.0519117278636581\\
0.06 0.0813702175602021\\
0.08 0.116580142194395\\
0.1 0.16010273444499\\
0.12 0.267229464542799\\
0.14 0.388820985342869\\
0.16 0.521100896022936\\
0.18 0.662229122397964\\
0.2 0.810792879463155\\
0.31 2.61548442534273 \\
0.4 2.76013822085281\\
0.6 2.98954911108231\\
1 3.19688451315341\\
};
\end{axis}

\end{tikzpicture}
	\end{subfigure}   
    \caption{Exemplary alteration of the relative velocity due to a modulated flow field as modelled by the IMSC. Input parameters (like $k$, $\varepsilon$, $\epsilon_p$, $\epsilon_g$, etc.) correspond to case G-1-30 in table~\ref{tab:sim_cases}. Only $r_p$ was varied.}
    \label{fig:disturbance_large_regime}
\end{figure}
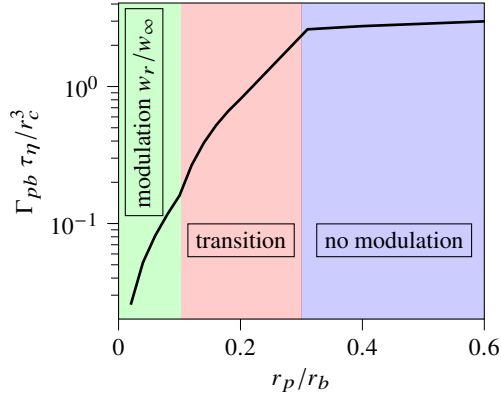

The deviation of the particle from its original trajectory and the subsequent change of the radial relative velocity are illustrated in figure \ref{fig:disturbance_large_sketch}.
The previously combined radial relative velocity for Mechanism II and gravity, $w_{II,\mathrm{G}}$, is taken as the far-field approach velocity $w_\infty$ \citep{Kostoglou2020b}.
\citet{Nguyen1999} devised a description of the flow field surrounding a bubble.
Assuming that the larger collision partner $j$ is a bubble with an immobile bubble surface, the local radial relative velocity at the collision radius $w_r$ results in
\begin{equation}
	\label{eq:RadCompNguyen}
	\frac{w_r}{w_\infty}  = \frac{2X\cos(\Phi) + 3 Y \cos^2(\Phi)- Y}{2} \left( \frac{r_i}{r_j}\right)^2,  
\end{equation}
\noindent where
\begin{equation}
	\label{eq:RadCompNguyen_comp}
	X = \frac{3}{2}+\frac{9}{32}\frac{Re_{j}}{1+0.309Re_{j}^{0.694}} \hspace{0.3cm},\hspace{0.3cm}Y = \frac{3}{8} \frac{Re_{j}}{1+0.217Re_{j}^{0.518}},
\end{equation}
valid for $r_i/r_j<(r_i/r_j)_m=0.1$.
The DNS data utilized for subsequent validation suggest that the assumption regarding the independence of collision partners on their relative velocity, due to their coupling with the fluid, holds true for the condition $r_i/r_j>(r_i/r_j)_n=0.3$ (cf. sections \ref{sec:valid_pb_coll_kern} and \ref{sec:valid_coarse} below).
To bridge the gap, the following piece-wise correction factor $(w_r/w_\infty)^2$ for $\langle w_{II,\mathrm{G}}^2 \rangle$ in (\ref{eq:wrms_gravity}) is introduced
\begin{equation}
	\label{eq:RadialComponentCorrection}
	\begin{split}
		\left( \frac{w_r}{w_\infty} \right)^2 =& \begin{cases}
			\left( \frac{1}{4} Y\left(\frac{r_i}{r_j}\right)^2 \right)^2 \hspace{1.9cm}&\mathrm{if}\hspace{0.4cm}r_i/r_j\le (r_i/r_j)_m \\
			\left( Z + \frac{1-Z}{(r_i/r_j)_n-(r_i/r_j)_m}\left(\frac{r_i}{r_j} - (r_i/r_j)_m\right )  \right)^2 &\mathrm{if}\hspace{0.4cm}(r_i/r_j)_m< r_i/r_j\le (r_i/r_j)_n \\
			1  &\mathrm{else} 
		\end{cases} \\
	\end{split}
\end{equation}
\noindent where
\begin{equation}
	Z =  \left(\left(\frac{r_i}{r_j}\right)_m\right)^2\frac{1}{4} Y= 0.0025 \ Y.
\end{equation}
The final expression of the radial relative velocity, then, is
\begin{equation}
	w_{r,\mathrm{rms}} = \sqrt{\frac{2}{\pi}\sigma_I^2  + \langle w_{II,\mathrm{G}}^2\rangle \left( \frac{w_r}{w_\infty}\right)^2}.
    \label{eq:final_w_r}
\end{equation}
Figure~\ref{fig:disturbance_large_regime} exemplarily illustrates the behaviour of the IMSC using this correction for varying $r_p/r_b$.
To this end, all statistical quantities were taken over from G-1-30 defined in table~\ref{tab:sim_cases} below and $r_p$ was changed.
The resulting data illustrate the impact of the size ratio, as accounted for by the model.

\subsection{Summary of the model}
\label{sec:summary_IMSC}

Including all these components, the IMSC is a model for the collisions between particles and bubbles, pairs of particles, and pairs of bubbles.
A guideline for the necessary steps of its implementation is shown in figure \ref{fig:overviewIMSC} above to be read from left to right.

The following input parameters are required for the IMSC.
For the fluid: turbulent kinetic energy $k$, turbulent dissipation rate $\varepsilon$, fluid density $\rho_f$, kinematic viscosity $\nu_f$.
For each of the two dispersed phases $\alpha=i,j=p,b$, i.e. particles or bubbles: radius of the elements $r_\alpha$, density of the elements $\rho_\alpha$, volume fraction of the phase $\epsilon_\alpha$, Reynolds number $Re_\alpha$.
Finally, the surface tension $\sigma_{bf}$ is needed to model the rise velocity of the bubbles.

Based on these input parameters, the corrections for the swarm effects and for the deviation from Stokes drag can be determined according to (\ref{eq:drag_correction}) and (\ref{eq:swarm correction}).
The next step is to determine the individual contributions for Mechanism I and Mechanism II.
The contribution of Mechanism I is given in (\ref{eq:MechI_complete}), using the piecewise definition of the fluid structure function (\ref{eq:blend_sin}), the velocity correlations in (\ref{eq:particle_velocity}),  (\ref{eq:particle_velocity_correlation}), and the parabolic exponential autocorrelation function of \citet{Williams1980} in (\ref{eq:correlation_function}).
The combined effects of Mechanism II and gravity are described by the integral in (\ref{eq: MechIIGrav}).
This integral must be solved numerically.
Suitable limits of integration for $w$ are $[-3\sigma_{II}, 3\sigma_{II}]$.
If there is a significant size difference between particles and bubbles, the disturbance of the flow field due to the presence of the bubble is accounted for in particle-bubble collisions.
The correction factor is given in (\ref{eq:RadialComponentCorrection}), where the larger collision partner, usually the bubble, is assumed to be collision partner $j$ and the smaller one collision partner $i$.
The individual components of the modelled radial relative velocity are then combined in (\ref{eq:final_w_r}), and the collision kernel is determined based on the spherical formulation (\ref{eq:colkern_short}).
A concise summary of all the equations needed to implement the model is given in table \ref{tab:model} of appendix \ref{sec:summary}.

\section{Validation with DNS for bubbles and fine particles}
\label{sec:fines}

\subsection{Numerical setup and computational method of DNS}
\label{sec:numMethod}

An important issue with earlier models for the collision frequency is their inaccuracy in predicting collision rates obtained from DNS
 \citep{Chan2023, Tiedemann2024a, Tiedemann2024b}.
Hence, this section and the subsequent section \ref{sec:further_valid} focus on validating the IMSC with well-controlled numerical simulations and compare it with a range of other models.
Beyond validating the model as a whole, the data are used to assess fundamental modelling assumptions and to provide hints for the selection of appropriate models of specific sub-processes.

\begin{figure}
	\centering
	\includegraphics[width=0.7\textwidth]{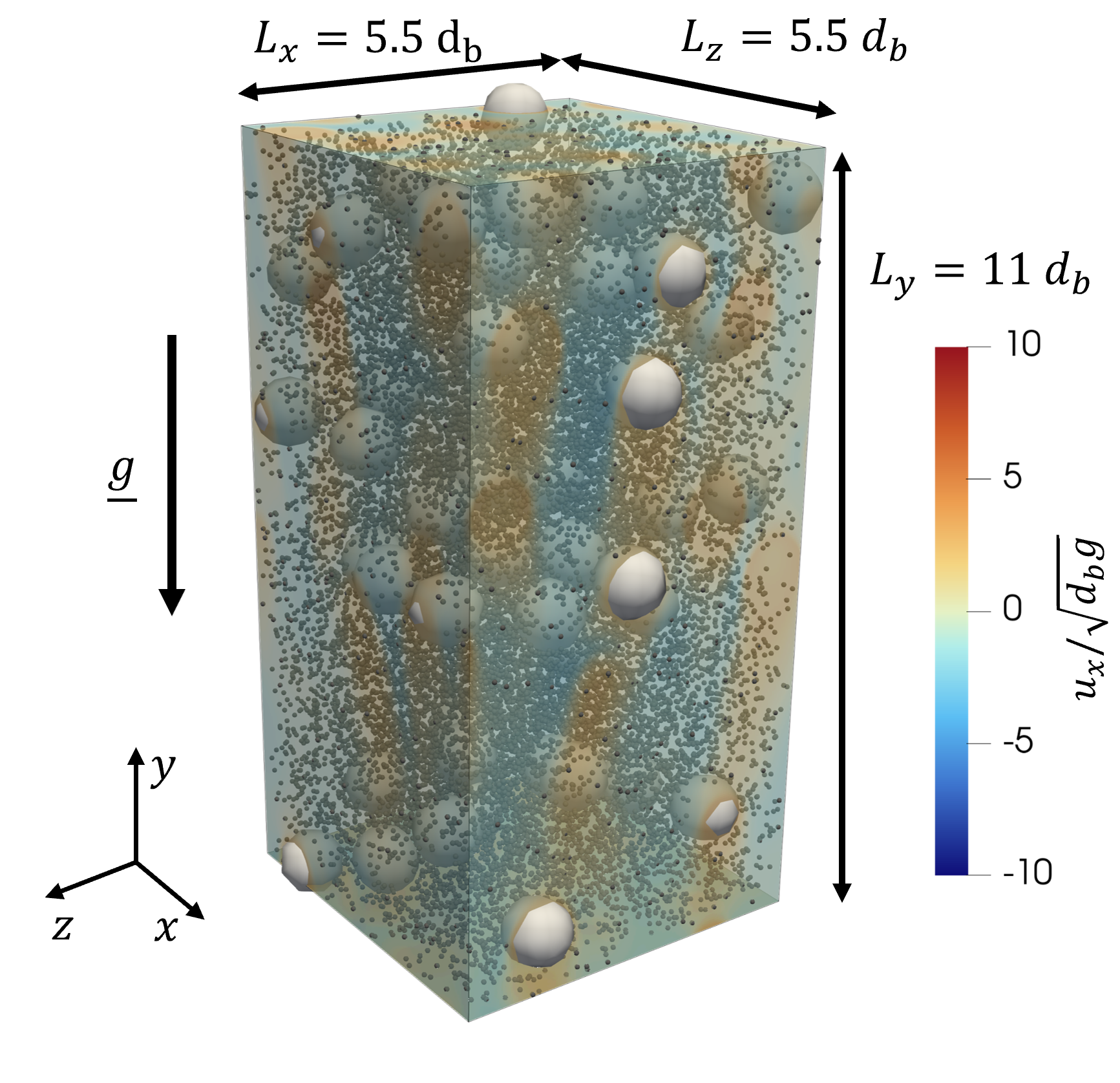}
	\caption{Instantaneous snapshot from a typical simulation to investigate the collision behaviour of bubbles (light) and particles (dark). Particles were enlarged and their concentration was reduced by a factor of 200 for better visibility. Domain extensions and coordinate system are shown as well. The contour plot on the left and right side facing the reader shows the instantaneous vertical velocity component normalised with the buoyancy-related scale $\sqrt{d_bg}$. Data from case G-1-30 defined in table \ref{tab:sim_cases}. Picture taken from \citet{Tiedemann2024b}.}
	\label{fig:domain}
\end{figure}

A comprehensive description of these simulations and the results obtained can be found in \citet{Tiedemann2024a, Tiedemann2024b}.
Here, a brief summary is provided for convenience.
The motion of the fluid phase was described by the unsteady, three-dimensional Navier-Stokes equations for incompressible fluids discretised with a second-order finite-volume scheme on a staggered, Cartesian grid.
The bubbles were modelled as fully-resolved rigid spheres coupled to the fluid by the Immersed Boundary Method of \citet{Tschisgale2018}.
Due to their small size, the fine particles were modelled as two-way coupled Lagrangian point particles.
Particle-bubble collisions were counted once the particle approached the bubble and reached the collision radius $r_c$.
The particle was then removed and reseeded in the fluid domain at an arbitrary position.
This way, statistically stationary conditions are achieved, since the bubbles conserve their buoyancy and the number of particles in the domain remains constant.
The computational domain was defined as a representative volume element of a flotation cell with side lengths of $L_x \times L_y \times L_z=(5.5\times11\times5.5)d_b^3$.
Triple-periodic boundary conditions were used.
The size was chosen after comprehensive tests varying the domain size.
On one hand, the domain size should be large enough to yield reliable flow structures at the particle and bubble scale.
On the other hand, it should be small enough to avoid large-scale clustering as typically observed in bubble swarms \citep{Santarelli2015} or particle swarms \citep{Uhlmann2014}.
An impression of the setup is given in figure \ref{fig:domain}.

The majority of the simulations focused on the separation and concentration zones of the flotation cell, where low turbulence intensity levels prevail \citep{Tiedemann2024b}.
In this situation, the main driver of particle as well as bubble motion is gravity.
Additionally, a case exhibiting medium turbulence levels was obtained using the same method with the additional forcing of the background flow to emulate the effects of the rotor and stator in mechanical flotation cells \citep{Tiedemann2024a}.
This was devised according to \citet{Chouippe2015}, who employed the method of \citet{Eswaran1988}.
Between the different simulation cases, the physical parameters influencing the bubble collision process were varied, such as the bubble diameter $d_b$, the particle diameter $d_p$, particle density $\rho_p$, as well as the gas hold-up $\epsilon_g$ (i.e. the bubble volume fraction), and the particle volume fraction $\epsilon_p$.
The parameters were chosen to cover the range of realistic flotation conditions \citep{Deglon2000, Tabosa2016, Norori2017, Hadler2019, Ran2019, Mesa2020, Ostadrahimi2020}.
Furthermore, a case emulating particle-laden bubbles, i.e. with higher bubble density, was simulated.

The simulation cases are labelled as follows:
First, a letter to indicate the driving mechanism, with G for gravity-driven alone and RXX for gravity and turbulent background forcing as the driver of the flow.
In the latter case, RXX represents the Taylor-Reynolds number of the single-phase flow obtained by the forcing.
The second and third items in the label indicate the bubble diameter and the  particle diameter, respectively. Other parameters different from the reference case G-1-30 are denoted in the last part.
The simulation data obtained over a wide range of flotation conditions enable a comprehensive comparison of the IMSC with other models.
The complete list of cases with the main physical parameters can be found in table \ref{tab:sim_cases}.
 
\begin{table}
	\centering
	\caption{Simulated cases and their physical parameters with the nomenclature defined in the text. Bold values mark the parameters different from the reference case G-1-30 [adapted from \citet{Tiedemann2024b}].}
	\label{tab:sim_cases}
	\begin{tabular}{l|r|r|r|r|r|r|}
		\toprule
		\textbf{Case label} & $d_b$ [$\si{\milli\m}$] & $\rho_b$ [$\si{\kilogram\per\cubic\meter}$] & $\epsilon_g$  [$\si{\percent}$]& $d_p$ [$\si{\micro\m}$] & $\rho_p$ [$\si{\kilogram\per\cubic\meter}$] & $\epsilon_p$ [$\si{\percent}$]\\
		
		\midrule
		
		G-1-30 & $1$ & $1.225$ & $8.8$ & $30$ & $3000$& $10.0$\\
        \midrule
		G-1-50 & $1$ & $1.225$ & $8.8$ & $\bf{50}$ & $3000$& $10.0$\\
		\midrule
		G-0.6-30 & $\bf{0.6}$ & $1.225$ & $8.8$ & $30$ & $3000$& $10.0$\\
		G-1.4-30 & $\bf{1.4}$ & $1.225$ & $8.8$ & $30$ & $3000$& $10.0$\\
		G-1.4-50 & $\bf{1.4}$ & $1.225$ & $8.8$ & $30$ & $3000$& $10.0$\\
		G-2.4-50 & $\bf{2.4}$ & $1.225$ & $8.8$ & $\bf{50}$ & $3000$& $10.0$\\
		G-2.4-70 & $\bf{2.4}$ & $1.225$ & $8.8$ & $\bf{70}$ & $3000$& $10.0$\\
		\midrule
		G-1-30-eg6 & $1$ & $1.225$ & $\bf{6}$ & $30$ & $3000$& $10.0$\\
		G-1-30-eg16 & $1$ & $1.225$ & $\bf{16}$ & $30$ & $3000$& $10.0$\\
		\midrule
		G-1-30-rb25 & $1$ & $\bf{25}$ & $8.8$ & $30$ & $3000$& $10.0$\\
		\midrule
		G-1-30-rp6k & $1$ & $1.225$ & $8.8$ & $30$ & $\bf{6000}$& $10.0$\\
		\midrule
		G-1-30-ep5.0 & $1$ & $1.225$ & $8.8$ & $30$ & $3000$& $\bf{5.0}$\\
		G-1-30-ep7.5 & $1$ & $1.225$ & $8.8$ & $30$ & $3000$& $\bf{7.5}$\\
        \midrule
        R53-1-30 & $1$ & $1.225$ & $8.8$ & $30$ & $3000$& $10.0$\\
		\bottomrule
	\end{tabular}
\end{table}

\subsection{Fluid isotropy}
\label{sec:isotropy}

A main assumption made for the IMSC in section \ref{sec:assumptions} is the presence of homogeneous and isotropic turbulence.
The adequacy of this assumption is validated using the present DNS data.
The anisotropy of the fluid velocities is fully given by the normalised anisotropy tensor as stated by \citet{Pope2000}.
Only regarding the main diagonal elements of the normalised anisotropy tensor, disregarding the isotropic stresses, and using the fact that the fluid velocity fluctuations in $x$- and $y$-direction exhibit almost no difference, the remaining entry of the normalised anisotropy tensor is \citep{Tiedemann2024a, Tiedemann2024b} 
    \begin{equation}
        \label{eq:isotropy}
        a_a = \sqrt{\frac{\langle u^\prime_{y} u^\prime_{y} \rangle}{\langle u^\prime_{x} u^\prime_{x} \rangle}}.
    \end{equation}
\noindent In this anisotropy factor, the fluid velocity fluctuations in the direction of gravity, $u^\prime_{f,y}$, are related to the fluid velocity fluctuations perpendicular to it, $u^\prime_{f,x}$.

For the gravity-driven cases with a bubble diameter of $d_b=\qty{1}{\milli\meter}$, the anisotropy factor is $a_a \approx 2.4$.
For larger bubbles of $d_b=\qty{2.4}{\milli\meter}$, creating a higher turbulence intensity, it reduces to $a_a \approx 1.9$.
In the moderately turbulent case R53-1-30, the anisotropy factor is only $a_a=1.05$, indicating very weak anisotropy.
This is attributable to the larger unsteady sideward components of the bubble motion exhibited, resulting from the turbulent agitation.
As a result, the underlying homogeneous and isotropic forced turbulence is overshadowing the bubble-induced turbulence, thus reducing the overall anisotropy.

Since the most relevant conditions in flotation are those near the rotor and stator assembly, where collisions are turbulence-driven.
Especially for these turbulence-driven cases, the assumption of homogeneous and isotropic turbulence can be made.

\subsection{Velocity distribution}
\label{sec:velo_distribution}

An assumption of the IMSC made in section \ref{sec:decomposition} and in other collision frequency models is that the velocity distribution of particles and bubbles is Gaussian.
Although it is well known that this is not entirely true, the literature suggests that a Gaussian distribution of particle and bubble velocity can be a good approximation for the first two moments of the probability distribution \citep{Wang1996}.
The validity of this assumption and the modelled velocity distributions is now validated by the data published in \citet{Tiedemann2024a, Tiedemann2024b}.
Under the conditions addressed there, the Kolmogorov length scale is of the order of the particle diameter, while the bubble diameter is larger than the Taylor length scale for all cases.
In this situation, the model sets $S_{\ell\ell}^\mathrm{(IMSC)}(r_p)>0$ and $S_{\ell\ell}^\mathrm{(IMSC)}(r_b)=0$, according to (\ref{eq:blend_sin}).

\begin{figure}
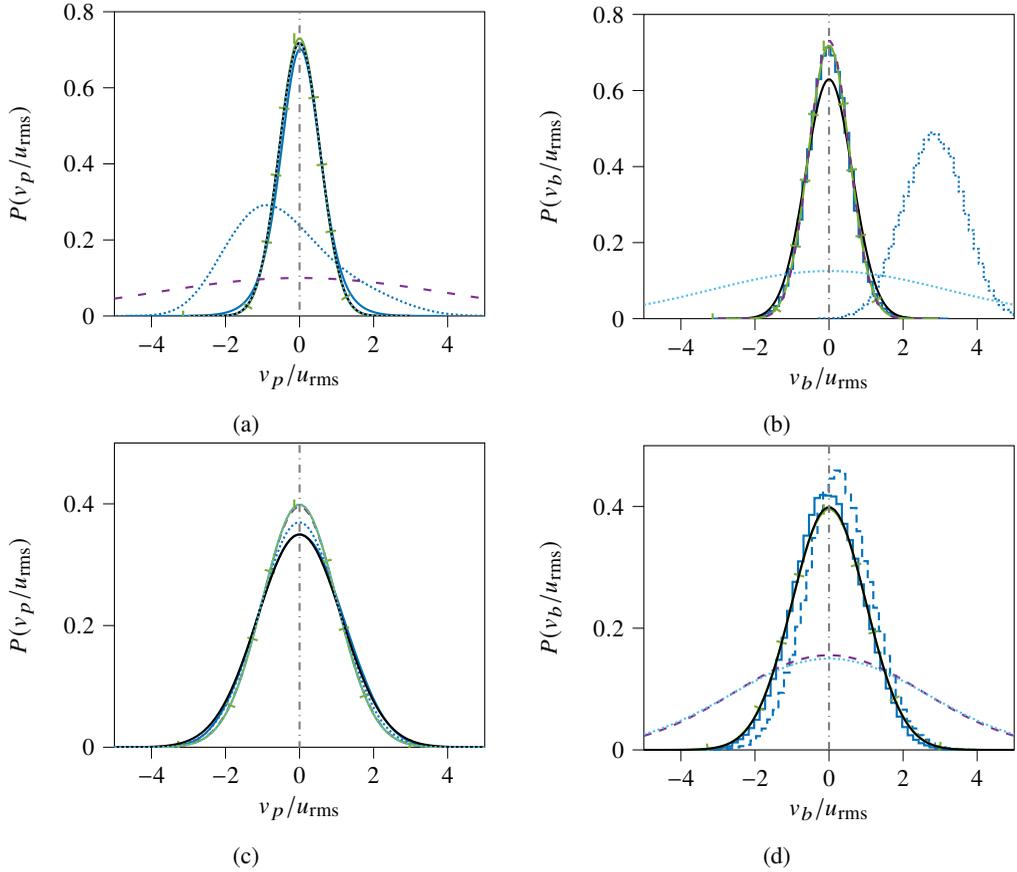

	\centering
    \begin{subfigure}{0.48\textwidth}
		\setlength\fwidth{\textwidth}
		\input{figures/hist-std-P-all.tex}
		\caption{}
		\label{fig:hist-std-P-all}
	\end{subfigure}
    \hfill
    \begin{subfigure}{0.48\textwidth}
		\input{figures/hist-std-B-all.tex}
		\caption{}
		\label{fig:hist-std-B-all}
	\end{subfigure}
    \begin{subfigure}{0.48\textwidth}
		\input{figures/hist_R53_P.tex}
		\caption{}
		\label{fig:hist-std-P-all}
	\end{subfigure}
    \hfill
	\begin{subfigure}{0.48\textwidth}
		\input{figures/hist_R53_B.tex}
		\caption{}
		\label{fig:hist-std-B-all}
	\end{subfigure}

	\caption{PDF of particle and bubble velocity for simulation cases R53-1-30 and G-1-30, as defined in table \ref{tab:sim_cases}, and comparison with several models. (a) Particle velocity for G-1-30, (b) bubble velocity for G-1-30, (c) particle velocity for R53-1-30, (d) bubble velocity for R53-1-30. With $\alpha=b,p$ the curves show \ref{plt: SIMline} $v_{\alpha x}$ from DNS, \ref{plt: SIMzline} $v_{\alpha y}$ from DNS, \ref{plt: NMline} $v_\alpha$ with the ISMC, \ref{plt: Aline} $v_\alpha$ with the model of \citet{Abrahamson1975}, \ref{plt: CbbKK} $v_\alpha$ with the model of \citet{Kruis1997}, \ref{plt: Zline} $v_\alpha$ with the model of \citet{Zaichik2010}. The vertical line denotes $v_\alpha=0$. In a) the curves of DNS and IMSC are on top of each other. }
	\label{fig:veloHist}
\end{figure}

Figure \ref{fig:veloHist} shows the probability density function of the particle and bubble velocities for the gravity-driven case G-1-30 and the turbulence-driven case R53-1-30. 
Detailed information on the first four moments of the particle and bubble velocity distribution for all cases is provided in tables \ref{tab:veloP} and \ref{tab:veloB} of appendix \ref{sec:velo_dist_app}.
The corresponding IMSC results are also given for comparison.

For the gravity-driven case G-1-30, the mean velocity perpendicular to gravity is close to zero for both  particles and bubbles.
Perpendicular to gravity, the skewness and kurtosis of the bubble velocity distribution are close to the values of a Gaussian probability distribution with a skewness of zero and a kurtosis of three.
While the particle velocity distribution in these directions is also only barely skewed, it has positive excess kurtosis.
In the direction of gravity, the mean velocity of the bubbles is positive and the mean velocity of the particles is negative, indicating their rising and settling, respectively.
Furthermore, the variance of the velocity distribution obtained from the simulation in the direction of gravity is higher than those perpendicular to gravity, indicating more velocity fluctuations of particles and bubbles in this direction.
The skewness and kurtosis of the bubble velocity are still close to those of a Gaussian distribution.
However, the particle velocity distribution in the direction of gravity is negatively skewed.
This is due to the fact that in cases where gravity is the main driving force, the motion of the particle in the vertical direction is strongly influenced by the wakes of the bubbles.
Also, in the direction of gravity, the particle velocity distribution has positive excess kurtosis.

For the turbulence-driven case R53-1-30, the mean of both the particle and the bubble velocity distribution is close to zero in all spatial directions.
The offset in the direction of gravity is small compared to the gravity-driven cases, since the isotropic turbulence-induced motion of particles and bubbles dominates over the anisotropic gravity-induced motion.
The third and fourth moments of both velocity distributions are well matched by a Gaussian probability distribution in all directions.
As highlighted above, the data show that for the purpose of modelling, where only the first two moments of the velocity distribution are needed, a Gaussian probability distribution is a suitable choice.

Figure \ref{fig:veloHist} also compares the velocity distributions with different models from the literature.
It should be noted that only those models capable of independently modelling the velocity distributions of the collision partners are shown here.
The velocities perpendicular to the direction of gravity are the focus of this analysis.
For case R53-1-30, the particle velocity distribution is almost exactly matched by the IMSC.
The models by \citet{Abrahamson1975}, \citet{Kruis1997}, and \citet{Zaichik2010} provide similar predictions also closely matching the DNS data.
For case G-1-30, the particle velocity distribution is well matched by the IMSC and the models of Kruis and Kusters, and Zaichik.
The model of \citet{Abrahamson1975} does not agree with the particle velocity distribution, as it was designed for particles with $St\rightarrow\infty$ and no gravity, which does not align with the conditions of the simulations.

The bubble velocity distribution for both cases, G-1-30 and R53-1-30, is well matched by the IMSC and the model of Kruis and Kusters, which provide similar predictions.
The model of Abrahamson captures only the bubble velocity distribution for case G-1-30.
Conversely, the bubble velocity distribution is not well predicted by the model of \citet{Zaichik2010} in both cases.
For case R53-1-30, it provides a prediction close to the model of Abrahamson.
A more detailed discussion of the models and their match to the simulation data can be found in the following sections.

\subsection{Collision kernel}

\subsubsection{Particle-bubble collision kernel}
\label{sec:valid_pb_coll_kern}

\begin{figure}
	\centering
	\setlength\fwidth{0.5\textwidth}
%

%
\begin{tikzpicture}

\begin{axis}[%
width=0.951\fwidth,
height=0.75\fwidth,
at={(0\fwidth,0\fwidth)},
scale only axis,
xmin=0,
xmax=15,
ymode=log,
ymin=0.001,
ymax=10,
yminorticks=true,
ylabel style={font=\color{white!15!black}},
ylabel={$\Gamma{}_{pb}(\tau_\eta /r_c^3)$},
axis background/.style={fill=white},
title style={font=\bfseries},
xmajorgrids,
ymajorgrids,
yminorgrids,
xtick={1,2,...,14},
xticklabel style={rotate=45.0,anchor=east},
xticklabels={
	G-1-30,
	G-1-50,
	G-1.4-30,
	G-1.4-50,
	G-2.4-50,
	G-2.4-70,
	G-0.6-30,
	G-1-30-eg6,
	G-1-30-eg16,
	G-1-30-rb25,
	G-1-30-rp6k,
	G-1-30-ep7.5,
	G-1-30-ep5.0,
	R53-1-30,
},
]

\addplot [color=Zcol, line width=1pt, densely dashed, forget plot, mark=*, mark options={solid, Zcol}, mark size = 1pt]
  table[row sep=crcr]{%
1 2.0331584038192756\\
2 2.002363859735696\\
3 2.1277829929795407\\
4 2.1062267273881248\\
5 1.8616995015569004\\
6 1.8562857766368597\\
7 1.4320469796174122\\
8 2.1052878305206018\\
9 1.8090007386899989\\
10 1.9751834459712783\\
11 2.021919284436846\\
12 2.0244597376243822\\
13 2.0202158690868557\\
14 2.6572826381013863\\
};\label{plt:CpbZ}

\addplot [color=Kcol, line width=1pt, mark=triangle*,  mark size=2pt, mark options={solid, Kcol}]
  table[row sep=crcr]{%
1 0.06631091782661976\\
2 0.1394741202136171\\
3 0.047046587644843145\\
4 0.09936422158303869\\
5 0.05356972111689405\\
6 0.08905215592978676\\
7 0.10384492264827021\\
8 0.0672606304382423\\
9 0.06397966655353064\\
10 0.06478143954379761\\
11 0.0660256425151427\\
12 0.066124524123895\\
13 0.06617922614903866\\
14 0.11562775062413043\\
};\label{plt:CpbK}

\addplot [color=STcol, line width=1pt, densely dashdotted, mark=*, mark options={solid, STcol}, mark size = 1pt,   forget plot]
  table[row sep=crcr]{%
1 1.2944172750371328\\
2 1.294417275037133\\
3 1.2944172750371328\\
4 1.2944172750371328\\
5 1.2944172750371328\\
6 1.294417275037133\\
7 1.2944172750371328\\
8 1.2944172750371328\\
9 1.2944172750371328\\
10 1.2944172750371326\\
11 1.294417275037133\\
12 1.294417275037133\\
13 1.2944172750371328\\
14 1.2944172750371328\\
};\label{plt:CpbST}

\addplot [color=Acol, line width=1pt, loosely dashed,mark=*, mark options={solid, Acol}, mark size = 1pt, forget plot]
  table[row sep=crcr]{%
1 2.9617155743696335\\
2 2.9053864158556304\\
3 2.672938574044012\\
4 2.631212120486066\\
5 2.1081675904935704\\
6 2.0963011353362084\\
7 3.3222140892584373\\
8 3.1469369263398552\\
9 2.5768698868948894\\
10 2.912608982884004\\
11 2.9459792276580967\\
12 2.948986900431763\\
13 2.92682660246682\\
14 3.491538040067521\\
};\label{plt:CpbA}

\addplot [color=SIMcol, line width=1.0pt, mark=o, mark options={solid, SIMcol}]
  table[row sep=crcr]{%
1	0.035649853467125\\
2	0.14154660416138537\\
3	0.012236722884816875\\
4	0.039908204337231676\\
5	0.007923535026498083\\
6	0.06558871443194216\\
7	0.11612555631302433\\
8	0.03900231856850735\\
9	0.03373129388033387\\
10	0.03856058779537775\\
11	0.11788870088343002\\
12	0.03675863570799846\\
13  0.03647620075724349\\
14  0.01811648287230616\\
};\label{plt:CpbSIM}

\addplot [color=NMcol, line width=1pt, mark=o,  mark size=1pt, mark options={solid, NMcol}]
table {%
	1 0.0380081693029321
	2 0.07985452322615397
	3 0.02396885182744528
	4 0.044176342767246
	5 0.019910373777420804
	6 0.0491308140927911
	7 0.0684075857810366
	8 0.038438142714551
	9 0.037645752750108
	10 0.038060254879791
	11 0.04824474962999049
    12 0.0380027161086684
    13 0.037978986954087
    14 0.0354826518125928
};\label{plt:CpbIMSC}

\addplot [color=KKcol, line width=1pt, mark=*, mark options={solid, KKcol}, mark size = 1pt, forget plot]
table {%
1 0.0777047293991377
2 0.137889143560277
3 0.0544342562169823
4 0.0936121639807566
5 0.0515551397530874
6 0.0722359046529484
7 0.13619108284273
8 0.0771056456898603
9 0.0765223614667019
10 0.0779962782394018
11 0.109271462730376
12 0.077760658925689
13 0.077826071229568
14 0.0719838320374515
};\label{plt:CpbKK}

\addplot [color=Dcol, line width=1pt, mark=diamond, mark size = 1pt, forget plot]
table {%
1 0.044186123633696
2 0.18579246184431
3 0.00473181386588771
4 0.0955366075066166
5 0.00301613532456002
6 0.070269317513272
7 0.143057734205736
8 0.0495664836773769
9 0.03187237410424967
10 0.0460317780272028
11 0.0460317780272028
12 0.044124061519486
13 0.0431023469279624
14 0.005087155870713874
};\label{plt:CpbD}

\addplot [color=Effcol, line width=1pt, mark=otimes, mark size = 1pt, forget plot]
table {%
1 6.31000477807275
2 6.27507692334774
3 6.11987866103992
4 6.06437443948937
5 5.92078742714671
6 5.90947594204009
7 6.8139862727537
8 6.39590600716596
9 6.12158203984765
10 6.4172206211746
11 6.36424890929397
12 6.31646249987106
13 6.28598614360518
14 4.85439605538602
};\label{plt:CpbNGC}
\end{axis}

\begin{axis}[%
width=1.227\fwidth,
height=0.92\fwidth,
at={(-0.16\fwidth,-0.101\fwidth)},
scale only axis,
xmin=0,
xmax=1,
ymin=0,
ymax=1,
axis line style={draw=none},
ticks=none,
axis x line*=bottom,
axis y line*=left
]
\end{axis}
\end{tikzpicture}%
	\caption{Comparison of non-dimensional bubble-particle collision kernel obtained from DNS and various models. The cases are defined in Table \ref{tab:sim_cases}.\ref{plt:CpbSIM} DNS, \ref{plt:CpbIMSC} IMSC, \ref{plt:CpbA} Abrahamson, \ref{plt:CpbK} Kostoglou, \ref{plt:CpbKK} Kruis \& Kusters, \ref{plt:CpbST} Saffman \& Turner, \ref{plt:CpbD} Dodin \& Elperin, \ref{plt:CpbZ} Zaichik, \ref{plt:CpbNGC} Ngo-Cong.}
	\label{fig:ColKernPB}
\end{figure}
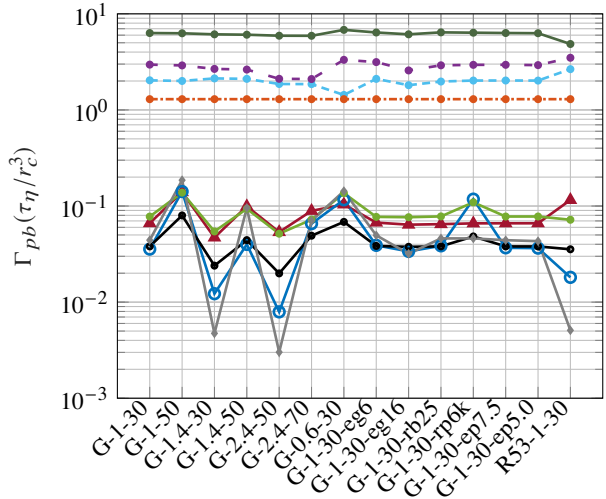

Due to its direct influence on the flotation performance and recovery, an accurate prediction of the particle-bubble collision kernel is of high interest.
Figure \ref{fig:ColKernPB} provides a comparison of the results obtained from the DNS with those from the IMSC and different models for the collision kernel from the literature.
The input parameters for the models were the same as for the simulations listed in table \ref{tab:sim_cases} and described in section \ref{sec:numMethod}.
The values for $k$, $\varepsilon$, and $Re_\alpha$ were taken from the DNS results.
The line connecting the data points is for orientation only and does not imply a physical connection.
In the simulations, several parameters like the turbulence forcing, the bubble diameter, or the particle diameter were varied.
On this basis, conclusions can be drawn which effects are well captured by the models.

The models of \citet{Saffman1956}, \citet{Abrahamson1975}, \citet{Zaichik2010}, and \citet{NgoCong2018} significantly overpredict the simulation results by almost two orders of magnitude.
All these models were designed for collision partners of similar size, do not take into account the changes in the flow field in the vicinity of the bubble, and omit the influence of gravity.
In particular, the models of Saffman and Turner, and Abrahamson are based on the limiting cases for $St_\alpha\rightarrow0$ and $St_\alpha\rightarrow\infty$, respectively, without the influence of gravity.
Therefore, the underlying modelling assumptions do not match the conditions in the flotation and those present in the simulation.
\citet{Dodin2002} started from the formulation of Saffman and Turner and incorporated the effect of gravity.
Using this formulation, significantly better results are obtained.
Their model, when compared to that of Saffman and Turner, shows that the combined influence of turbulence and gravity results in a lower collision kernel than would be found in a purely turbulent environment.
The deviations observed with the model by Zaichik can be attributed to an overestimation of the variance in the bubble velocity. 

The models incorporating the effects of gravity, such as those of \citet{Kostoglou2020b}, \citet{Kruis1997}, \citet{Dodin2002}, as well as the IMSC, provide collision kernels in the same order of magnitude as the DNS data, with the IMSC yielding the best overall fit to the simulation data.
The model of \citet{Kostoglou2020b} was developed specifically for the case of fine particles, also including the flow distortion due to the bubble.
Hence, its good match is not surprising.
However, different models for the bubble and particle terminal velocity, drag modification and differences in the fluid structure function make the IMSC provide a better fit.
The Kruis and Kusters model is largely in agreement with the Kostoglou model.
However, as previously mentioned, it has an inconsistent use of reference frames and employs the cylindrical collision kernel.
This prevents it from being generally applicable.

While many cases are well predicted by the IMSC, the increase in collision frequency with increasing particle density is not well represented (case G-1-30-rp6k).
This finding is consistent with the other models, which also demonstrate only a marginal increase over their baseline for the case G-1-30.
One potential explanation for this could be a more uneven distribution of collision angles along the bubble circumference for gravitationally driven bubble motion \citep{Tiedemann2024b}.

Case G-1-30-rb25 emulates a loaded bubble with a large portion of the bubble surface area covered with particles.
This was obtained by using a higher bubble density.
In particular, the Kostoglou model and the IMSC include sub-models for the bubble rise velocity based on unladen bubbles without $\rho_b$ as input parameter.
However, as the bubble density is still low compared to the fluid density, only a little difference in bubble motion and collision behaviour was observed in the DNS \citep{Tiedemann2024b}.
Hence, the model by Kostoglou and the IMSC are still able to accurately predict the respective collision kernel.

Most other cases with $d_b=\SI{1}{mm}$ or $d_b=\SI{0.6}{mm}$ are well matched by the IMSC.
The largest discrepancies are noticed for cases G-1.4-30 and G-2.4-50.
These cases exhibit the most pronounced size disparity between particles and bubbles within this study.
For all cases with $d_b=\SI{2.4}{mm}$ the bubble rise velocity obtained from the simulations and the ones resulting from the employed sub-model by \citet{Rodrigue2001} differ as shown in figure \ref{fig:Re_b} and discussed in section \ref{sec:grav} above.
Furthermore, the slight onset of particle clustering was observed in \citep{Tiedemann2024b}, which may have implications for the particle-bubble collision kernel, not captured by any model \citep{Chan2023}.

\subsubsection{Particle-particle collision kernel}

\begin{figure}
	\centering
	\setlength\fwidth{0.5\textwidth}
%
%
\definecolor{mycolor1}{rgb}{0.00000,0.44700,0.74100}%
\definecolor{mycolor2}{rgb}{0.85000,0.32500,0.09800}%
\definecolor{mycolor3}{rgb}{0.49400,0.18400,0.55600}%
\definecolor{mycolor4}{rgb}{0.30100,0.74500,0.93300}%
\definecolor{mycolor5}{rgb}{0.46600,0.67400,0.18800}%
\definecolor{mycolor6}{rgb}{0.92900,0.69400,0.12500}%
\begin{tikzpicture}

\begin{axis}[%
width=0.951\fwidth,
height=0.75\fwidth,
at={(0\fwidth,0\fwidth)},
scale only axis,
xmin=0,
xmax=15,
xlabel style={font=\color{white!15!black}},
ymode=log,
ymin=0.1,
ymax=10,
yminorticks=true,
ylabel style={font=\color{white!15!black}},
ylabel={$\Gamma{}_{pp}(\tau_\kolmlen /r_c^3)$},
axis background/.style={fill=white},
title style={font=\bfseries},
xmajorgrids,
ymajorgrids,
yminorgrids,
xtick={1,2,...,14},
xticklabel style={rotate=45.0,anchor=east},
xticklabels={
	G-1-30,
	G-1-50,
	G-1.4-30,
	G-1.4-50,
	G-2.4-50,
	G-2.4-70,
	G-0.6-30,
	G-1-30-eg6,
	G-1-30-eg16,
	G-1-30-rb25,
	G-1-30-rp6k,
	G-1-30-ep7.5,
	G-1-30-ep5.0,
	R53-1-30},
]
\addplot [color=SIMcol, line width=1.0pt, mark=o, mark options={solid, SIMcol}]
  table{%
1 1.08154777915934
2 1.14357034740347
3 1.11513336960435
4 1.3137342255912
5 1.68363418980195
6 2.3955253206656
7 1.0410226173427
8 0.966593702148408
9 1.37222362521303
10 1.12546688696884
11 1.11586283057157
12 1.07679443084234
13 1.05567708760609
14 2.13687436243495
};

\addplot [color=STcol, line width=1pt, densely dashdotted, mark=*, mark options={solid, STcol}, mark size = 1pt,   forget plot]
  table{%
1 1.29441727503713
2 1.29441727503713
3 1.29441727503713
4 1.29441727503713
5 1.29441727503713
6 1.29441727503713
7 1.29441727503713
8 1.29441727503713
9 1.29441727503713
10 1.29441727503713
11 1.29441727503713
12 1.29441727503713
13 1.29441727503713
14 1.29441727503713
};

\addplot [color=Acol, line width=1pt, loosely dashed,mark=*, mark options={solid, Acol}, mark size = 1pt, forget plot]
  table{%
1 74938.3792432862
2 16215.492141441
3 184466.091736209
4 39818.9884694953
5 158891.550674402
6 58101.664630933
7 18447.2291468341
8 79469.6458468136
9 65056.416350439
10 73732.1668907552
11 74639.8529300104
12 74625.9108715585
13 74081.5967098059
14 87813.2241818779
};

\addplot [color=Zcol, line width=1pt, densely dashed, forget plot, mark=*, mark options={solid, Zcol}, mark size = 1pt]
  table{%
1 1.29664039240744
2 1.31444100099259
3 1.3083834205581
4 1.34710160794688
5 1.43623023190995
6 1.53940754075011
7 1.28481546925748
8 1.29606589219769
9 1.29647734300834
10 1.29521218983535
11 1.3801769762684
12 1.29652792533581
13 1.29682270856952
14 1.45709006909568
};

\addplot [color=KKcol, line width=1pt, mark=|,  mark size=2pt, mark options={solid, KKcol}]
  table{%
1 0.495123048820928
2 0.522201619745355
3 0.499193200376006
4 0.533691292714095
5 0.550270204173427
6 0.602700448667592
7 0.486262973521223
8 0.493409740861184
9 0.497106644923181
10 0.494107509140799
11 0.589194090231909
12 0.495108764899409
13 0.49552936867715
14 0.542308725739936
};

\addplot [color=NMcol, line width=1pt, mark=*,  mark size=1pt, mark options={solid, NMcol}]
  table{%
1 1.28696988478126
2 1.28180775708592
3 1.2880454795744
4 1.28517639780237
5 1.28967678788614
6 1.29124491605826
7 1.28321214153625
8 1.28747672499209
9 1.28602941626423
10 1.28685745529471
11 1.29427725574235
12 1.28789414088562
13 1.28939800407047
14 1.29110299510973
};

\end{axis}

\begin{axis}[%
width=1.227\fwidth,
height=0.92\fwidth,
at={(-0.16\fwidth,-0.101\fwidth)},
scale only axis,
xmin=0,
xmax=1,
ymin=0,
ymax=1,
axis line style={draw=none},
ticks=none,
axis x line*=bottom,
axis y line*=left
]
\end{axis}
\end{tikzpicture}%
	\caption{Comparison of non-dimensional particle-particle collision kernel obtained from DNS and various models. The cases are defined in Table \ref{tab:sim_cases}.\ref{plt:CpbSIM} DNS, \ref{plt:CpbIMSC} IMSC, \ref{plt:CpbKK} Kruis \& Kusters, \ref{plt:CpbST} Saffman \& Turner, \ref{plt:CpbZ} Zaichik. Data of IMSC and Saffman \& Turner are on top of each other.}
	\label{fig:ColKernPP}
\end{figure}
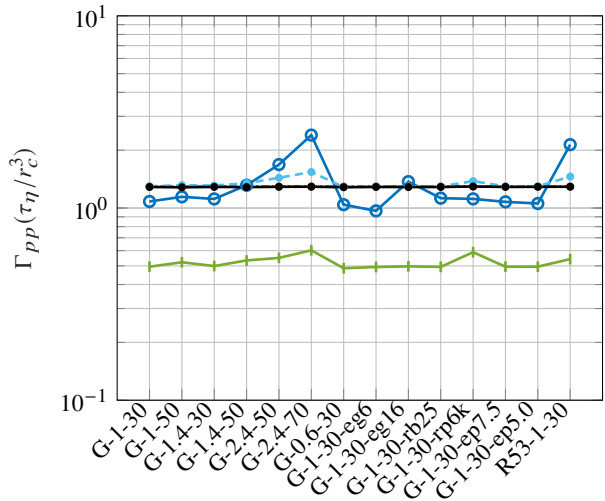

In addition to the important particle-bubble collision kernel, also collisions between pairs of particles and pairs of bubbles take place in a flotation cell.
In figure \ref{fig:ColKernPP} the particle-particle collision kernel obtained from the DNS in \citet{Tiedemann2024a, Tiedemann2024b} is compared with the results of the IMSC and the models of \citet{Kruis1997}, \citet{Saffman1956}, and \citet{Zaichik2010}.
Note the significantly narrower range of the vertical axis compared to the other graphs reporting collision kernels.
Models that assume coarse particles with $St_p\rightarrow\infty$ or that assume a significant size difference between the collision partners such as those of \citet{Abrahamson1975} and \citet{Kostoglou2020b}, are not included, because the assumptions in these models are very different from the situation to be described resulting in large discrepancies with the simulation results.
All models retained  are very close to the simulation results, with the IMSC and the model of Zaichik almost reducing to the limiting case of Saffman and Turner.
The errors of the predictions for $\Gamma_{pp}$ are comparatively small when evaluated against the errors obtained for the other collision kernels.
This is to be expected, since the particles in the DNS are mostly of very low Stokes numbers and $Re<1$, thus matching the respective assumptions.

The collision kernels obtained by the DNS fluctuate between cases, which is not properly captured by the models.
The movement of particles in the simulation is found to be strongly influenced by the local flow field.
Particularly in the vicinity of the bubbles and in regions of high shear the particle motion increases in magnitude and becomes anisotropic.
These conditions have a significant impact on the collision kernel.
However, they are not captured by the models, which, instead, assume a constant turbulence intensity over the entire sub-domain considered.

The largest differences exist for the cases with $d_b=\SI{2.4}{mm}$ and R53-1-30.
These cases have the largest differences in the local fluid velocity fluctuations compared to the average ones \citep{Tiedemann2024a, Tiedemann2024b}.
Furthermore, the particle Stokes numbers are highest for these cases.
Thus, there is a slight beginning of particle accumulation in certain regions of the flow \citep{Tiedemann2024a, Tiedemann2024b}.
This particle clustering has been shown to increase the rate of particle-particle collisions \citep{Vosskuhle2014, Ireland2016}.
While the majority of models predict collision kernels that align closely with the DNS data and the limiting case of Saffman and Turner, the collision kernels predicted by the Kruis and Kusters model are found to be inadequate.
This finding agrees with the earlier analyses by \citet{Wang1998} who showed that the cylindrical collision kernel used is not suitable in this situation.

\subsubsection{Bubble-bubble collision kernel}

\begin{figure}
	\centering
	\setlength\fwidth{0.5\textwidth}
%

%
\begin{tikzpicture}

\begin{axis}[%
width=0.951\fwidth,
height=0.75\fwidth,
at={(0\fwidth,0\fwidth)},
scale only axis,
xmin=0,
xmax=15,
xlabel style={font=\color{white!15!black}},
ymode=log,
ymin=0.09,
ymax=3,
yminorticks=true,
ylabel style={font=\color{white!15!black}},
ylabel={$\Gamma{}_{bb}(\tau_\kolmlen /r_c^3)$},
axis background/.style={fill=white},
title style={font=\bfseries},
xmajorgrids,
ymajorgrids,
yminorgrids,
xtick={1,2,...,14},
xticklabel style={rotate=45.0,anchor=east},
xticklabels={
	G-1-30,
	G-1-50,
	G-1.4-30,
	G-1.4-50,
	G-2.4-50,
	G-2.4-70,
	G-0.6-30,
	G-1-30-eg6,
	G-1-30-eg16,
	G-1-30-rb25,
	G-1-30-rp6k,
	G-1-30-ep7.5,
	G-1-30-ep5.0,
	R53-1-30},
]

\addplot [color=SIMcol, line width=1.0pt, mark=o, mark options={solid, SIMcol},forget plot]
table{%
1 0.4651062529496585
2 0.486588167334862
3 0.5268856829688963
4 0.4847489567260639
5 0.4298208130813382
6 0.45055130349738537
7 0.28884762130405556
8 0.4149505961708756
9 0.4587418008857164
10 0.5139306428724101
11 0.4974651104237759
12 0.46354222607382656
13 0.43836451229975015
14 0.7352496535169
};\label{plt: CbbSIM}

\addplot [color=STcol, line width=1pt, densely dashdotted, mark=*, mark options={solid, STcol}, mark size = 1pt,   forget plot]
  table{%
1 1.29441727503713
2 1.29441727503713
3 1.29441727503713
4 1.29441727503713
5 1.29441727503713
6 1.29441727503713
7 1.29441727503713
8 1.29441727503713
9 1.29441727503713
10 1.29441727503713
11 1.29441727503713
12 1.29441727503713
13 1.29441727503713
14 1.29441727503713
};\label{plt: CbbST}

\addplot [color=Acol, line width=1pt, loosely dashed,mark=*, mark options={solid, Acol}, mark size = 1pt, forget plot]
  table{%
1 0.000672950949498938
2 0.00184522790778624
3 0.000301986943225948
4 0.000829609727286004
5 0.000217676674055257
6 0.00042452122033627
7 0.00218948241716383
8 0.000725164506648811
9 0.00059492054435989
10 0.000659401270194172
11 0.000662786665085107
12 0.000669427384178183
13 0.000663312061752729
14 0.000827409217785858
};\label{plt: CbbA}

\addplot [color=Zcol, line width=1pt, densely dashed, forget plot, mark=*, mark options={solid, Zcol}, mark size = 1pt]
  table{%
1 1.47986166484447
2 1.47980471511501
3 1.56655512129625
4 1.56792555240351
5 1.39777158978299
6 1.40155044537697
7 1.0708037783021
8 1.53329173447658
9 1.3200526144789
10 1.43577885951102
11 1.47117483134252
12 1.47332300742108
13 1.47048151780383
14 1.96089857352524
};\label{plt: CbbZ}

\addplot [color=KKcol, line width=1pt, mark=|,  mark size=2pt, mark options={solid, KKcol},forget plot]
  table{%
1 1.46907582084479
2 1.46868298223709
3 1.50079832861355
4 1.5010778151244
5 1.49745841051311
6 1.49765729300661
7 1.35979953318838
8 1.46461268194425
9 1.40555037842778
10 1.46506533106294
11 1.46845576912825
12 1.46885945788836
13 1.4699169950895
14 1.56989716020482
};\label{plt: CbbKK}

\addplot [color=NMcol, line width=1pt, mark=*,  mark size=1pt, mark options={solid, NMcol}]
  table{%
1 0.398300549589287
2 0.390667025071043
3 0.518098389986084
4 0.511666218987644
5 0.564841756914138
6 0.556969418369927
7 0.195778689438525
8 0.444131279378273
9 0.293566687910309
10 0.38188700566797
11 0.393491976822688
12 0.397033128517647
13 0.399497531833011
14 0.756286817013369
};\label{plt: CbbNM}

\end{axis}

\begin{axis}[%
width=1.227\fwidth,
height=0.92\fwidth,
at={(-0.16\fwidth,-0.101\fwidth)},
scale only axis,
xmin=0,
xmax=1,
ymin=0,
ymax=1,
axis line style={draw=none},
ticks=none,
axis x line*=bottom,
axis y line*=left
]
\end{axis}
\end{tikzpicture}%
	\caption{Comparison of non-dimensional bubble-bubble collision kernel obtained from DNS and various models. The cases are defined in Table \ref{tab:sim_cases}.\ref{plt:CpbSIM} DNS, \ref{plt:CpbIMSC} IMSC, \ref{plt:CpbKK} Kruis \& Kusters, \ref{plt:CpbST} Saffman \& Turner, \ref{plt:CpbZ} Zaichik.}
	\label{fig:ColKernBB}
\end{figure}
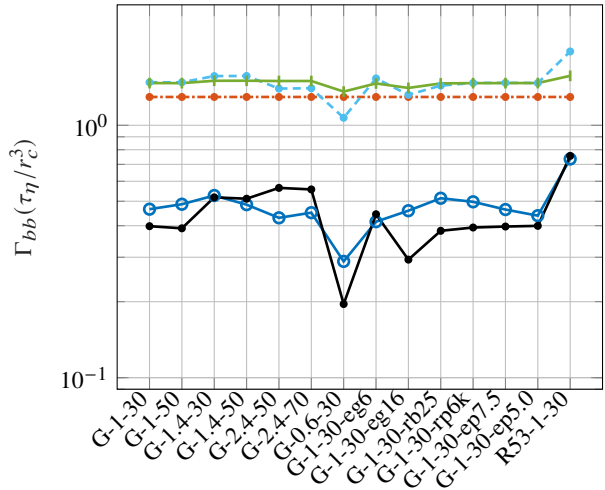

The third type of collisions is constituted by bubble-bubble collisions.
In figure \ref{fig:ColKernBB}, the collision kernels obtained in the DNS are compared with the IMSC and other models.
Comparisons are made with the models of \citet{Zaichik2010}, \citet{Saffman1956}, and \citet{Kruis1997}.
The model by \citet{Kostoglou2020b} is specifically designed for the collisions of fine particles with bubbles, hence not appropriate for equal-sized bubbles.
The model by \citet{Abrahamson1975} significantly underpredicts $\Gamma_{bb}$, as it assumes point particles with $St_p\rightarrow\infty$ not matched by the bubbles.
Hence, neither of the two is shown here.
The model by \citet{Dodin2002} is not shown, as the incorporation of gravitational effects into the model of Saffman and Turner has no added benefit for collision partners of the same class, $i=j$, as no relative velocity due to gravitational settling is obtained.

The remaining three models from the literature in figure \ref{fig:ColKernBB} are found to significantly overpredict the bubble-bubble collision kernel.
This discrepancy can be attributed, at least in part, to the Taylor gradient (\ref{eq:Sll_Taylor}) employed for the fluid velocity.
As shown in section \ref{sec:structure_func} above, this description increases monotonically with $r$, leading to an excessive contribution of Mechanism I at high collision radii, $r_c$.
Also, the model by Zaichik overpredicts the collision kernel.
This is attributable to an overestimation of the contribution of Mechanism I.
The motion of larger bubbles is less sensitive to small-scale fluid motions.
Therefore, the contribution of Mechanism I should decrease with increasing $r_b$.
Furthermore, swarm effects reducing the bubble velocity are not taken into account.

The IMSC incorporates all these effects, thus yielding a more accurate prediction of the collision kernel.
In particular, the cases with $d_b=\qty{1.4}{mm}$, G-1-30-eg6 and R53-1-30 are matched closely.
The slight alterations of the bubble-bubble collision kernel with changing particle parameters are not captured by the IMSC or any other model (cases G-1-30-dp50, G-1-30-ep7.5, and G-1-30-ep5.0).
With decreasing particle concentration, a better match of the bubble-bubble collision kernel is obtained (case G-1-30-ep5.0).
Like all models, the IMSC is only concerned with the two collision partners and disregards any other dispersed phase present in the flow. Hence, the particles and their small influence on the bubble-bubble collision kernel are not taken into account.
A further investigation of the influencing factor is performed in the next section by examining the radial relative velocity.

\subsubsection{Bubble-bubble radial relative velocity}

\begin{figure}
	\centering
	\setlength\fwidth{0.5\textwidth}
%
%

\definecolor{mycolor7}{rgb}{0,0,0}%
\begin{tikzpicture}

\begin{axis}[%
width=0.951\fwidth,
height=0.75\fwidth,
at={(0\fwidth,0\fwidth)},
scale only axis,
xmin=0,
xmax=15,
xlabel style={font=\color{white!15!black}},
ymin=0,
ymax=4,
ylabel style={font=\color{white!15!black}},
ylabel={$w_{r,bb} / \urms$},
axis background/.style={fill=white},
,
ymajorgrids,
xtick={1,2,...,14},
xticklabel style={rotate=45.0,anchor=east},
xticklabels={
	G-1-30,
	G-1-50,
	G-1.4-30,
	G-1.4-50,
	G-2.4-50,
	G-2.4-70,
	G-0.6-30,
	G-1-30-eg6,
	G-1-30-eg16,
	G-1-30-rb25,
	G-1-30-rp6k,
	G-1-30-ep7.5,
	G-1-30-ep5.0,
	R53-1-30},
]

\addplot [color=SIMcol, line width=1.0pt,mark=o,  mark options={solid, SIMcol},forget plot]
table[row sep=crcr]{%
	1	0.7942\\
	2	0.807022682\\
	3	0.81451559516\\
	4	0.797313\\
	5	0.75572155\\
	6	0.778589\\ 
	7	0.7401\\
	8	0.814\\
	9	0.704607\\
	10	0.8521\\
	11	0.81246\\
	12	0.7947\\
    13 0.77408\\
	14 0.756869997660779\\
};\label{plt: wrBBSIM}

\addplot [color=STcol, line width=1pt, densely dashdotted, mark=*, mark options={solid, STcol}, mark size = 1pt,   forget plot]
  table{%
1 1.93608696024449
2 1.93183663574824
3 2.20776697409298
4 2.20945381487199
5 2.88471122729691
6 2.87387046220297
7 1.63883272781368
8 1.79884685322548
9 2.1847922318242
10 1.97516354707372
11 1.94565202490433
12 1.9461004258294
13 1.96372535272086
14 1.2944172750371328
};\label{plt: wrBBST}

\addplot [color=Zcol, line width=1pt, densely dashed, forget plot, mark=*, mark options={solid, Zcol}, mark size = 1pt]
  table{%
1 2.21346000824107
2 2.20851576809352
3 2.67192714945409
4 2.67630783364956
5 3.11504448836097
6 3.11172795970351
7 1.35572068666781
8 2.13080979745168
9 2.22806103822278
10 2.19086852412711
11 2.21133814017411
12 2.21507746182093
13 2.23082764183289
14 1.3554890392957955
};\label{plt: wrBBZ}

\addplot [color=Zcol, line width=1pt, dotted, mark=x,  mark size=2pt, mark options={solid, Zcol},forget plot]
  table{%
1 1.91869172820506
2 1.91245812968185
3 2.49715009493723
4 2.50259617819918
5 3.04520409342226
6 3.04086119646605
7 0.974136278718522
8 1.81474513875508
9 1.99289755792015
10 1.89055315752632
11 1.91608401410246
12 1.92078390976499
13 1.94069516447347
14 0.7572798488204749
};\label{plt:wrBBZI}

\addplot [color=NMcol, line width=1.0pt, dotted, mark=x, mark size=2pt, mark options={solid, NMcol}, forget plot]
  table{%
1 0
2 0
3 0
4 0
5 0
6 0
7 0
8 0
9 0
10 0
11 0
12 0
13 0
14 0.4785755445793462
};\label{plt:wrBBNMI}

\addplot [color=NMcol, line width=1.0pt,mark=*, mark size=1pt, mark options={solid, NMcol},forget plot]
  table{%
1 0.595746453009838
2 0.583046043934609
3 0.883672164147539
4 0.873368195314719
5 1.25879450872562
6 1.23658575227048
7 0.247871014894421
8 0.617207580380672
9 0.495498809879839
10 0.582724989261959
11 0.591461869563434
12 0.596922148195532
13 0.606066874059131
14 0.922745127135157
};\label{plt: wrBBNM}

\end{axis}
\end{tikzpicture}%
	\caption{Comparison of radial relative velocity for bubble pairs closer than $r<2d_b$ obtained from DNS and various models. The cases are defined in Table \ref{tab:sim_cases}. The contributions originating from Mechanism I are shown. \ref{plt:CpbSIM} DNS, \ref{plt:CpbIMSC} IMSC, \ref{plt:wrBBNMI} Mechanism I of IMSC, \ref{plt:CpbST} Saffman \& Turner, \ref{plt:CpbZ} Zaichik, \ref{plt:wrBBZI} Mechanism I of Zaichik.}
	\label{fig:wrBB}
\end{figure}
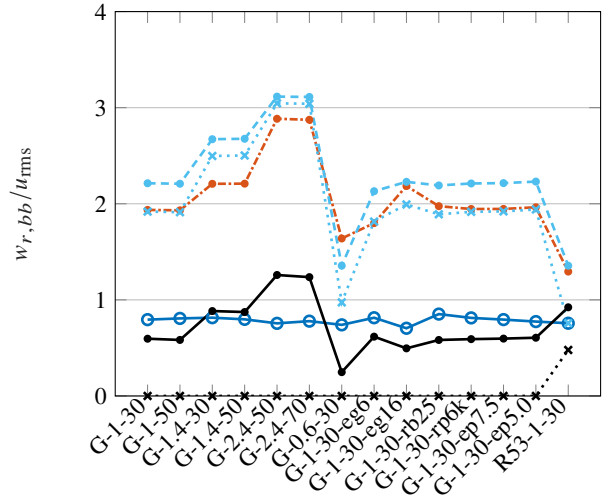

As the resulting collision kernel is mainly a function of the radial relative velocity.
The latter is examined here for the bubble-bubble case to shed further light on the findings in the preceding section.
Figure \ref{fig:wrBB} depicts the radial relative velocity between bubbles from the DNS in comparison to the model predictions of the IMSC, \citet{Saffman1956}, and \citet{Zaichik2010}.
For the IMSC, the contribution of Mechanism I to the overall relative velocity is shown.
The radial relative velocity from the DNS was obtained by averaging over the radial relative velocity between all bubble pairs with their centres closer than $r<2d_b$ during the averaging timespan.
The DNS data confirm that the collision kernel and the radial relative velocity are linked according to the spherical collision kernel formulation (\ref{eq:colkern_short}).
Both the models proposed by Saffman and Turner, as well as that of Zaichik, significantly overpredict the bubble-bubble relative velocity.
This observation underscores the overprediction of the associated collision kernel.
Bubbles in a gravity-driven flow are significantly larger than the viscous subscale and are hardly affected by any fluid motion at this scale.
Mechanism I is therefore not a relevant factor in determining their velocity.
This aspect is accurately represented by the IMSC.
Except for case R53-1-30, using forced background turbulence, there is no sizeable contribution from Mechanism I.
The relative velocities confirm the analysis in the previous section.
For all cases, the IMSC provides a good fit of the radial relative velocity.
The largest deviations exist for the cases with $d_b=\qty{2.4}{mm}$.
A contributing factor is that the bubble rise velocity obtained from the  simulations and the ones resulting from the employed sub-model by \citet{Rodrigue2001} differ as discussed in section \ref{sec:grav} above.

The relative velocity predicted by the model of Zaichik is already significantly higher than the total relative velocity obtained by DNS and consists almost entirely of the contribution from Mechanism I.
In this model, Mechanism I is not attenuated at large radii, which leads to high velocity contributions and, thus, too high collision kernels, as seen above.
The contribution of Mechanism I predicted by the model of Zaichik is very close to the one provided by Saffman and Turner.
This leads to large variations in the relative velocity between the cases, which are not seen in the DNS data or the IMSC.

The radial relative velocity predicted by the model of \citet{Kruis1997} is excessive, so it is not shown here.
This is the case since the contribution of Mechanism I is directly proportional to the bubble radius in their model, leading to this excessive prediction.
Despite the excessive relative velocity, the total collision kernel is only in the range of that of Saffman and Turner, alluding to the inconsistency of the cylindrical collision kernel formulation.

\section{Validation with further test cases}
\label{sec:further_valid}

\subsection{DNS with bubbles and coarse particles}
\label{sec:valid_coarse}

\subsubsection{Numerical setup and method}

\sisetup{detect-weight=true,detect-inline-weight=math}
\begin{table}
	\centering
	\caption{Simulated gravity-driven cases using coarse particles. The parameters varied in comparison to the case G-0.6-240 are highlighted.}
	\label{tab:simCoarse}
	\begin{tabular}{l|S|S[table-format=3.0]|S|S@{} c|S[table-format=2.1]@{} c|S}
		\toprule
		Simulation & {$d_b$ $[\si{mm}]$} & {$d_p$ $[\si{\micro\meter}]$} & {$d_p/d_b$} & {$\epsilon_g$}& & {$\epsilon_p$} & & {$\rho_p$ $[\si{\kilogram\per\cubic\meter}]$}\\
		\midrule
		G-0.6-240 & 0.6&240 & 0.4 & 8.8& $\si{\percent}$& 10 & $\si{\percent}$ & 4200\\
        \midrule
		G-0.6-240-rp3.2k & 0.6& 240& 0.4& 8.8 & $\si{\percent}$ & 10 & $\si{\percent}$ & \bfseries 3200\\
		G-0.6-240-rp5.2k & 0.6& 240& 0.4 & 8.8 & $\si{\percent}$ & 10 & $\si{\percent}$ & \bfseries 5200\\
        \midrule
		G-0.8-240 & \bfseries 0.8& 240& 0.3& 8.8 & $\si{\percent}$ & 10 & $\si{\percent}$ & 4200\\
        \midrule
		G-0.6-300 & 0.6& \bfseries 300& 0.5&8.8 & $\si{\percent}$ & 10 & $\si{\percent}$ & 4200\\
		G-0.6-180 & 0.6& \bfseries 180&0.3 &8.8 & $\si{\percent}$ & 10 & $\si{\percent}$ & 4200\\
        \midrule
        G-0.6-240-ep5 & 0.6 & 240&0.4& 8.8 & $\si{\percent}$ & \bfseries 5 & $\si{\percent}$ & 4200\\
		G-0.6-240-ep15 & 0.6& 240& 0.4&8.8 & $\si{\percent}$ & \bfseries 15 & $\si{\percent}$ & 4200\\
        \midrule
		G-0.6-240-eg11 & 0.6& 240& 0.4&\bfseries 11.0&$\si{\percent}$ & 10 & $\si{\percent}$ & 4200\\
		G-0.6-240-eg5 & 0.6& 240& 0.4&\bfseries 5.0& $\si{\percent}$& 10 & $\si{\percent}$ & 4200\\
		\bottomrule
	\end{tabular}
	
\end{table}

One particularity of flotation is the wide range of particle sizes encountered in applications.
Typically, particle diameters span an order of magnitude, ranging from the finest particles with a diameter smaller than $\qty{10}{\micro\meter}$ to the largest particles with a diameter larger than  $\qty{200}{\micro\meter}$ \citep{Norori2017,Ran2019}.
On this background, further simulations with coarse particles were conducted employing the same simulation method as before.
The only difference in the method is that the particles were not modelled as point particles but, instead, geometrically resolved spheres coupled to the fluid using the same IBM as used for the bubbles.
The simulated cases are listed in table \ref{tab:simCoarse}.
The same numerical setup was used as for the cases listed in table \ref{tab:sim_cases}.
The main driver of bubble and particle motion in these simulations is gravity.

\newpage

\subsubsection{Collision kernel and relative velocity}

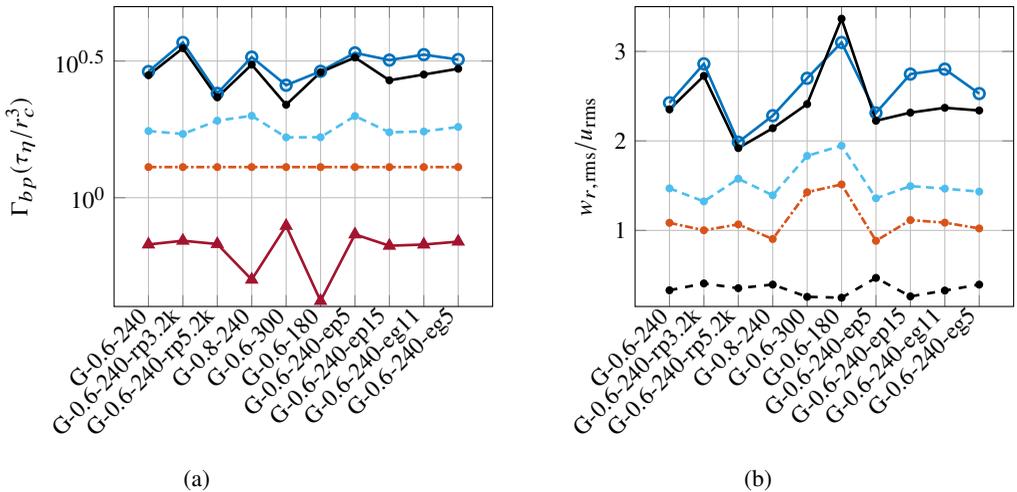
\begin{figure}
	
	\centering
	\begin{subfigure}{0.39\textwidth}
		
%
%
%
\begin{tikzpicture}

\begin{axis}[%
width=0.951\textwidth,
height=0.75\textwidth,
scale only axis,
xmin=0,
xmax=11,
ymode=log,
ymin=0.4,
ymax=5,
yminorticks=true,
ylabel={$\Gamma_{bp}(\tau_\eta/r_c^3)$},
xmajorgrids,
ymajorgrids,
yminorgrids,
scaled ticks=false, 
xtick={1,2,...,10},
xticklabel style={rotate=45.0,anchor=east},
xticklabels={
	G-0.6-240,
	G-0.6-240-rp3.2k,
	G-0.6-240-rp5.2k,
	G-0.8-240,
	G-0.6-300,
	G-0.6-180,
	G-0.6-240-ep5,
	G-0.6-240-ep15,
	G-0.6-240-eg11,
	G-0.6-240-eg5}
]
\addplot [color=SIMcol, line width=1pt, mark=o, mark options={solid, SIMcol}]
  table[row sep=crcr]{%
1	2.89484044457123\\
2	3.6976887130785\\
3	2.40865556517929\\
4	3.26888005070223\\
5	2.57937190154458\\
6	2.9\\
7	3.38757644985052\\
8	3.18794770164335\\
9	3.3376224618207\\
10	3.20241529723486\\
};

\addplot [color=Zcol, line width=1pt, densely dashed, forget plot, mark=*, mark options={solid, Zcol}, mark size = 1pt]
  table[row sep=crcr]{%
1	1.75381371849676\\
2	1.71082893137974\\
3	1.91355362632202\\
4	1.9944976646358\\
5	1.66275288099376\\
6	1.66441512967402\\
7	1.99035582388158\\
8	1.73526595503119\\
9	1.74731285139208\\
10	1.8152572223095\\
};

\addplot [color=Kcol, line width=1pt, mark=triangle*, mark options={solid, Kcol}, mark size = 2pt]
  table[row sep=crcr]{%
1	0.67450256632752\\
2	0.696084817503488\\
3	0.676858364560344\\
4	0.501487736979812\\
5	0.788914161970691\\
6	0.419098540492202\\
7	0.734360631189125\\
8	0.667185952623639\\
9	0.674066342342946\\
10	0.690890190122628\\
};

\addplot [color=STcol, line width=1pt, densely dashdotted, mark=*, mark options={solid, STcol}, mark size = 1pt,   forget plot]
  table[row sep=crcr]{%
1	1.29441727503713\\
2	1.29441727503713\\
3	1.29441727503713\\
4	1.29441727503713\\
5	1.29441727503713\\
6	1.29441727503713\\
7	1.29441727503713\\
8	1.29441727503713\\
9	1.29441727503713\\
10	1.29441727503713\\
};

\addplot [color=NMcol, line width=1pt, mark=*, mark options={solid, NMcol}, mark size = 1pt]
  table{%
1 2.80587198297664
2 3.52383272951896
3 2.32872908930933
4 3.0685936960012
5 2.18773194344421
6 2.87742720255346
7 3.25873154315602
8 2.6884282253056
9 2.82302027987042
10 2.9634042315989
};

\end{axis}

\begin{axis}[%
width=1.227\fwidth,
height=0.92\fwidth,
at={(-0.16\fwidth,-0.101\fwidth)},
scale only axis,
xmin=0,
xmax=1,
ymin=0,
ymax=1,
axis line style={draw=none},
ticks=none,
axis x line*=bottom,
axis y line*=left
]
\end{axis}
\end{tikzpicture}%
		\caption{}
		\label{fig:Cbp_coarse}
	\end{subfigure}
    \hspace{2cm}
	\begin{subfigure}{0.39\textwidth}
%
%
\begin{tikzpicture}

\begin{axis}[%
width=0.951\textwidth,
height=0.75\textwidth,
scale only axis,
xmin=0,
xmax=11,
ymin = 0.15,
ymax=3.5,
xmajorgrids,
ymajorgrids,
ylabel={$\wrms / \urms$},
ylabel near ticks,
xtick={1,2,...,10},
xticklabel style={rotate=45.0,anchor=east},
xticklabels={
	G-0.6-240,
	G-0.6-240-rp3.2k,
	G-0.6-240-rp5.2k,
	G-0.8-240,
	G-0.6-300,
	G-0.6-180,
	G-0.6-240-ep5,
	G-0.6-240-ep15,
	G-0.6-240-eg11,
	G-0.6-240-eg5}
]
]
\addplot [color=SIMcol, line width=1.0pt, mark=o, mark options={solid, SIMcol}, forget plot]
  table[row sep=crcr]{%
1	2.42620996981335\\
2	2.8611544594364\\
3	1.98576416268573\\
4	2.28183511980096\\
5	2.7\\
6	3.1\\
7	2.31294478243167\\
8	2.74634589631554\\
9	2.80201822576795\\
10	2.52856774286171\\
};

\addplot [color=STcol, line width=1pt, densely dashdotted, mark=*, mark options={solid, STcol}, mark size = 1pt,   forget plot]
  table[row sep=crcr]{%
1	1.08487087904386\\
2	1.00157910690125\\
3	1.06715442153255\\
4	0.903565365520914\\
5	1.42689323140695\\
6	1.51378434577297\\
7	0.883792801995262\\
8	1.11511163422965\\
9	1.08669594535997\\
10	1.02204788063808\\
};

\addplot [color=Zcol, line width=1pt, densely dashed, forget plot, mark=*, mark options={solid, Zcol}, mark size = 1pt]
  table[row sep=crcr]{%
1	1.46989805154616\\
2	1.32378526322046\\
3	1.57758804100525\\
4	1.3922550680773\\
5	1.8329258092792\\
6	1.94648635857856\\
7	1.35896065703034\\
8	1.49489294700768\\
9	1.46691320295357\\
10	1.4332934461348\\
};

\addplot [color=NMcol, line width=1pt, mark=*, mark options={solid, NMcol}, mark size = 1pt]
  table{%
1 2.35164414394045
2 2.72663020354063
3 1.91987050245186
4 2.14202563426307
5 2.41163337544592
6 3.36506962579109
7 2.22497299519862
8 2.31602100006248
9 2.36999671664314
10 2.33984903692874
};
    
\addplot [color=NMcol, line width=1pt, mark=*,dashed, mark options={solid, NMcol}, mark size = 1pt]
  table{%
1 0.331052005424959
2 0.407425581106567
3 0.354398476157477
4 0.39474174830571
5 0.257708444993025
6 0.248202010994135
7 0.469456911957528
8 0.262668143148808
9 0.32803490342435
10 0.394466882023185
};\label{plt: L-IMSC_mechI}

\end{axis}

\begin{axis}[%
width=1.227\fwidth,
height=0.92\fwidth,
at={(-0.16\fwidth,-0.101\fwidth)},
scale only axis,
xmin=0,
xmax=1,
ymin=0,
ymax=1,
axis line style={draw=none},
ticks=none,
axis x line*=bottom,
axis y line*=left
]
\end{axis}
\end{tikzpicture}%
		\caption{}
		\label{fig:wrbp_coarse}
	\end{subfigure}
	
	\caption{Comparison of results obtained from DNS of flotation with coarse particles and various models. (a) Particle-bubble collision kernel (b) Root mean square radial relative velocity between particles and bubbles over the entire domain. \ref{plt:CpbSIM} DNS, \ref{plt:CpbIMSC} IMSC, \ref{plt:CpbST} Saffman \& Turner, \ref{plt:CpbZ} Zaichik, \ref{plt:CpbK} Kostoglou. For reference, the contribution of Mechanism I of the ISMC is noted in (b) as well (\ref{plt: L-IMSC_mechI}).}
	\label{fig:coarse}
	
\end{figure}

Figure \ref{fig:coarse} compares the simulation results for the collision kernel and $w_{r,\mathrm{rms}}$ with those obtained by the IMSC and other models.
In all cases, the IMSC and the simulation results are in very good agreement, both for the collision kernel and for the radial relative velocity.
The ratio between the particle and bubble diameter ranges from $\numrange{0.3}{0.5}$.
Hence, equation (\ref{eq:RadialComponentCorrection}) yields $\left( w_r/w_\infty\right)=1$.
As these simulations are gravity-driven, all the fluid velocity fluctuations are created by particles and bubbles and, therefore, relate to spatial scales at most the size of the particles and bubbles, respectively.
As in these simulations both particles and bubbles are of similar size, and the domain was chosen to be smaller than potential clusters, the smaller-sized fluid fluctuations have little influence on the relative velocities of the collision partners close to the collisions and, hence, the collision kernel.
Thus, setting $\left( w_r/w_\infty\right)=1$ is reasonable.

As previously discussed, the influence of Mechanism I diminishes for particles significantly larger than the Kolmogorov length scale.
The simulated radial relative velocities are in agreement with the correlation used for the radial relative velocity, as outlined in (\ref{eq:colkern_short}).
In addition to the total radial relative velocity reported in figure \ref{fig:wrbp_coarse} for all models, the graph also highlights the contribution of Mechanism I to the IMSC.
It is only in the range of about $\SI{15}{\percent}$ of the total relative velocity.
Hence, Mechanism I is not very effective here because of the substantial size of the bubbles and particles relative to the Kolmogorov length scale.
This is confirmed by the fact that its contribution falls well below the theoretical limit of the collision kernel for $St=0$ obtained by the model of Saffman and Turner.

Other models differ significantly from the DNS results.
For instance, the model of \citet{Kostoglou2020b} provides a collision kernel that is too low.
Specifically designed for the case of fine particles, this model always applies the correction for the modulation of the flow field by the bubble.
However, this correction is not suitable for coarse particles.

The model of \citet{Zaichik2010} underpredicts the collision kernel and the radial relative velocity, since the influence of gravity is not included in this model.
However, the IMSC demonstrates that the combined effect of $w_{II,\mathrm{G}}$ provides a substantial fraction of the total relative velocity deduced from the fact that the contribution of Mechanism I is small for the cases presented.
The aforementioned underprediction of the DNS data by the model of \citet{Zaichik2010} is somewhat compensated by an over-prediction of Mechanism I compared to the IMSC.
Other factors, such as the assumption of Stokes drag and the omission of swarm effects, contribute to this discrepancy.

\subsection{Bubbles and fine particles in highly turbulent flow}

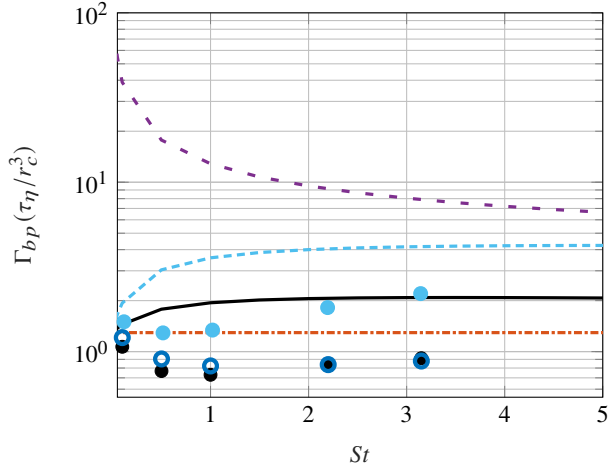
\begin{figure}
	\centering
	\setlength\fwidth{0.5\textwidth}
%
%
\begin{tikzpicture}

\begin{axis}[%
width=0.951\fwidth,
height=0.75\fwidth,
at={(0\fwidth,0\fwidth)},
scale only axis,
xmin=0.05,
xmax=5,
xlabel style={font=\color{white!15!black}},
xlabel={$St$},
ymode=log,
ymin=0.537935445387483,
ymax = 1e2,
yminorticks=true,
ylabel style={font=\color{white!15!black}},
ylabel={$\Gamma_{bp} (\tau_\kolmlen/r_c^3)$},
axis background/.style={fill=white},
xmajorgrids,
ymajorgrids,
yminorgrids
]

\addplot [color=Zcol, line width=1.2pt, densely dashed, forget plot]
  table[row sep=crcr]{%
9.94250632543173e-10 1.294411329635\\
9.99705247380195e-09 1.29440402954669\\
5.00357411614429e-08 1.2943883942667\\
9.94250632543173e-08 1.29437683894105\\
4.98570680256363e-07 1.29432917560442\\
9.99705247380195e-07 1.29429530931045\\
5.00357411614429e-06 1.29417045296609\\
9.94250632543173e-06 1.29409593974983\\
9.99705247380195e-05 1.2940394236806\\
0.000994250632543173 1.29955717649657\\
0.00498570680256363 1.33102303298325\\
0.00999705247380195 1.37152270175665\\
0.0200190175027184 1.44983111044131\\
0.0300293615146718 1.52346209797116\\
0.0399882098952078 1.59237545859177\\
0.0500357411614429 1.65794004004187\\
0.0994250632543173 1.93404077010665\\
0.498570680256363 3.03841053542392\\
0.999705247380195 3.57846600193199\\
1.49704845771984 3.84495188168626\\
2.00190175027184 4.00134760475117\\
2.50265137120877 4.09633907385177\\
3.00293615146718 4.15667877563255\\
3.49818757875348 4.19401742574701\\
3.99882098952078 4.21756407213489\\
4.49856103460905 4.22961160290105\\
5.00357411614429 4.23483810959017\\
5.49746733707303 4.23281721846885\\
6.91868865910911 4.20770841357854\\
8.78839407829089 4.14770002947347\\
20.0036300249501 3.69089227437648\\
};\label{plt:Z}

\addplot [color=NMcol, line width=1.2pt, forget plot]
  table[row sep=crcr]{%
9.94250632543173e-10 1.2944402975828\\
9.99705247380195e-09 1.29441009992396\\
5.00357411614429e-08 1.29439848643629\\
9.94250632543173e-08 1.29439087351282\\
4.98570680256363e-07 1.29435869492295\\
9.99705247380195e-07 1.29433491783166\\
5.00357411614429e-06 1.29423896991358\\
9.94250632543173e-06 1.29417227224009\\
9.99705247380195e-05 1.29378869971175\\
0.000994250632543173 1.2939068410683\\
0.00498570680256363 1.29919140975817\\
0.00999705247380195 1.30731761433867\\
0.0200190175027184 1.32442689449744\\
0.0300293615146718 1.34155504287421\\
0.0399882098952078 1.35827126017831\\
0.0500357411614429 1.37469057443784\\
0.0994250632543173 1.44837890937608\\
0.498570680256363 1.77970733089043\\
0.999705247380195 1.94280718754449\\
1.49704845771984 2.01752193312248\\
2.00190175027184 2.05605072822311\\
2.50265137120877 2.07526919031771\\
3.00293615146718 2.08395584154453\\
3.49818757875348 2.08609922635561\\
3.99882098952078 2.08402954482883\\
4.49856103460905 2.07912970995246\\
5.00357411614429 2.07237583118497\\
5.49746733707303 2.06464442222206\\
6.91868865910911 2.03756211525575\\
8.78839407829089 1.99799344685588\\
20.0036300249501 1.77993190818112\\
99.4474299275942 0.975333968757103\\
};\label{plt:IMSC}

\addplot [color=NMcol, only marks, mark=*, mark options={solid, NMcol,scale=1.2}, forget plot]
  table[row sep=crcr]{%
0.10	1.068\\
0.5	0.768\\
1.	0.73\\
2.2	0.826\\
3.15	0.914\\
};\label{plt:IMSCRDF}

\addplot [color=STcol, line width=1.2pt, densely dashdotted,  forget plot]
  table[row sep=crcr]{%
0.000211847978829714	1.29441727503713\\
0.0999207396402741	1.29441727503713\\
0.500149085748962	1.29441727503713\\
1.0001961106703	1.29441727503713\\
1.49654661999131	1.29441727503713\\
1.99050507866498	1.29441727503713\\
2.49801849100685	1.29441727503713\\
3.00090727209722	1.29441727503713\\
3.51631445157529	1.29441727503713\\
4.00078444268121	1.29441727503713\\
4.51651056224559	1.29441727503713\\
4.98343884860443	1.29441727503713\\
5.51519170750592	1.29441727503713\\
6.93894025279681	1.29441727503713\\
8.84173680772635	1.29441727503713\\
19.9637425471159    1.29441727503713\\
};\label{plt:STline}

\addplot [color=Acol, line width=1.2pt, loosely dashed, forget plot]
  table[row sep=crcr]{%
9.94250632543173e-10 386976.885142058\\
9.99705247380195e-09 122102.023447653\\
5.00357411614429e-08 54606.7383859242\\
9.94250632543173e-08 38697.6887261803\\
4.98570680256363e-07 17284.6655712436\\
9.99705247380195e-07 12210.203017671\\
5.00357411614429e-06 5460.675345698\\
9.94250632543173e-06 3869.77099236135\\
9.99705247380195e-05 1221.02703078436\\
0.000994250632543173 386.998295416592\\
0.00498570680256363 172.894605528468\\
0.00999705247380195 122.169953267216\\
0.0200190175027184 86.4436216077671\\
0.0300293615146718 70.5779342739255\\
0.0399882098952078 61.1866298131923\\
0.0500357411614429 54.7585194224861\\
0.0994250632543173 38.9105475429676\\
0.498570680256363 17.7506246967026\\
0.999705247380195 12.8521847587925\\
1.49704845771984 10.743876992775\\
2.00190175027184 9.49751925902388\\
2.50265137120877 8.65912957593277\\
3.00293615146718 8.0505244174732\\
3.49818757875348 7.58155191980842\\
3.99882098952078 7.21154513234293\\
4.49856103460905 6.90053889067112\\
5.00357411614429 6.64612851136898\\
5.49746733707303 6.42028654125246\\
6.91868865910911 5.92758136636118\\
8.78839407829089 5.4680990418506\\
20.0036300249501 4.19734580257853\\
};\label{plt:A}

\addplot [color=SIMcol, only marks, mark=o, ultra thick, mark options={solid, SIMcol, scale=1.2}, forget plot]
  table[row sep=crcr]{%
0.1	1.21\\
0.5	0.9051\\
1	0.823\\
2.2	0.838\\
3.15 0.88\\
};\label{plt:Chan}

\addplot [color=Zcol, only marks, mark=*, mark options={solid, Zcol, ultra thick}, forget plot]
  table[row sep=crcr]{%
0.117347007	1.5\\
0.51530616	1.29\\
1.020408162	1.34\\
2.193877302	1.82\\
3.14285682	2.2\\
};\label{plt:ZRDF}

\end{axis}

\begin{axis}[%
width=1.227\fwidth,
height=0.92\fwidth,
at={(-0.16\fwidth,-0.101\fwidth)},
scale only axis,
xmin=0,
xmax=1,
ymin=0,
ymax=1,
axis line style={draw=none},
ticks=none,
axis x line*=bottom,
axis y line*=left
]
\end{axis}
\end{tikzpicture}%
 	\caption{Collision kernels for the setup of \citet{Chan2023} featuring turbulent flow with $Re_\lambda=175$ and comparison to various models. \ref{plt:Chan} DNS results from Chan et al., \ref{plt:IMSC} IMSC, \ref{plt:IMSCRDF} IMSC corrected with $g(r_c)$, \ref{plt:A} Abrahamson, \ref{plt:STline} Saffman \& Turner, \ref{plt:Z} Zaichik, \ref{plt:ZRDF} Zaichik corrected with $g(r_c)$.}
	\label{fig:Cbp-Chan}
\end{figure}

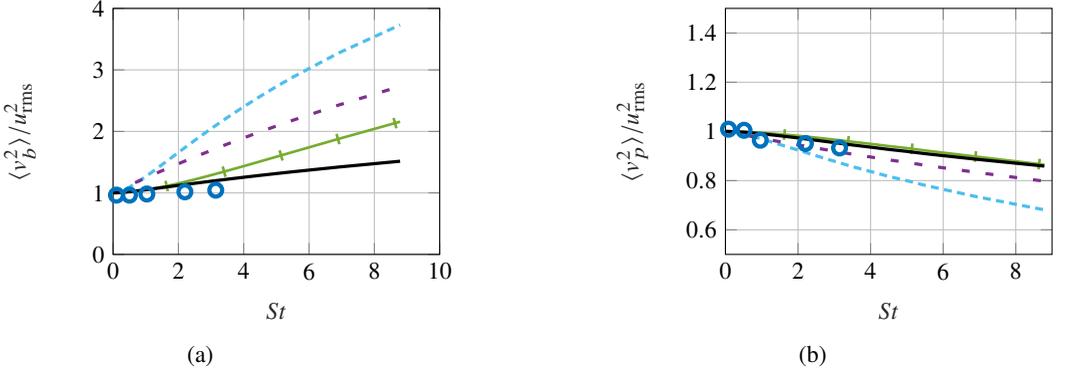
\begin{figure}
	
	\centering
	\begin{subfigure}{0.40\textwidth}

%
%
\definecolor{mycolor1}{rgb}{0.92900,0.69400,0.12500}%
\definecolor{mycolor2}{rgb}{0.30100,0.74500,0.93300}%
\definecolor{mycolor3}{rgb}{0.46600,0.67400,0.18800}%
\definecolor{mycolor4}{rgb}{0.49400,0.18400,0.55600}%
\definecolor{mycolor5}{rgb}{0.00000,0.44700,0.74100}%
\begin{tikzpicture}

\begin{axis}[%
width=0.8\textwidth,
height=0.6\textwidth,
scale only axis,
xmin=0,
xmax=10,
xlabel style={font=\color{white!15!black}},
xlabel={$St$},
ylabel style={font=\color{white!15!black}},
ylabel={$\langle v_b^2\rangle / \urms^2 $},
ymin=0,
ymax=4,
axis background/.style={fill=white},
title style={font=\bfseries},
xmajorgrids,
ymajorgrids
]

\addplot [color=Zcol, line width=1.2pt, densely dashed, forget plot]
  table[row sep=crcr]{%
  	0.00839618475578865	1.00004006374502\\
0.0994250632543173	1.0051081132875\\
0.498570680256363	1.09105568959122\\
0.999705247380195	1.26409568913649\\
1.49704845771984	1.45803015999976\\
2.00190175027184	1.65892398210681\\
2.50265137120877	1.85509488804146\\
3.00293615146718	2.0450420182293\\
3.49818757875348	2.22593061980658\\
3.99882098952078	2.40114853890402\\
4.49856103460905	2.56838647995912\\
5.00357411614429	2.72980406733468\\
5.49746733707303	2.88058379205284\\
6.91868865910911	3.27844549753513\\
8.78839407829089	3.73123406894301\\
};\label{plt: ZvBChan}

\addplot [color=KKcol, line width=1pt,
			decoration={ markings,
						mark=between positions 0 and 1 step 0.2
						with  {\draw (-2pt,-2pt) -- (-2pt,2pt);}},
		postaction=decorate]
  table[row sep=crcr]{%
0.00839618475578865	1.00000435861682\\
0.0994250632543173	1.00059555117311\\
0.498570680256363	1.01344685656381\\
0.999705247380195	1.04779291152966\\
1.49704845771984	1.09585371963663\\
2.00190175027184	1.15442684303473\\
2.50265137120877	1.21926830778114\\
3.00293615146718	1.2886775755637\\
3.49818757875348	1.36049316275705\\
3.99882098952078	1.43516000521788\\
4.49856103460905	1.51097782888511\\
5.00357411614429	1.58829344627933\\
5.49746733707303	1.66414218786098\\
6.91868865910911	1.88101300916067\\
8.78839407829089	2.15762702307937\\
};\label{plt: KKsmallvBChan}

\addplot [color=Acol, line width=1.2pt, loosely dashed, forget plot]
  table[row sep=crcr]{%
  	0.00839618475578865	1.00211561491753\\
0.0994250632543173	1.02498052304793\\
0.498570680256363	1.12370801272865\\
0.999705247380195	1.24423887191087\\
1.49704845771984	1.3602490037582\\
2.00190175027184	1.47449846432855\\
2.50265137120877	1.58447756210464\\
3.00293615146718	1.6911757827806\\
3.49818757875348	1.79380441753571\\
3.99882098952078	1.8946459581425\\
4.49856103460905	1.99251787536865\\
5.00357411614429	2.08870749478698\\
5.49746733707303	2.18024599699528\\
6.91868865910911	2.43055910584895\\
8.78839407829089	2.73302844407322\\
};\label{plt: AvBChan}

\addplot [color=SIMcol, only marks, mark=o, ultra thick, mark options={solid, SIMcol, scale=1.2}, forget plot]
  table[row sep=crcr]{%
0.104477384	0.96412549\\
0.50248741	0.96412549\\
1.034825567	0.982062847\\
2.199004671	1.017937153\\
3.144278455	1.044842984\\
};\label{plt: SIMvBChan}

\addplot [color=NMcol, line width=1.2pt, forget plot]
  table[row sep=crcr]{%
4.98570680256363e-07 1.00000000000003\\
9.99705247380195e-07 1.00000000000013\\
5.00357411614429e-06 1.00000000000337\\
9.94250632543173e-06 1.00000000001332\\
9.99705247380195e-05 1.00000000134673\\
0.000994250632543173 1.0000001330438\\
0.00498570680256363 1.00000332647824\\
0.00999705247380195 1.00001327761237\\
0.0200190175027184 1.00005247101909\\
0.0300293615146718 1.0001163614603\\
0.0399882098952078 1.00020339524747\\
0.0500357411614429 1.00031390544811\\
0.0994250632543173 1.00115736048383\\
0.498570680256363 1.01873729894075\\
0.999705247380195 1.05125858583223\\
1.49704845771984 1.08627512723049\\
2.00190175027184 1.12196096163867\\
2.50265137120877 1.15658900254308\\
3.00293615146718 1.19027104228263\\
3.49818757875348 1.22259553271514\\
3.99882098952078 1.25424777736045\\
4.49856103460905 1.28485680306147\\
5.00357411614429 1.3148405083704\\
5.49746733707303 1.34352108190938\\
6.91868865910911 1.42120228732066\\
8.78839407829089 1.51504209295157\\
};\label{plt: NMvBChan}

\end{axis}

\begin{axis}[%
width=1.227\fwidth,
height=0.92\fwidth,
at={(-0.16\fwidth,-0.101\fwidth)},
scale only axis,
xmin=0,
xmax=1,
ymin=0,
ymax=1,
axis line style={draw=none},
ticks=none,
axis x line*=bottom,
axis y line*=left
]
\end{axis}
\end{tikzpicture}%
		\caption{}
		\label{fig:vb_Chan}
	\end{subfigure}
	\hfill
	\begin{subfigure}{0.4\textwidth}
%
%
\definecolor{mycolor1}{rgb}{0.92900,0.69400,0.12500}%
\definecolor{mycolor2}{rgb}{0.30100,0.74500,0.93300}%
\definecolor{mycolor3}{rgb}{0.46600,0.67400,0.18800}%
\definecolor{mycolor4}{rgb}{0.49400,0.18400,0.55600}%
\definecolor{mycolor5}{rgb}{0.00000,0.44700,0.74100}%
\begin{tikzpicture}

\begin{axis}[%
width=0.8\textwidth,
height=0.6\textwidth,
scale only axis,
xlabel style={font=\color{white!15!black}},
xlabel={$St$},
ylabel style={font=\color{white!15!black}},
ylabel={$\langle v_p^2\rangle / \urms^2 $},
xmin=0,
xmax=9,
ymin=0.5,
ymax=1.5,
axis background/.style={fill=white},
xmajorgrids,
ymajorgrids
]

\addplot [color=Zcol, line width=1.2pt, densely dashed, forget plot]
  table[row sep=crcr]{%
  	0.00839618475578865	0.999999997008966\\
0.0994250632543173	0.999400678675161\\
0.498570680256363	0.989362591552625\\
0.999705247380195	0.96928424713728\\
1.49704845771984	0.946789115472706\\
2.00190175027184	0.923942258808212\\
2.50265137120877	0.90082556367282\\
3.00293615146718	0.8786289989985\\
3.49818757875348	0.856764502120533\\
3.99882098952078	0.837074920999172\\
4.49856103460905	0.817034163332968\\
5.00357411614429	0.799686839911416\\
5.49746733707303	0.7808169358143\\
6.91868865910911	0.734572990236596\\
8.78839407829089	0.681183582874371\\
};

\addplot [color=KKcol, line width=1pt,
			decoration={ markings,
						 mark=between positions 0 and 1 step 0.2
						with  {\draw (-2pt,-2pt) -- (-2pt,2pt);}
						},
		postaction=decorate]
  table[row sep=crcr]{%
  	0.00839618475578865	0.999999999676739\\
0.0994250632543173	0.999930100594275\\
0.498570680256363	0.998427883759295\\
0.999705247380195	0.994440504151524\\
1.49704845771984	0.988865770856775\\
2.00190175027184	0.98221542877835\\
2.50265137120877	0.974588279534646\\
3.00293615146718	0.966482312164141\\
3.49818757875348	0.957791442924538\\
3.99882098952078	0.949389616038925\\
4.49856103460905	0.940293807057492\\
5.00357411614429	0.93198534925021\\
5.49746733707303	0.922494446085345\\
6.91868865910911	0.897249712105299\\
8.78839407829089	0.864561297102172\\
};

\addplot [color=Acol, line width=1.2pt, loosely dashed, forget plot]
  table[row sep=crcr]{%
  	0.00839618475578865	0.999993794354575\\
0.0994250632543173	0.997082236899465\\
0.498570680256363	0.985577360594061\\
0.999705247380195	0.971600090062722\\
1.49704845771984	0.958143848591433\\
2.00190175027184	0.945147276068742\\
2.50265137120877	0.932186264135614\\
3.00293615146718	0.919718406568863\\
3.49818757875348	0.907310742551333\\
3.99882098952078	0.895975420877973\\
4.49856103460905	0.884242970048427\\
5.00357411614429	0.873905973671722\\
5.49746733707303	0.862450643297541\\
6.91868865910911	0.833335766156968\\
8.78839407829089	0.797625380892801\\
};
\addplot [color=SIMcol, only marks, mark=o, ultra thick, mark options={solid, SIMcol, scale=1.2}, forget plot]
  table[row sep=crcr]{%
0.0845769625991135	1.00896867890422\\
0.512437507287779	1.00448423681451\\
0.960198701317631	0.964125489658338\\
2.20397994767374	0.950672573939615\\
3.14427845513643	0.932735421406386\\
};

\addplot [color=NMcol, line width=1.2pt, forget plot]
  table[row sep=crcr]{%
.99705247380195e-05 0.999999999843312\\
0.000994250632543173 0.999999984369975\\
0.00498570680256363 0.999999609282662\\
0.00999705247380195 0.999998441915695\\
0.0200190175027184 0.999993864200279\\
0.0300293615146718 0.999986135057105\\
0.0399882098952078 0.999975498155594\\
0.0500357411614429 0.999962198496878\\
0.0994250632543173 0.999852315591506\\
0.498570680256363 0.997001419587609\\
0.999705247380195 0.990403129267032\\
1.49704845771984 0.982215657622811\\
2.00190175027184 0.973322156546113\\
2.50265137120877 0.963871887268806\\
3.00293615146718 0.954445533296581\\
3.49818757875348 0.944866191288748\\
3.99882098952078 0.936012011233431\\
4.49856103460905 0.926791133013002\\
5.00357411614429 0.91864511204488\\
5.49746733707303 0.909613663850424\\
6.91868865910911 0.886726959710369\\
8.78839407829089 0.858962341337489\\
};\label{plt: NMvBChan}

\end{axis}

\begin{axis}[%
width=1.227\fwidth,
height=0.92\fwidth,
at={(-0.16\fwidth,-0.101\fwidth)},
scale only axis,
xmin=0,
xmax=1,
ymin=0,
ymax=1,
axis line style={draw=none},
ticks=none,
axis x line*=bottom,
axis y line*=left
]
\end{axis}
\end{tikzpicture}%
		\caption{}
		\label{fig:vp_Chan}
	\end{subfigure}
	
	\caption{Velocity fluctuations of bubbles and particles for the setup of \citet{Chan2023} with $Re_\lambda=175$ in comparison to model predictions. \ref{plt: SIMvBChan} DNS results from Chan et al., \ref{plt: NMvBChan} IMSC, \ref{plt: AvBChan} Abrahamson, \ref{plt: ZvBChan} Zaichik, \ref{plt:CpbKK} Kruis \& Kusters.}
\label{fig:ChanParticleVel}
	
\end{figure}

Apart from the own simulations, there are other simulations investigating collision events in the literature.
A recent study conducted by \citet{Chan2023} is particularly noteworthy.
DNS of a multi-particle, multi-bubble system under the influence of turbulence were performed.
Point particles were employed for both the particles and the bubbles, with varying Stokes numbers, where $St=St_p=St_b$.
In light of the prior validation cases that predominantly focused on low and medium turbulence, their case with high turbulence, defined by a Taylor Reynolds number of $Re_\lambda=175$, is used here.
One-way coupling was implemented for particles and bubbles.
As a result, the hindrance effect of bubble and particle swarms does not exist in the simulations.
Therefore, the corresponding correction factor was set to $c_\epsilon=1$ here.
Furthermore, the influence of gravity is disregarded as it was not considered in the simulations.
Figure \ref{fig:Cbp-Chan} compares a variety of collision models under the given conditions as a function of $St$ with the results of \citet{Chan2023}.
All the models analysed in this study lie within the theoretical limits of Saffman and Turner and Abrahamson for $St=0$ and $St\rightarrow\infty$, respectively.
With varying Stokes number, the largest changes in the collision kernel occur at low Stokes numbers for all models.
The IMSC and the model of Zaichik predict an increase in the collision kernel by a factor of $1.5$ to $2.5$ for Stokes numbers up to $St=2$.
Conversely, for higher Stokes numbers, the collision kernel is barely affected by an increase in $St$.
In contrast, the model by Abrahamson yields a decline in the collision kernel with rising Stokes numbers.

Overall, none of the models in the literature successfully captures the qualitative and quantitative trends of the simulation results.
This is due to the substantial clustering of particles and bubbles around $St=1$ identified by \citet{Chan2023}, which results in the separation of bubbles and particles and leads to a reduction of the number of collisions.
In fact, for small Stokes numbers, first a decay with increasing $St$ is seen, until a minimum around $St=1$ is met, with a subsequent increase for $St>1$.
With the present approach, the modelled collision kernels are, therefore, corrected with the numerically obtained radial distribution function $g(r_c)$ according to (\ref{eq:correctionRDF}) found by \citet{Chan2023}.
The resulting collision frequencies are also shown in figure \ref{fig:Cbp-Chan}.
The IMSC and the model of Zaichik predict the trend of the simulation data qualitatively, while the former also yields good quantitative agreement.

This can be supported by comparing the modelled velocities for particles and bubbles shown in figure \ref{fig:ChanParticleVel}.
The simulated bubble velocity by Chan~\textit{et~al.} is best matched by the IMSC.
The models of Kruis and Kusters, Abrahmason, and Zaichik overpredict the bubble velocity with higher deviations for larger $St$.
The IMSC yields a very good match with the simulation results for small $St$.
The particle velocities obtained by the IMSC and the model of \citet{Kruis1997} agree very well with the simulation data.
Due to the formulation with a cylindrical collision kernel and other choices, the overall collision kernel by the model of \citet{Kruis1997} significantly differs from the simulation results and is, for this reason, not shown in figure \ref{fig:Cbp-Chan}.
The deviation of the model by Abrahmson for bubbles is caused by assuming $St\rightarrow\infty$, which is not met here.
The model by Zaichik underpredicts the simulated particle velocities.
The differences in the modelling formulation by Zaichik are due to the previously mentioned shortcomings, such as the over-prediction of Mechanism I and the assumption of Stokes drag.
Taken together, these also lead to the discrepancies in the collision kernel shown in figure \ref{fig:Cbp-Chan}.

\section{Conclusion}

The accuracy of CFD simulations and process simulations highly depends on the applicability and quality of the underlying models for the unresolved sub-processes.
Many of the existing collision models are too strongly simplified to capture the complexities of the flotation process.
Significant challenges arise from the range of turbulence intensities, the difference in bubble and particle diameter, and finite Stokes number effects.
Adequately representing the impacts of turbulence and gravity is another challenge.
Furthermore, some models suffer from mathematical or physical inconsistencies as pointed out in previous publications.

In this paper, the novel Integrated Multi-Size Collision model (IMSC) is proposed, which is specifically designed to address these issues.
The model incorporates models for turbulent motion, as well as for the impact of gravitation on the relative motion of particles, including swarm effects.
The IMSC is based on the approach of a spherical collision kernel formulation.
The radial relative velocity is decomposed into a contribution of Mechanism I and a combined contribution of Mechanism II and gravity.
For Mechanism I, the coupling of the velocity of the dispersed elements to the fluid velocity fluctuations is achieved by a piecewise longitudinal fluid structure function based on that of \citet{Borgas2004} and the fluid autocorrelation function of \citet{Williams1980}.
Drag corrections accounting for swarm effects and deviations from Stokes drag are included.
The contributions of Mechanism II and gravity are combined and corrected to account for the effects of flow distortion introduced by the bubbles.
The IMSC can be implemented following the schematic in figure \ref{fig:overviewIMSC} and the summarising description in section \ref{sec:summary_IMSC} with the complete set of equations compiled in appendix \ref{sec:summary}.

The collision kernels obtained from the IMSC were compared with data from own Direct Numerical Simulations and simulations from the literature.
The findings demonstrate that the IMSC provides a very good match of the collision kernel, generally much better than other existing collision models, while also offering a comprehensive coverage of the full range of process parameters relevant for flotation.
Furthermore, the validity of underlying modelling assumptions was substantiated.

In the present work, the longitudinal structure function employed is that of single-phase flow.
This was decided after an in-depth screening of the literature.
The available models were found to partly contradictory or only applicable over a limited range only.
It was attempted to extract this quantity from the own DNS database, but this was unsuccessful.
Hence, for reasons of caution, the simplest, single-phase variant was employed in the final model.
For the same reasons, a single-phase fluid autocorrelation function is also used.
Any improvement in this respect could readily be included in the present approach.

Nonetheless, the IMSC provides a good approximation of the collision frequency, not only between particles and bubbles.
It is now ready for further use and may be employed in Euler-Euler frameworks or process simulations.

\backsection[Acknowledgements]{The work was conducted in the frame of the EU ITN "FlotSim", No. 955805, joining academia and industry concerned with flotation. The project partners are gratefully acknowledged for the stimulating discussions. The authors also gratefully acknowledge the computing time made available to them at the NHR Center of TU Dresden and at the NHR Center NHR4CES at RWTH Aachen University (project number p0020495). These are funded by the Federal Ministry of Education and Research, and the state governments participating in the NHR and on the basis of the resolutions of the GWK for national high performance computing at universities.}

\backsection[Funding]{This project has received funding from the European Union’s Horizon 2020 Marie Sklodowska-Curie Actions (MSCA), Innovative Training Networks (ITN), H2020-MSCA-ITN-2020 under grant agreement No. 955805.}

\backsection[Declaration of interests]{The authors report no conflict of interest.}

\backsection[Data availability statement]{The data that support the findings of this study are available upon request.}

\backsection[Author ORCIDs]{B. Tiedemann, https://orcid.org/0009-0000-0425-3337; M. Kreuseler, https://orcid.org/0009-0009-1853-0839; J. Fröhlich, https://orcid.org/0000-0003-1653-5686}

\appendix

\newpage
\setcounter{table}{0}
\renewcommand{\thetable}{A.\arabic{table}}
\section{Model summary of the IMSC}
\label{sec:summary}

For the convenience of the reader, table \ref{tab:model} collects the equations employed by the IMSC in full detail, as they were implemented.

\begin{adjustwidth}{-1cm}{}

\small\begin{longtblr}[
	caption ={Technical description of the IMSC, its sub-models, and the publications they are modelled after.},
	label = {tab:model}, ]{ll}

		\hline
		\underline{Collision kernel} & \underline{Modelled after}\\
		 $\Gamma_{ij}=2\pi (r_i+r_j)^2\sqrt{\frac{2}{\pi}\sigma_I^2  + \langle w_{II,\mathrm{G}}^2\rangle \left( \frac{w_r}{w_\infty}\right)^2}$ & \citet{Yuu1984} \& present \\
		\hline
			\underline{Mechanism I}  &\\
			
			  $
			\sigma_I^2 = S_{\mathrm{\ell\ell}}^{\mathrm{(IMS)}}(r_i) \frac{\langle \Tilde{v}_i^2\rangle}{\urms^2} + S_{\mathrm{\ell\ell}}^{\mathrm{(IMS)}}(r_j) \frac{\langle \Tilde{v}_j^2\rangle}{\urms^2} + 2\sqrt{S_{\mathrm{\ell\ell}}^{\mathrm{(IMS)}}(r_i)S_{\mathrm{\ell\ell}}^{\mathrm{(IMS)}}(r_j)} \frac{\langle \Tilde{v}_i\Tilde{v}_j\rangle}{\urms^2}$ & present\\
			
			  ~~~~~~~~~~$\frac{\langle \Tilde{v}_\alpha^2 \rangle }{ \urms^2 } = \frac{\gamma}{\gamma-1} \left( \frac{T_{\mathrm{L}} a_\alpha + b_\alpha^2}{T_{\mathrm{L}} a_\alpha +1} - \frac{T_{\mathrm{L}} a_\alpha + \gamma b_\alpha^2}{\gamma(T_{\mathrm{L}} a_\alpha +\gamma)} \right)\label{eq:particle_velocity_app}$ & present\\
			
			  ~~~~~~~~~~$\frac{\langle \Tilde{v}_i\Tilde{v}_j \rangle }{  \urms^2 } = \frac{\gamma}{\gamma -1} (I^{\mathrm{(PE)}}_1 - I^{\mathrm{(PE)}}_2)$ & present\\
			
			  ~~~~~~~~~~$ a_\alpha = \frac{9\mu_f c_\alpha c_\epsilon^{-2}}{r_\alpha^2(2\rho_\alpha + \rho_f)}\hspace{0.5cm},\hspace{1cm}
			b_\alpha = \frac{3\rho_f}{2\rho_\alpha + \rho_f}$ & \citet{Abrahamson1975} \& present\\
			
			  ~~~~~~~~~~$I_1 =   \dfrac{(((T_\mathrm{L}a_i+1)a_j+a_i)b_i+T_\mathrm{L}a_i^2)b_j + T_\mathrm{L}a_j^2b_i + T_\mathrm{L}^2a_ia_j^2}
				{\left(T_{\mathrm{L}}a_i+1\right)\left(a_j+a_i\right)
					\left(T_{\mathrm{L}}a_j+1\right)} $ & \\
			  ~~~~~~~~~~$	\;\;\;\; +  \;\dfrac{T_\mathrm{L}a_ia_j(T_\mathrm{L}a_i+1)+T_\mathrm{L}a_ia_j(b_i-1)(b_j-1)}             {\left(T_{\mathrm{L}}a_i+1\right)\left(a_j+a_i\right)\left(T_{\mathrm{L}}a_j+1\right)}$ & present\\
			
			  ~~~~~~~~~~$I_2 = \dfrac{(a_i+a_j)b_ib_j\gamma^2 + (T_\mathrm{L}a_ia_jb_i+T_\mathrm{L}a_i^2)b_j\gamma + T_\mathrm{L}a_j^2b_i\gamma}
				{\gamma (a_j+a_i)(T_{\mathrm{L}}a_i+\gamma)(T_{\mathrm{L}}a_j)+\gamma}$ & \\
			  ~~~~~~~~~~$\;\;\;\;+\;\dfrac{T_\mathrm{L}a_ia_j\gamma+T_\mathrm{L}^2a_ia_j^2+T_\mathrm{L}^2a_i^2a_j + a_ia_jT_\mathrm{L}(b_i-1)(b_j-1)}{\gamma(a_j+a_i)(T_{\mathrm{L}}a_i+\gamma)(T_{\mathrm{L}}a_j)+\gamma}$ & present\\
		\hline	

		 	\underline{Mechanism II \& gravity }  &\\
		 	 $ \sqrt{\langle w_{II,\mathrm{G}}^2 \rangle} = \int_{\phi = 0}^{\pi/2} \int_{w=w_a}^{w_b} \sin(\phi) 
		 	\sqrt{(w-\cos (\phi) \Delta w_{\mathrm{G}} )^2}$\\
            ~~~~~~~~~~~~~~~~~~~$\times \frac{1}{\sqrt{2\pi\sigma^2_{II}}} \exp \left( - \frac{w^2}{2\sigma^2_{II}} \right) \diff w \diff \phi$ & \citet{Dodin2002}\\ 
		 	
             ~~~~~~~~~~$\sigma_{II}^2 = \langle\Tilde{v}_{i}^2 \rangle + \langle\Tilde{v}_{j}^2 \rangle - 2 \langle \Tilde{v}_i u \rangle- 2 \langle \Tilde{v}_j u \rangle + 2 \langle u^2 \rangle$ & \citet{Yuu1984} \& present\\
			
			 ~~~~~~~~~~$\frac{ \langle \Tilde{v}_\alpha u \rangle }{\urms^2} 
			=   \frac{\gamma}{\gamma-1} \biggl( \frac{b_\alpha+T_\mathrm{L}a_\alpha}{T_\mathrm{L}a_\alpha+1} - 
			\frac{b_\alpha\gamma+T_\mathrm{L}a_\alpha}{\gamma(T_\mathrm{L}a_\alpha+\gamma)} \biggr)$ & present\\

            ~~~~~~~~~~$\langle \Tilde{v}_\alpha^2 \rangle  = \frac{\urms^2 \gamma}{\gamma-1} \left( \frac{T_{\mathrm{L}} a_\alpha + b_\alpha^2}{T_{\mathrm{L}} a_\alpha +1} - \frac{T_{\mathrm{L}} a_\alpha + \gamma b_\alpha^2}{\gamma(T_{\mathrm{L}} a_\alpha +\gamma)} \right)$ & present \\
		 	
			 ~~~~~~~~~~$\Delta \Tilde{w}_{\mathrm{G}} = \Tilde{v}_{\mathrm{G}i} -  \Tilde{v}_{\mathrm{G}j}$ & \\

             ~~~~~~~~~~$\langle u^2 \rangle = \urms^2 = \frac{2}{3}k$ & \\
\hline		
\pagebreak
		
		\underline{Fluid disturbance by large bubble} &\\
		 \SetCell[c=2]{l}The larger collision partner (usually the bubble) is denoted $j$ here and the smaller one $i$ \\
	
		$
				\left( \frac{w_r}{w_\infty} \right)^2 = 
				\begin{cases}
					\left( \frac{1}{4} Y\left(\frac{r_i}{r_j}\right)^2 \right)^2 \hspace{2cm}\mathrm{if}\hspace{0.2cm}r_i/r_j\le 0.1 \\
					\left( Z + \frac{1-Z}{0.3-0.1}\left(\frac{r_i}{r_j} - 0.1\right )  \right)^2 \hspace{0.35cm}\mathrm{if}\hspace{0.2cm}0.1< r_i/r_j\le 0.3 \\
					1  \hspace{3.575cm}\mathrm{else} 
				\end{cases}$ &  present\\

			~~~~~~~~~~$Z = \frac{0.01}{4} Y, \hspace{0.3cm} X = \frac{3}{2}+\frac{9}{32}\frac{Re_{\infty j}}{1+0.309Re_{\infty j}^{0.694}},\hspace{0.3cm}Y = \frac{3}{8} \frac{Re_{\infty j}}{1+0.217Re_{\infty j}^{0.518}}$ & \citet{Nguyen1999}\\
		
		\hline
			\underline{Description of fluid} &\\
					
				$S_{\mathrm{\ell\ell}}^{\mathrm{(IMSC)}}(r)  = 
					\begin{cases}
                        S_{\mathrm{\ell\ell}}^{\mathrm{(BY)}} & \hspace{0.5cm},r\leq r_\eta\\
                        S_{\mathrm{\ell\ell}}^{\mathrm{(BY)}}\left(r\cos^2\left( \frac{\pi}{2}\frac{r-r_\eta}{r_\lambda-r_\eta} \right) \right)	 & \hspace{0.5cm}, r_\eta<r\leq r_\lambda\\
                        0 & \hspace{0.5cm},r>r_\lambda\\
					\end{cases}
			$ & present\\
			
			~~~~~~~~~~$S^{\mathrm{(BY)}}_{\mathrm{\ell\ell}}(r) = 2Re_\lambda \sqrt{\frac{\varepsilon \kinvisc}{15}} \left[1- \exp\left(-\frac{r}          {30^{3/4}\kolmlen}\right)\right]^{4/3} \left[\frac{15^3 r^4}{15^3 r^4 + \kolmlen^4 Re_\lambda^6}\right]^{1/6}$ & \citet{Borgas2004}\\
			
			~~~~~~~~~~$T_L =  \frac{2(Re_\lambda+32)}{7\sqrt{15} }\sqrt{\frac{\kinvisc}{\tdr}}$ & \citet{Sawford1991} \\

            ~~~~~~~~~~$\lambda = \sqrt{\frac{10\kinvisc\tke}{\tdr}}$, $Re_\lambda = \sqrt{\frac{2\tke}{3}}\frac{\lambda}{\kinvisc}$ & \citet{Sawford1991}\\
			
			\hline

            \underline{Drag correction} & \\
            $c_{\alpha} = \begin{cases}
                1+0.15 Re_\alpha^{0.687} & \hspace{0.5cm} Re_\alpha<136 \\
                0.95 \frac{Re_\alpha}{24} & \hspace{0.5cm} Re_\alpha\ge 136,
            \end{cases}$ & {\citet{Clift1978}\\ ~\\ \citet{Karamanev1992}}\\
            
            \hline
            
			\underline{Bubble drag and velocity models} & \\

			$v_{\mathrm{G}b} = v_{\mathrm{G}\alpha}^{(R)}c_{\epsilon b}$ & \\
			
			~~~~~~~~~~$v_{\mathrm{G}b}^{(R)} = \frac{1}{12} g \left( \frac{(2r_b)^8 \rho_f}{\sigma_{bf} \nu^4} \right)^{1/3} \left( 1+ 0.049 \left(g \left( \frac{(2r_b)^8 \rho_f}{\sigma_{bf} \nu^4} \right)^{1/3}\right)^{3/4} \right)^{-1} \left( \frac{(2r_b)^2 \rho_f}{\sigma_{bf} \nu} \right)^{-1/3}$ & \citet{Rodrigue2001}\\

			~~~~~~~~~~$c_{\epsilon b} = (1-\epsilon_b^{1/3})$ & \citet{Garnier2002}\\

            \hline

            \underline{Particle drag and velocity models} & \\

            $v_{\mathrm{G}p} = v_{\mathrm{G}\alpha}^{(NS)}c_{\epsilon p}$ & \\

            ~~~~~~~~~~$v_{\mathrm{G}p}^{(NS)} = \frac{2 r_p^2 (\rho_p-\rho_f) g}{9 \nu \rho_f}
				\left(
				1 + \frac{Ar_p}{96}(1+0.079Ar_p^{0.749})^{-0.755}
				\right)^{-1}$ & \citet{Nguyen2004}\\

            ~~~~~~~~~~$c_{\epsilon p} = (1-\epsilon_p)^{n}$ & \citet{Richardson1954}\\
        \hline		
        \pagebreak
            \SetCell[c=2]{l}~~~~~~~~~~Values of $n$ extracted from \citet{Richardson1954} and deemed suitable for the present study\\

            ~~~~~~~~~~$n = \begin{cases}
        		4.65 & \hspace{0.5cm} Re_p=0\\
        		4.64 & \hspace{0.5cm} 0<Re_p<0.2\\
        		4.35Re_p^{-0.03} & \hspace{0.5cm} 0.2\leq Re_p<1\\
        		4.45Re_p^{-0.1} & \hspace{0.5cm} Re_p\geq 1\\
        	\end{cases}$ & \citet{Richardson1954}\\

            \hline

\end{longtblr}
\end{adjustwidth}

\section{Detailed velocity}
\label{sec:velo_dist_app}
\setcounter{table}{0}
\renewcommand{\thetable}{B.\arabic{table}}

Tables \ref{tab:veloP} and \ref{tab:veloB} provide detailed information on the statistical measures of the particle and bubble velocity distribution, respectively.
First, the results of DNS are reported.
The respective simulations for the gravity-driven cases are described in \citet{Tiedemann2024b}.
In tables 4 and 5 of this reference only a selected subset of simulations was considered for conciseness.
Here, the entire set of simulations was used. as needed for evaluating the IMSC.
An own simulation with additional turbulent forcing was reported in \citet{Tiedemann2024a} providing the values reported here for R53-1-30.
The mean velocity $u_\alpha$, the velocity variance variance $\sigma^{2}_\alpha$, skewness $\gamma_\alpha$, and kurtosis $\beta_\alpha$ obtained from the DNS are shown.

Second, a comparison of the overall variance obtained from the IMSC is made. 
From the simulations, three one-dimensional velocity variances corresponding to all spatial directions are obtained.
However, the IMSC provides only a single variance that combines all three spatial directions.
Hence, the result of the IMSC and the variances obtained from the DNS are not directly comparable.
As in the IMSC homogenous and isotropic turbulence is used, the single combined variance of the IMSC $\sigma^2_{\alpha 3\mathrm{D}}$ can be subdivided into three equal contributions, each being $\sigma_{\alpha 1\mathrm{D}}^2=\sigma^2_{\alpha 3\mathrm{D}}/3$.
As the velocity distributions in $x$- and $z$-direction are identical up to more than $2$ digits, only the quantities for the $x$-direction are reported.
Recall that for a Gaussian distribution $\gamma=0$ and $\beta=3$.

    \begin{table}
		\setlength\extrarowheight{-15pt}
		\centering
		\caption{Statistical measures of the particle velocity distribution $P_{\bm{u}_p}$ in each spatial direction obtained from the DNS for the cases in table \ref{tab:sim_cases}. Reported are the mean particle velocity $\bm{u}_p$ in terms of its horizontal and vertical components, variance $\sigma^2_p$, skewness $\gamma_p$, and kurtosis $\beta_p$. As the velocity distributions in $x$- and $z$-direction are identical up to more than two digits, only the quantities for the $x$-direction are reported (data partially reproduced from \citet{Tiedemann2024a, Tiedemann2024b}).}
        \begin{adjustwidth}{-0.8cm}{}

		\begin{tabular}{l|l|S[tight-spacing=true,table-format=-1.5]|S[tight-spacing=true,table-format=-1.5]|S[tight-spacing=true,table-format=-1.5]|S[tight-spacing=true,table-format=-1.5]|S[tight-spacing=true,table-format=-1.5]|S[tight-spacing=true,table-format=-1.5]|S[tight-spacing=true,table-format=-1.5]}
			\toprule
			\multicolumn{2}{c|}{} &{\begin{sideways}G-1-30 \end{sideways} } 
			&{\begin{sideways}G-1-50 \end{sideways} }
			&{\begin{sideways}G-1.4-30 \end{sideways}}
			&{\begin{sideways}G-1.4-50 \end{sideways}  }  
			&{\begin{sideways}G-2.4-50 \end{sideways} }   
			&{\begin{sideways}G-2.4-70 \end{sideways}  }  
			&{\begin{sideways}G-0.6-30 \end{sideways}  }  
			\\ \midrule
			\multirow{8}{*}{DNS} & $\langle u_{px} \rangle$ &0.00024&0.00050&-0.00027&0.00072&-0.00098&0.0054&0.0039\\ 
			& $\langle u_{py} \rangle /\urms$ &-0.32&-0.69&-0.41&-0.59&-0.57&-0.77&-0.51\\
			\cmidrule{2-9}
            & $ \sigma^{2}_{px}/\urms^2$ &0.40&0.44&0.45&0.45&0.55&0.58&0.35\\
			& $ \sigma^{2}_{py}/\urms^2$ &2.1&2.0&1.9&1.8&1.4&1.3&2.1\\
			\cmidrule{2-9}
			& $\gamma_{px}$ &-0.00039&0.0059&-0.00088&-0.00092&0.0080&0.00233&0.020\\ 
			& $\gamma_{py}$ &0.47&0.45&0.60&0.60&0.81&0.80&0.33\\ 
			 \cmidrule{2-9}
			& $\beta_{px}$ &4.0&4.0&3.8&3.8&3.6&3.6&3.9\\ 
			& $\beta_{py}$ &3.0&4.4&3.3&3.4&3.9&4.1&2.9\\
			\midrule
			IMSC & ${\sigma^{2\mathrm{(ISMC)}}_{p3\mathrm{D}}}/(3\urms^2)$ &0.25&0.25&0.29&0.29&0.36&0.36&0.18\\ 
			\bottomrule
		\end{tabular}
        \end{adjustwidth}
        \begin{adjustwidth}{-0.8cm}{}

		\begin{tabular}{l|l|S[tight-spacing=true,table-format=-1.5]|S[tight-spacing=true,table-format=-1.5]|S[tight-spacing=true,table-format=-1.5]|S[tight-spacing=true,table-format=-1.5]|S[tight-spacing=true,table-format=-1.5]|S[tight-spacing=true,table-format=-1.5]|S[tight-spacing=true,table-format=-1.5]}
			\toprule
			\multicolumn{2}{c|}{} &{\begin{sideways}G-1-30-eg6 \end{sideways} }   
			&{\begin{sideways}G-1-30-eg16 \end{sideways} }  
			&{\begin{sideways}G-1-30-rb25 \end{sideways}   } 
			&{\begin{sideways}G-1-30-rp6000 \end{sideways} }
			&{\begin{sideways}G-1-30-ep7.5 \end{sideways} }
			&{\begin{sideways}G-1-30-ep5.0 \end{sideways} }
			&{\begin{sideways}R53-1-30 \end{sideways} }  

			\\ \midrule
			\multirow{8}{*}{DNS} & $\langle u_{px} \rangle$ &0.00047&0.00045&0.0018&-0.0012&0.00008&0.00031&-0.00088 \\ 
			& $\langle u_{py} \rangle /\urms$ &-0.37&-0.57 &-0.47&-0.79&-0.59&-0.43& -0.05\\
			\cmidrule{2-9}
            & $ \sigma^{2}_{px}/\urms^2$ &0.26&0.47 &0.43&0.40&0.40&0.38 &1.2\\
			& $ \sigma^{2}_{py}/\urms^2$ &2.1&1.5 &2.3&1.9&2.1&2.0 &1.1\\
			 \cmidrule{2-9}
			& $\gamma_{px}$ &0.016&0.00002 &-0.0012&-0.0096&-0.0076&-0.0055 &0.020\\ 
			& $\gamma_{py}$ &0.67&0.16 &0.47&0.46&0.47&0.50 &-0.009\\ 
			 \cmidrule{2-9}
			& $\beta_{px}$ &4.3&3.6 &4.0&4.0&4.0&4.0& 2.8\\ 
			& $\beta_{py}$ &3.4&2.6 &2.9&3.0&3.0&3.0 &2.9\\
			 \midrule
			IMSC & ${\sigma^{2\mathrm{(ISMC)}}_{p3\mathrm{D}}}/(3\urms^2)$ &0.24&0.25 &0.24&0.25&0.25&0.25&1.0 \\ 
			\bottomrule
		\end{tabular}
        \end{adjustwidth}
		\label{tab:veloP}
	\end{table}

	\newpage
	
	\begin{table}
	  	\setlength\extrarowheight{-15pt}
		\centering
		\caption{Statistical measures of the bubble velocity distribution $P_{\bm{u}_b}$ in each spatial direction obtained from DNS for the cases in table \ref{tab:sim_cases}. Reported are the mean bubble velocity $\bm{u}_b$ in terms of its horizontal and vertical components, variance $\sigma^{2}_b$, skewness $\gamma_b$, and kurtosis $\beta_b$. As the velocity distributions in $x$- and $z$-direction are identical up to more than two digits, only the quantities for the $x$-direction are reported (data partially reproduced from \citet{Tiedemann2024a, Tiedemann2024b}).}
        \begin{adjustwidth}{-0.8cm}{}

		\begin{tabular}{l|l|S[tight-spacing=true,table-format=-1.5]|S[tight-spacing=true,table-format=-1.5]|S[tight-spacing=true,table-format=-1.5]|S[tight-spacing=true,table-format=-1.5]|S[tight-spacing=true,table-format=-1.5]|S[tight-spacing=true,table-format=-1.5]|S[tight-spacing=true,table-format=-1.5]}
			\toprule
			\multicolumn{2}{c|}{} &{\begin{sideways}G-1-30 \end{sideways} } 
			&{\begin{sideways}G-1-50 \end{sideways} }
			&{\begin{sideways}G-1.4-30 \end{sideways}}
			&{\begin{sideways}G-1.4-50 \end{sideways}  }  
			&{\begin{sideways}G-2.4-50 \end{sideways} }   
			&{\begin{sideways}G-2.4-70 \end{sideways}  }  
			&{\begin{sideways}G-0.6-30 \end{sideways}  }  
			\\ \midrule
			\multirow{8}{*}{DNS} & $\langle u_{bx} \rangle $    &-0.0013&-0.00038& 0.0023&-0.00025&-0.00094&-0.0030&0.0017\\ 
			& $\langle u_{by} \rangle /\urms$ &2.8&2.8& 2.7&2.8&2.7&2.8&2.9\\
			\cmidrule{2-9}
            & $ \sigma^{2\,(\mathrm{(DNS)}}_{bx}/\urms^2$ &0.33&0.32& 0.35&0.36&0.41&0.43&0.22\\
			& $ \sigma^{2\,(\mathrm{(DNS)}}_{by}/\urms^2$ &0.66&0.68& 0.67&0.63&0.50&0.53&0.65\\
			\cmidrule{2-9}
			& $\gamma_{bx}$ &0.024&0.0043& 0.0059&-0.011&-0.012&0.014&0.024\\ 
			& $\gamma_{by}$ &-0.0014&0.027& 0.085&-0.0070&0.011&0.015&0.056\\ 
			\cmidrule{2-9}
			& $\beta_{bx}$ &3.3&3.1& 3.0&3.1&3.1&3.0&3.6\\ 
			& $\beta_{by}$ &2.9&2.9& 2.9&3.0&2.9&2.9&3.0\\
			\midrule
			IMSC & ${\sigma^{2\mathrm{(ISMC)}}_{b3\mathrm{D}}}/(3\urms^2)$ 
			&0.28 &0.28 & 0.37 &0.37 &0.48 &0.47 &0.18 \\ \bottomrule
		\end{tabular}
        \end{adjustwidth}
        \begin{adjustwidth}{-0.8cm}{}
		\begin{tabular}{l|l|S[tight-spacing=true,table-format=-1.5]|S[tight-spacing=true,table-format=-1.5]|S[tight-spacing=true,table-format=-1.5]|S[tight-spacing=true,table-format=-1.5]|S[tight-spacing=true,table-format=-1.5]|S[tight-spacing=true,table-format=-1.5]|S[tight-spacing=true,table-format=-1.5]}
			\toprule
			\multicolumn{2}{c|}{} &{\begin{sideways}G-1-30-eg6 \end{sideways} }   
			&{\begin{sideways}G-1-30-eg16 \end{sideways} }  
			&{\begin{sideways}G-1-30-rb25 \end{sideways}   } 
			&{\begin{sideways}G-1-30-rp6000 \end{sideways} }
			&{\begin{sideways}G-1-30-ep7.5 \end{sideways} }
			&{\begin{sideways}G-1-30-ep5.0 \end{sideways} }
			&{\begin{sideways}R53-1-30 \end{sideways} }  
			\\ \midrule
			\multirow{8}{*}{DNS} & $\langle u_{bx} \rangle$    &0.0036&0.0010&-0.0022&0.0054&-0.0022&-0.0029 &-0.0089\\ 
			& $\langle u_{by} \rangle /\urms$ &3.4&2.1&3.0&2.8&2.8&2.8 &0.27\\
			 \cmidrule{2-9}
           & $ \sigma^{2}_{bx}/\urms^2$ &0.36&0.26&0.37&0.33&0.32&0.30 &0.88\\
			& $ \sigma^{2}_{by}/\urms^2$ &0.69&0.51&0.78&0.69&0.67&0.63 &0.80\\
			\cmidrule{2-9}
			& $\gamma_{bx}$ &0.048&0.0031&-0.026&-0.029&0.0025&-0.0035 &-0.0038\\ 
			& $\gamma_{by}$ &0.098&-0.048&0.056&0.061&0.046&0.081 &0.052\\ 
			\cmidrule{2-9}
			& $\beta_{bx}$ &3.3&3.1&3.1&3.1&3.1&3.2 &2.9\\ 
			& $\beta_{by}$ &3.0&3.0&3.0&2.8&3.0&3.0 &3.1\\
			 \midrule
			IMSC & ${\sigma^{2\mathrm{(ISMC)}}_{b3\mathrm{D}}}/(3\urms^2)$ 
			&0.27 &0.28&0.26 &0.28 &0.28 &0.29  &1.3\\  \bottomrule
		\end{tabular}
        \end{adjustwidth}
		\label{tab:veloB}
	\end{table}

\clearpage
		
\newpage

\bibliographystyle{jfm}
\bibliography{bibliography}

\end{document}